  \documentclass{aa}
   \usepackage{graphicx}
\begin{document}

  %\thesaurus{03(....... 3C345)}

  \title{Flux evolution and kinematics of superluminal components 
   in blazar 3C345}
     
  \author{S.J.~Qian\inst{1}}
     %{1,2} \and S.Britzen\inst{1}
    %\and A.~Witzel\inst{1} }
   % \and S.~Britzen\inst{1} \and J.A.~Zensus\inst{1}}
  %\offprints{S.J.~Qian, email: rqsj@bao.ac.cn}
   \institute{
  %Max-Planck-Institut f\"ur Radioastronomie, Auf dem H\"ugel 69,
  %       53121 Bonn, Germany \and
    National Astronomical Observatories, Chinese Academy of Sciences,
    Beijing 100012, China}
   % \and    Max-Planck-Institut f\"ur Radioastronomie, Auf dem H\"ugel 69,
   %     53121 Bonn, Germany}
  % \and    National Astronomical Observatories, Chinese Academy of Sciences,
  %      Beijing 100012, China
  % \and University of Michigan, USA
  % \and  Max-Planck-Institut f\"ur Radioastronomie, Auf dem H\"ugel 69,
  %      53121 Bonn, Germany} 
 \date{Compiled by using A\&A latex}
 \abstract{The precessing jet-nozzle scenario previously proposed was 
    applied to model-fit the kinematics of five superluminal 
    components (C19, C20, C21, B5 and B7)  of jet-B in blazar 3C345.}
    {Based on a specific pattern for the precessing common trajectory of jet-B,
    the kinematic properties (including trajectory, coordinates,
     core separation and apparent velocity) were model-fitted and their 
    flux evolution could be studied.}{Through model-simulation of their
    kinematic behavior, the bulk Lorentz factor, viewing angle and
    Doppler factor were derived as continuous functions
    of time and the association of their flux evolution with their
      Doppler-boosting effect was investigated.}
   {The 43GHz light-curves of the five superluminal components
     can be well interpreted in terms of their Doppler effect. The close 
    association of  their flux evolution with the Doppler-boosting effect
     firmly validate our precessing nozzle scenario and support the traditional
   point-view that superluminal components are physical entities 
   (traveling shocks or plasmoids) participating relativistic motion toward us at small
   viewing angles.}{The model-simulation of kinematic behavior of 
    superluminal components by using our precessing nozzle scenario with
    specific patterns (helical or ballistic) assumed for the precessing common
    trajectories yields the model-derived bulk Lorentz factor, apparent speed,
     viewing angle and Doppler factor as continuous functions of time, 
    which are most applicable to study the connection of flux evolution with
     Doppler boosting effect for the superluminal components.}
  \maketitle
  \section{Introduction} 
   3C345 (z=0.595) is a prototypical quasar which has been detected in all
   wavebands of the entire electromagnetic spectrum from radio/mm, 
   infrared/optical/UV and  X-rays to high-energy $\gamma$ rays 
   (Biretta et al. \cite{Bi86}, Babadzhanyants et al. \cite{Ba95}, 
   Schramm et al. \cite{Sch93}, Moore \& Stockman \cite{Mo84}, 
   Wang et al. \cite{Wa04}, Malina et al. \cite{Ma94}, 
    Lobanov \& Zensus \cite{Lo99}, Schinzel et al. \cite{Sc11b}, 
   Zensus \cite{Ze97}, Ros et al. \cite{Ro00}). 
 %  It is also  one of the best-studied 
 % blazars (e.g., Biretta et al. cite{Bi86}, Hardee et al. cite{Ha87},
 % Steffen et al. cite{St96}, Unwin et al. cite{Un97}, Zensus et al.
 %  cite{Ze97}, Klare cite{Kl03}, Klare et al. cite{Kl05}, Lobanov \& Roland
 %  cite{Lo05}, Jorstad et al. cite{Jo05}, cite{Jo13},
 %   cite{Jo17}, Qian et al. cite{Qi91a}, cite{Qi96}, cite{Qi09},
 %   Schinzel et al. cite{Sc10}, Schinzel cite{Sc11a}, Schinzel et al.
 %   cite{Sc11b}, Homan et al. cite{Ho14}). 
    It exhibits  remarkable and violent multi-waveband variations and
   its spectral energy distribution has been extensively monitored and
    studied, yielding significant results on the physical properties of 
   the source.  Studies of the correlation between the variabilities at
    multi-frequencies (from radio to $\gamma$ rays) play an important role. \\
    3C345 is one of the firstly discovered quasar having a relativistic jet.
    VLBI-observations have revealed its parsec structure and 
    monitored the kinematic behavior of superluminal components ejected
    from the radio core (Qian et al. \cite{Qi91a}, \cite{Qi91b}, \cite{Qi96}, 
   \cite{Qi09}, Klare \cite{Kl03}, Klare et al. \cite{Kl05}, Steffen et al.
    \cite{St95}, Ros et al. \cite{Ro00}, Schinzel et al. \cite{Sc10a},
    Lobanov \& Roland \cite{Lo05},  Lobanov \& Zensus \cite{Lo94}, 
    Zensus et al. \cite{Ze95}, Hardee \cite{Ha87}). 
   The flaring activities of the source at 
    multi-frequencies (from radio to $\gamma$-rays) are closely connected with
    the ejection of superluminal components (e.g., Schinzel \cite{Sc10b}). \\
    VLBI-monitoring observations have also shown that the swing of the knot's 
    ejection position angle could be periodic and caused  by its 
   jet-precession. We have tried to interpret the VLBI-kinematics of 
   superluminal components in 3C345 in terms of a precessing jet-nozzle
   scenario since early years (Qian et al. \cite{Qi91a}, \cite{Qi91b},
    \cite{Qi96}, \cite{Qi09}). Recently, Qian (\cite{Qi22a}) has tried to 
    make model simulations of the kinematic behavior for twenty-seven 
    superluminal components (measured during a 38\,yr period) in detail. 
   It was found that these superluminal knots could be hypothetically separated
    into two groups, which were ejected from a precessing double nozzle
    (jet-A and jet-B), respectively. Based on this division a precession 
    period of $\sim$7.3\,yr was derived for both the jet-nozzles.\\
    In a recent paper (Qian \cite{Qi22b}), the flux evolution of five
    superluminal components (C4, C5, C9, C10 and C22) belonging to  jet-A 
    was well explained in terms of their  Doppler boosting effect combined with
    the intrinsic flux variations of the components.\\
    In this paper we present the results of modelfits to the flux evolution
     for five superluminal components (C19, C20, C21, B5 
    and B7) belonging to jet-B. Before doing so, we firstly give a brief 
    description of the results obtained in the previous work for the five
    components belonging to jet-A (Qian \cite{Qi22b}), taking knot C9 as 
    a representative example.\\
     \begin{figure*}
    \centering
    \includegraphics[width=6cm,angle=-90]{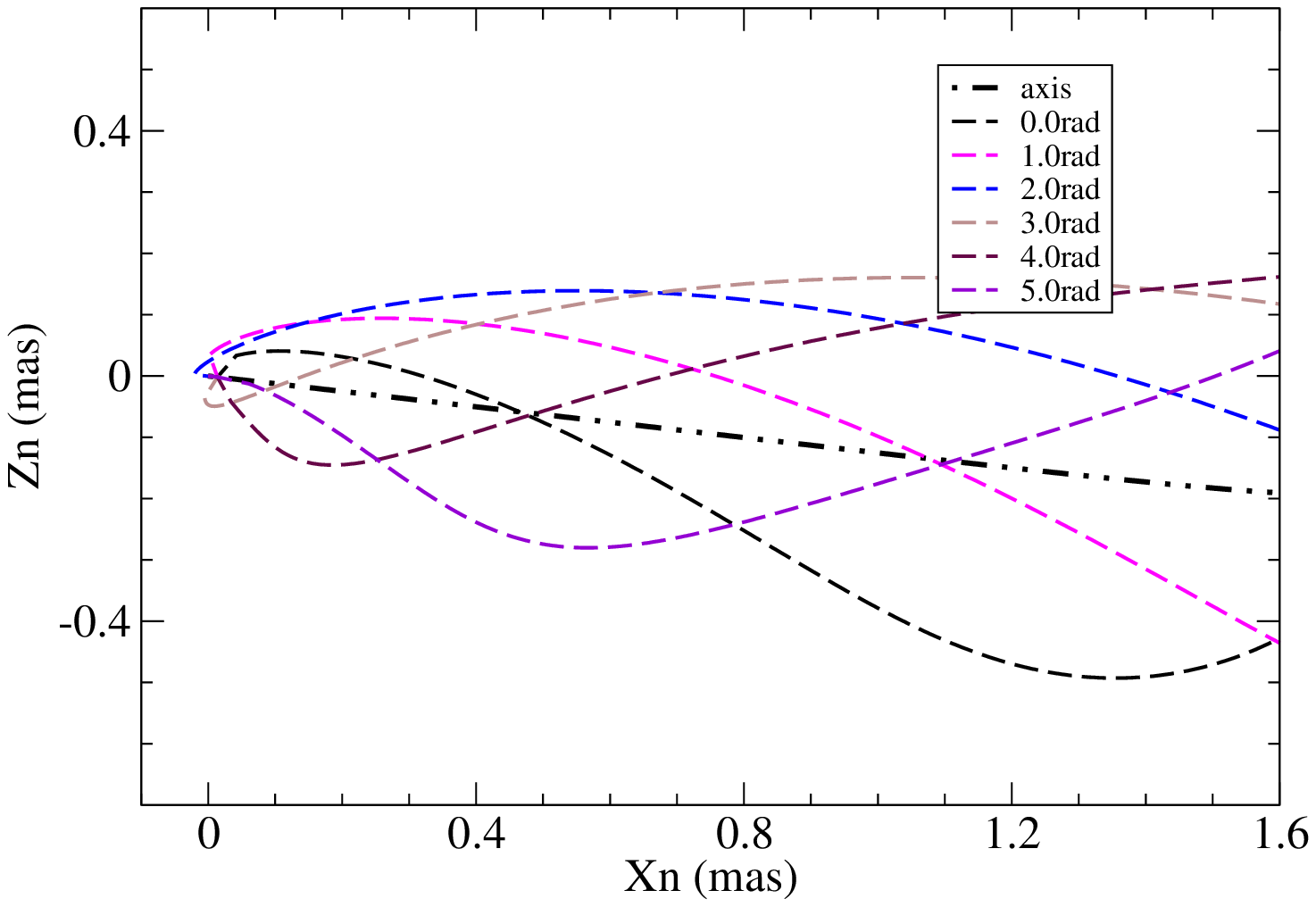}
    \includegraphics[width=6cm,angle=-90]{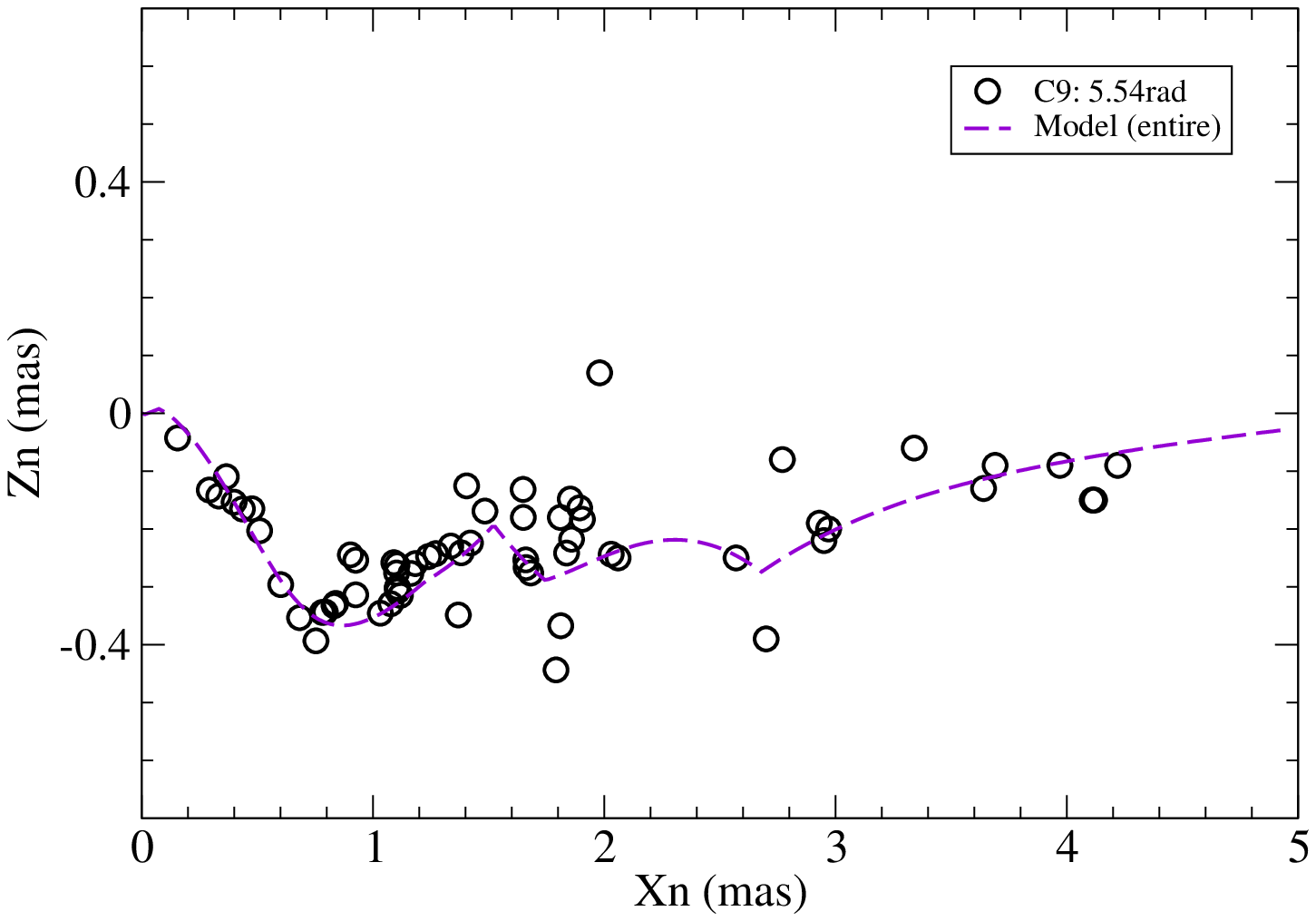}
    \caption{Left panel: Distribution of the precessing trajectories of 
    the superluminal components of jet-A at precession phases $\phi_0$=0.0, 
    1.0, 2.0, 3.0, 4.0 and 5.0\,rad. The jet axis is at position angle 
    of --$97.2^{\circ}$. Jet-A consists of 13 superluminal components (C4
    to C14, C22 and C23). The prominent curvatures of the trajectories are 
    caused by the helical pattern assumed for the precessing common trajectory.
     This distribution demonstrates both the swing of the ejection position angles
    of the components and different curved tracks for different knots ejected
    at different precession phases (or ejection times). Right panel: A 
   model-fit to the entire trajectory of knot C9 (ejection time $t_0$=1995.06
    and precession phase $\phi_0$=5.54+2$\pi$), extending to 
    $X_n{\simeq}$4.16\,mas, equivalent to a traveled  distance 
    Z=106.0\,mas=705\,pc.  Within $X_n$=1.22\,mas ($r_n$=1.25\,mas or 
    before 1999.94) knot C9 moved along the precessing common trajectory, 
   equivalent to its traveled distance Z=44.8\,mas=298\,pc.}
    \end{figure*}
    \begin{figure*}
     \centering
     \includegraphics[width=8cm,angle=-90]{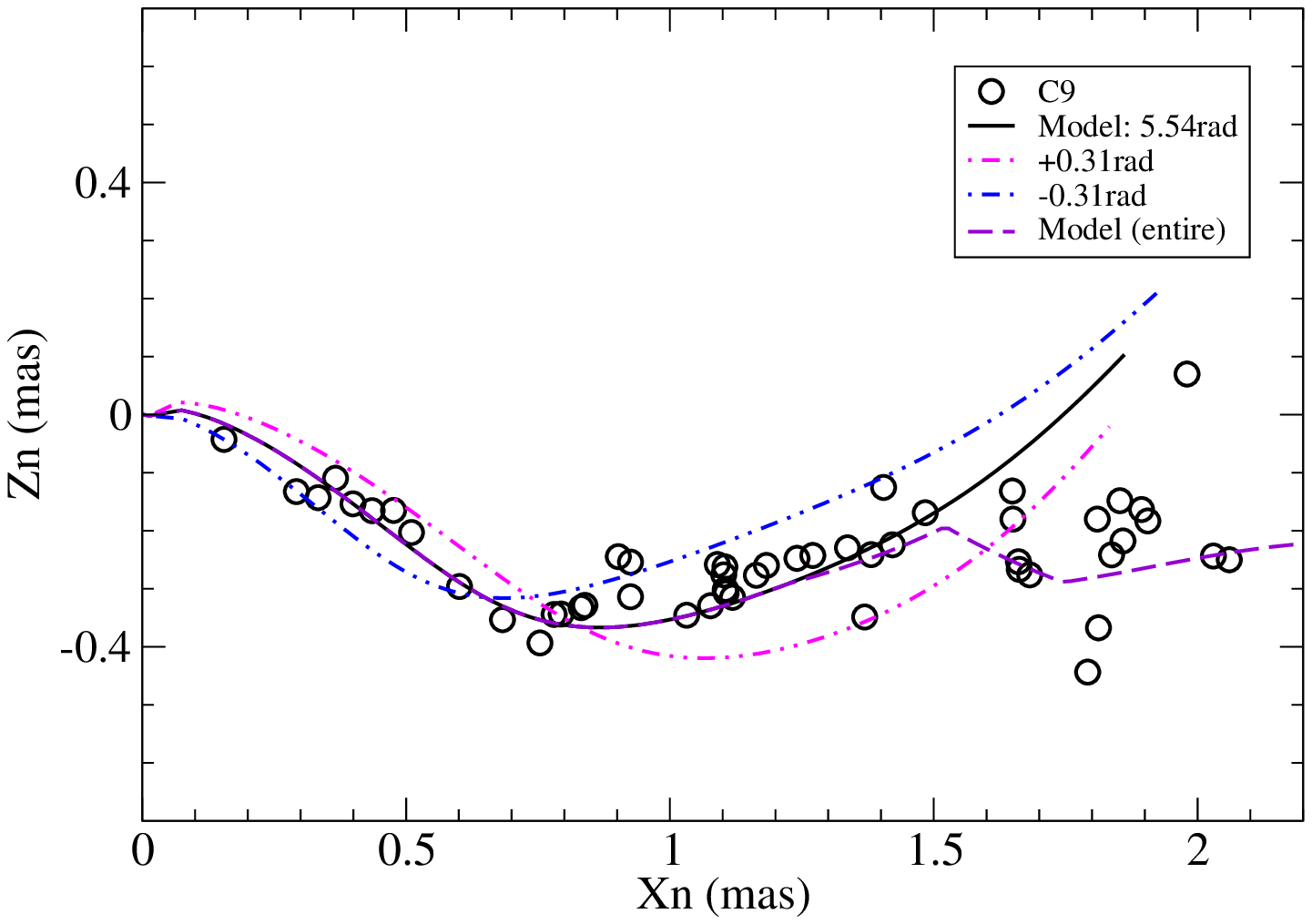}
     \caption{Knot C9: Its inner-jet trajectory ($X_n{\leq}$1.5\,mas) exhibits 
      a prominent curvature induced by the helical pattern assumed for the
      precessing common trajectory. Moreover, most of the data-points are
      very well fitted into an area delimited by the lines in magneta and blue
      which are trajectories calculated for precession phase
      $\phi_0$=5.54${\pm}$0.31\,rad, demonstrating that its precession phase was
      determined with an uncertainty $\pm$5\% of the precession period.}
    \end{figure*}
    \begin{figure*}
    \centering
    \includegraphics[width=6cm,angle=-90]{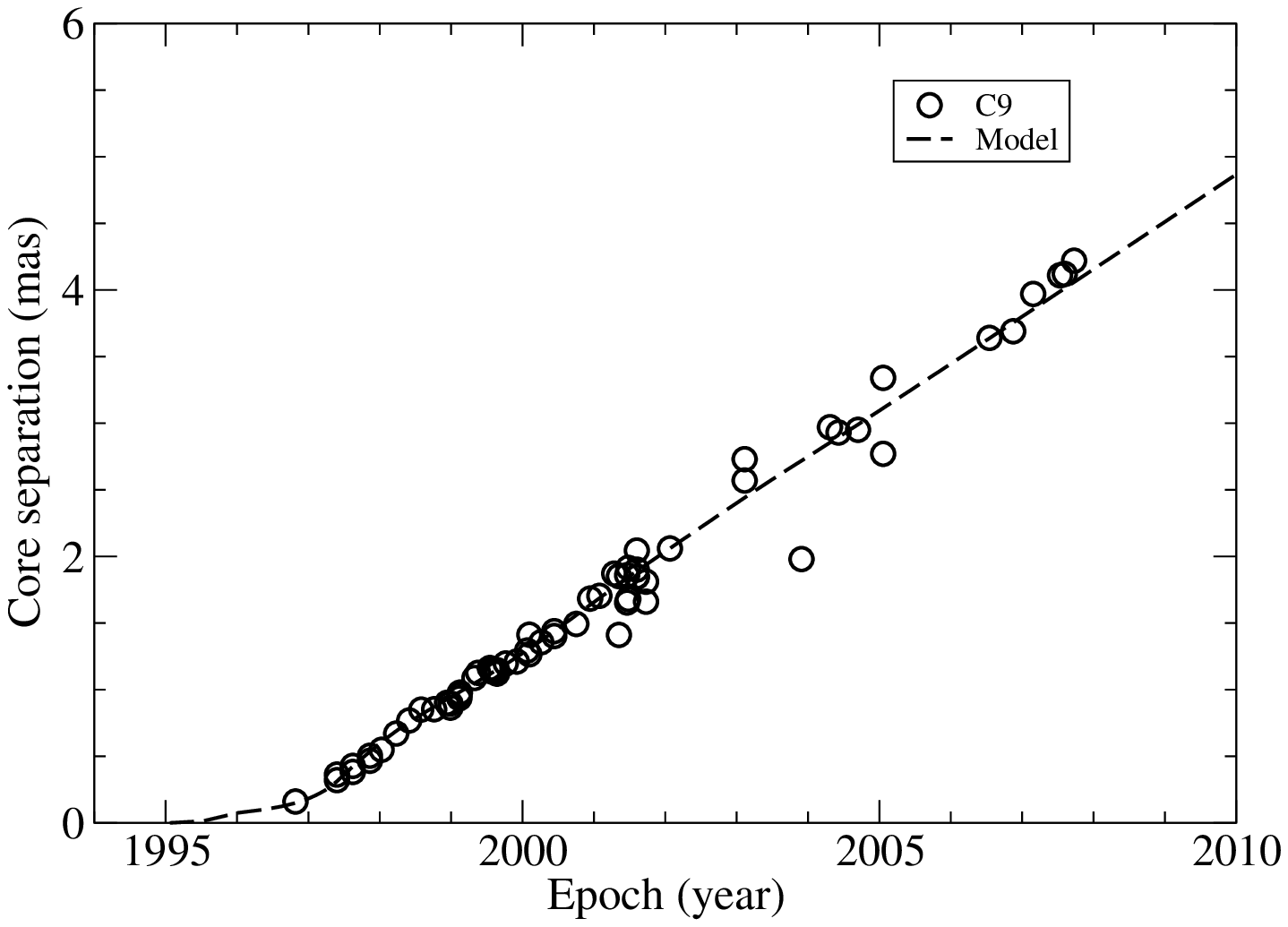}
    \includegraphics[width=6cm,angle=-90]{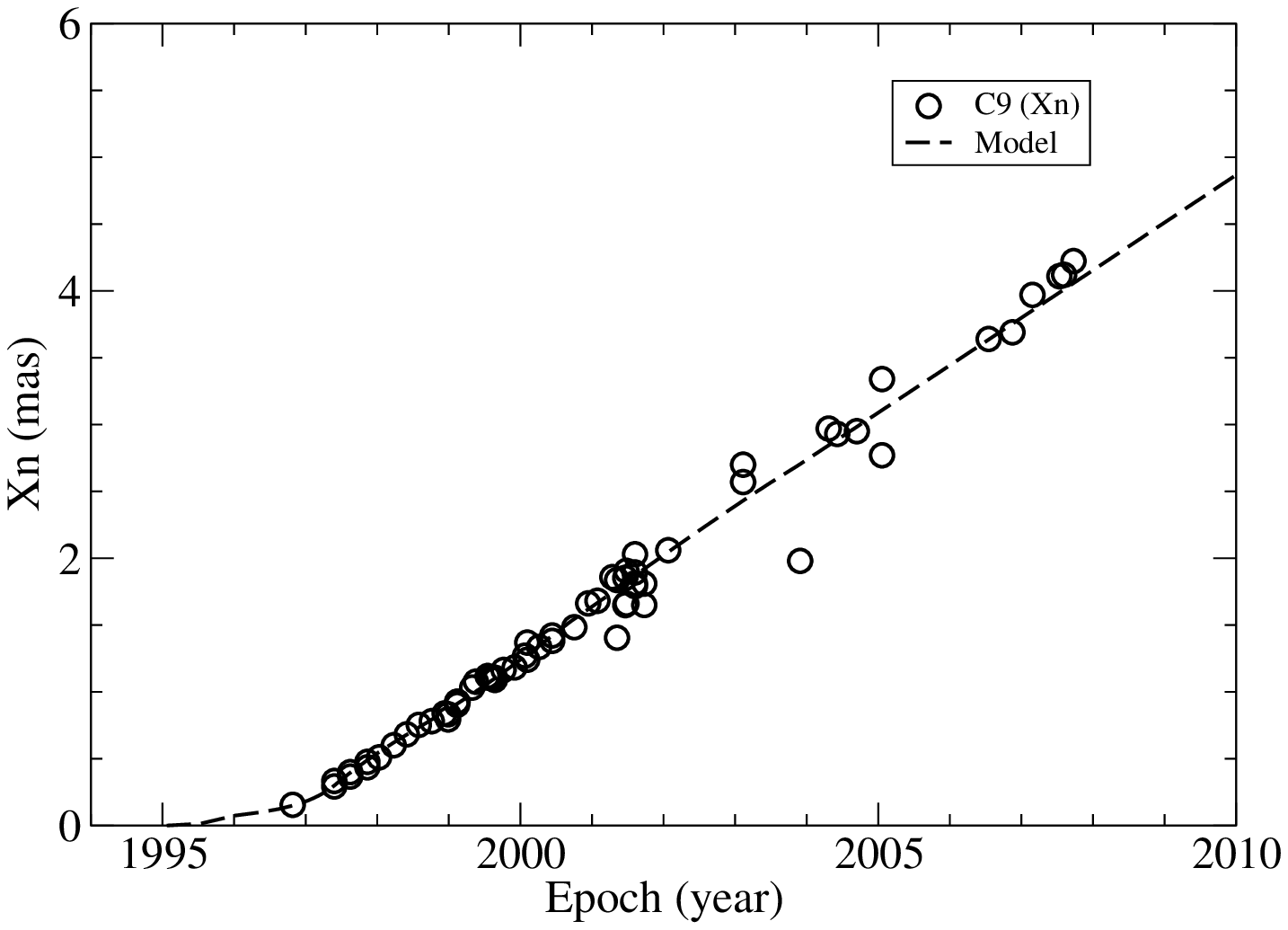}
    \includegraphics[width=6cm,angle=-90]{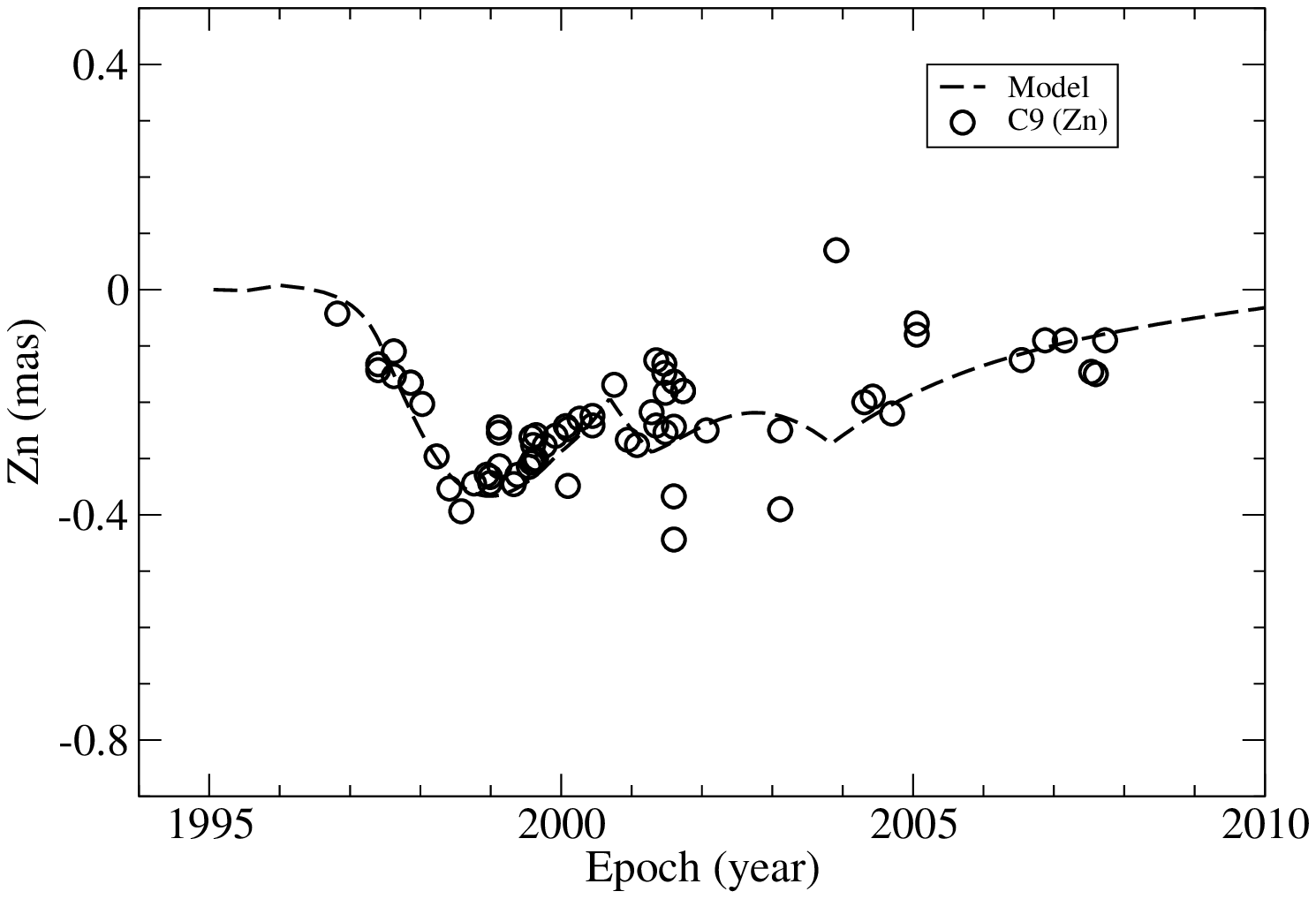}
    \includegraphics[width=6cm,angle=-90]{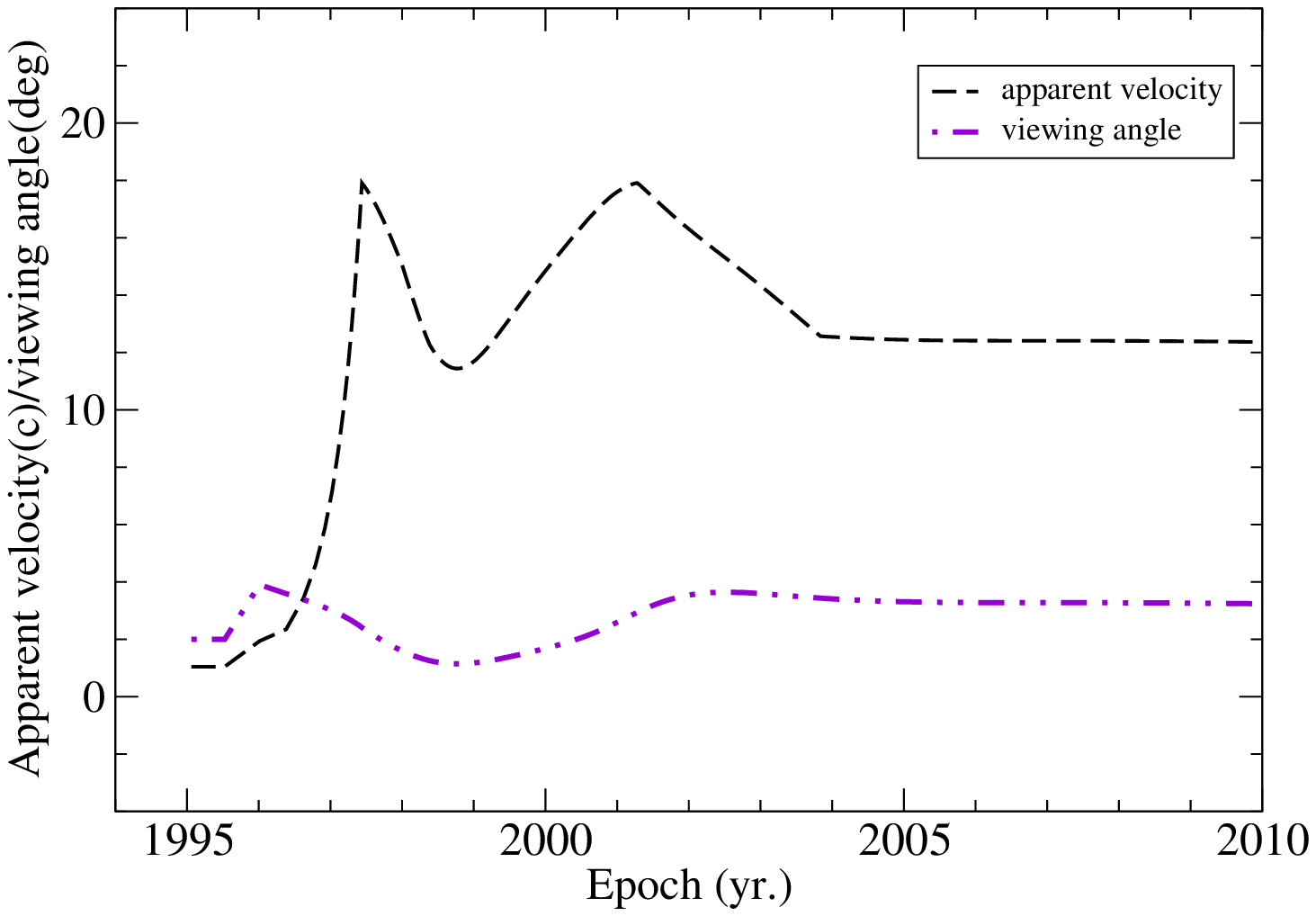}
    \caption{Knot C9: Model fitting of the  core separation $r_n(t)$,
     coordinates $X_n(t)$ and $Z_n(t)$ (upper two and bottom left panels),
      and the model-derived apparent speed $\beta_{app}(t)$ and viewing angle
     $\theta(t)$ (botton right panel) as continuous functions of 
    time. $r_n$, $X_n$ and $Z_n$ are very well fitted, especially the fit to
    the coordinate $Z_n(t)$. The model-derived apparent speed $\beta_{app}$ 
    has an oscillating structure, showing a distinctive pattern of motion:
    acceleration--deceleration--reacceleration--deceleration. The apparent 
    speed has two maxima at 1997.44 and 2001.28: both $\beta_{app,max}$=17.91 and 
    the corresponding viewing angles  $\theta$=$2.41^{\circ}$ and 
    $2.94^{\circ}$, respectively. The apparent speed has a minimum 
    $\beta_{app,min}$=11.44 at 1998.77, with a minimal viewing angle 
    $\theta_{min}$=$1.14^{\circ}$, corresponding to the maximal Doppler factor
     $\delta_{max}$=31.86. }
    \end{figure*}
    \begin{figure*}
    \centering
    \includegraphics[width=8cm,angle=-90]{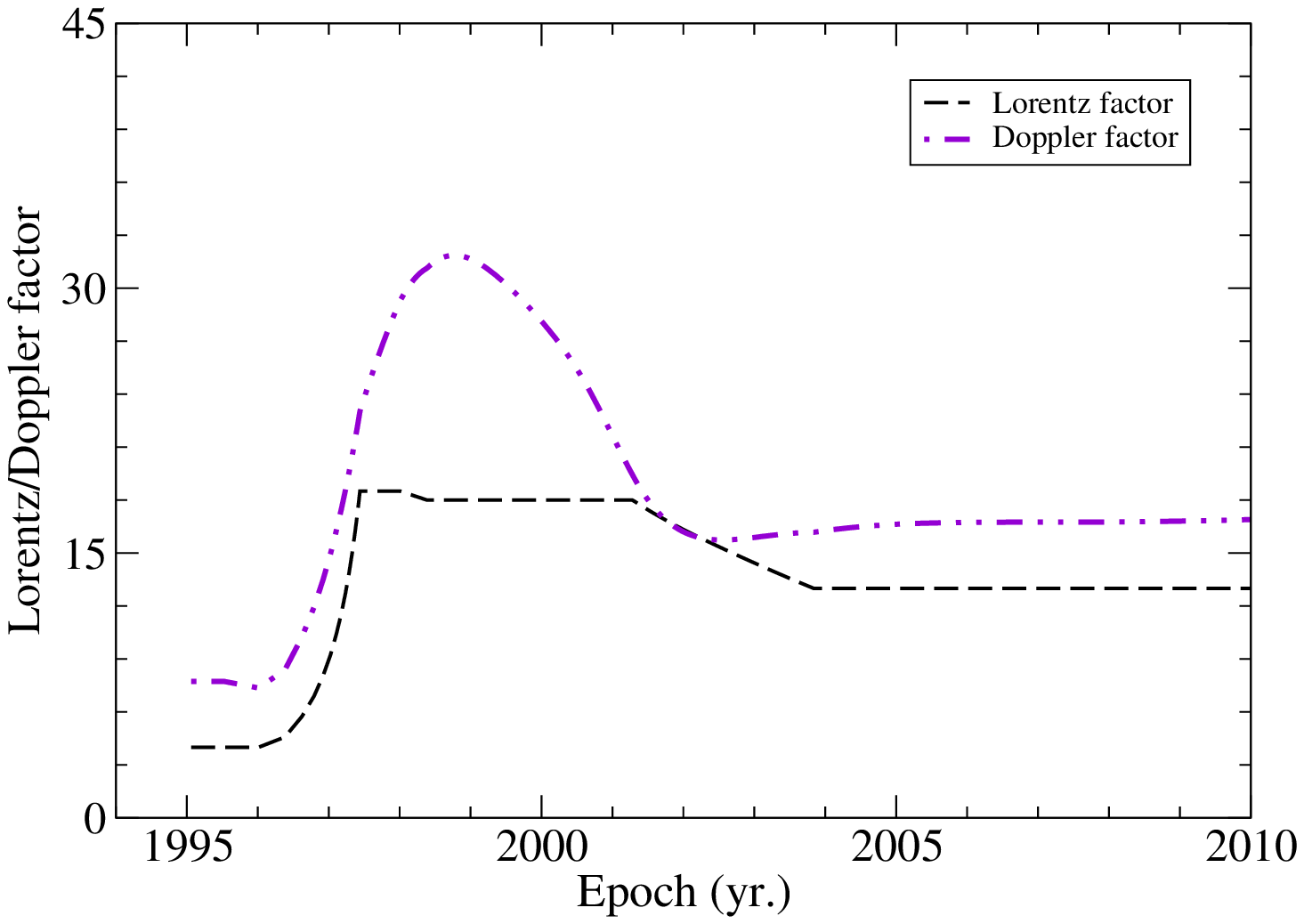}
    \caption{Knot C9: Model-derived bulk Lorentz factor $\Gamma(t)$ and
    Doppler factor $\delta(t)$ as continuous functions of time.
    $\Gamma(t)$ has a plateau structure ($\Gamma$$\sim$18.5--18.0) during 
    $\sim$1997.5--2001.5, while $\delta(t)$ has a bump structure with a peak 
    $\delta_{max}$= 31.86 at 1998.77, when $\Gamma$=18.00 and 
    $\theta$=$1.14^{\circ}$ (a minmum). It is worth noting that the bump 
    structure of Doppler factor is derived completely from the model-fitting 
    of its kinematics only, thus the Doppler boosting effect 
    ($[\delta(t)/\delta_{max}]^{3+\alpha}$) for knot C9 was
    independently derived or  predicted with respect to its
    flux evolution.}
    \end{figure*}
    \begin{figure*}
    \centering
    \includegraphics[width=7cm,angle=-90]{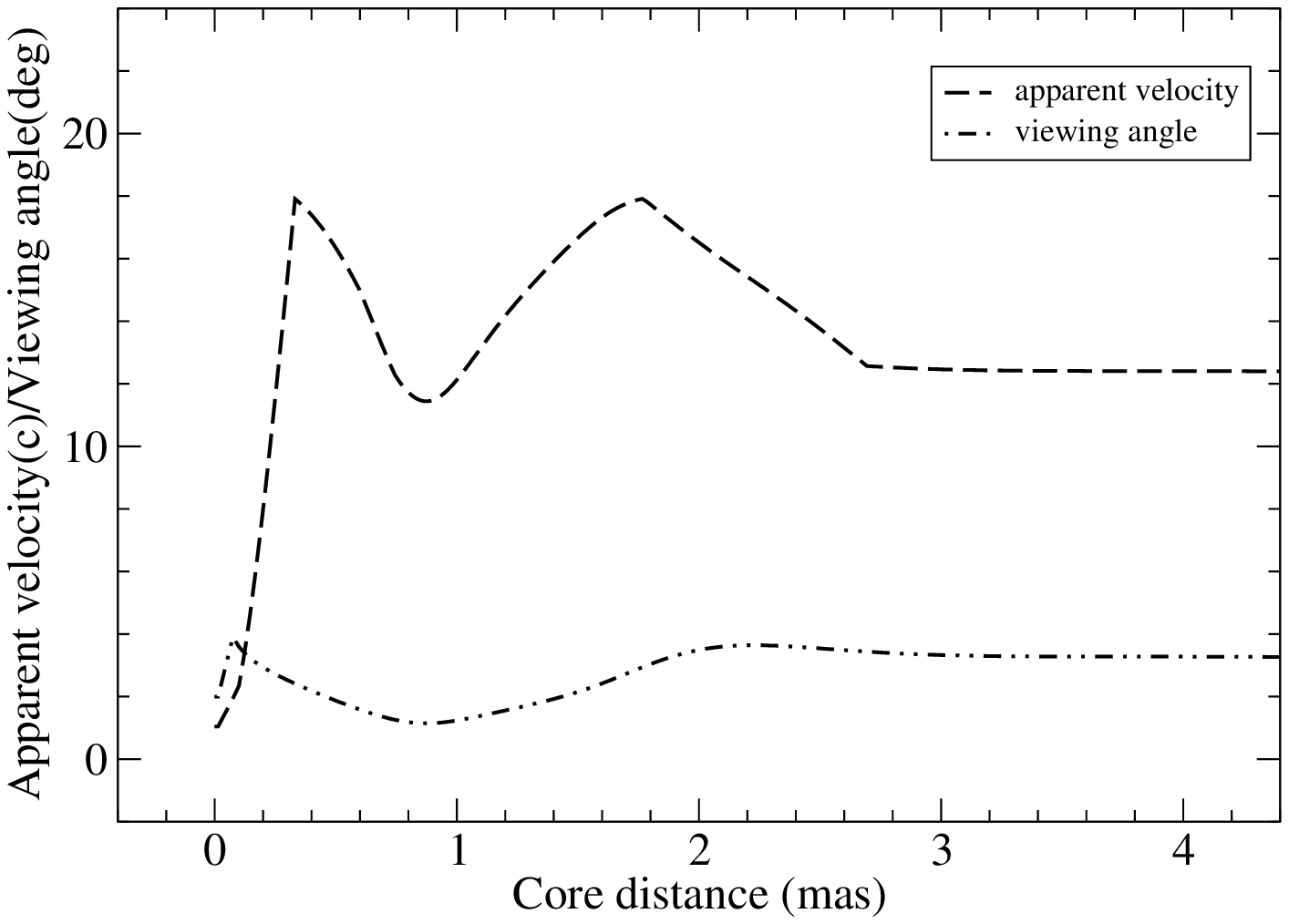}
    \caption{Knot C9: Model-derived apparent speed $\beta_{app}(t)$
     and viewing angle $\theta(t)$ as functions of core separation $r_n$. The
    two maxima in $\beta_{app}(t)$ are at $r_n$=0.33\,mas and $r_n$=1.77\,mas.
    Its minimum is at $r_n$=0.88\,mas. The pattern of motion of knot C9 has
    a remarkable oscillating structure: 
    acceleration--deceleration--reacceleration--deceleration,
    which is fully consistent with the results analysed by Jorstad et al.
    (\cite{Jo05}).} 
      \end{figure*}
    \begin{figure*}
   \centering
    \includegraphics[width=6cm,angle=-90]{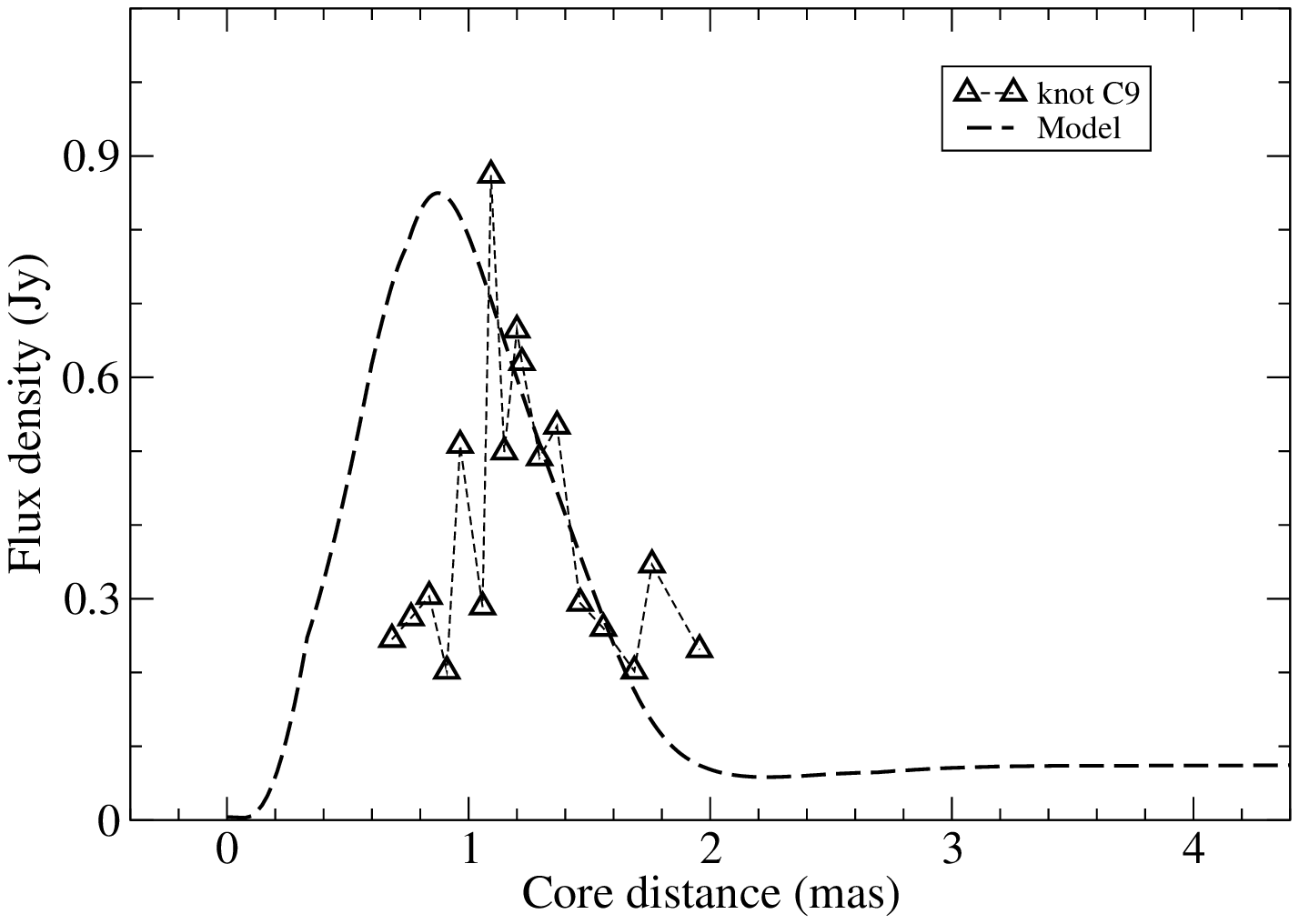}
    \includegraphics[width=6cm,angle=-90]{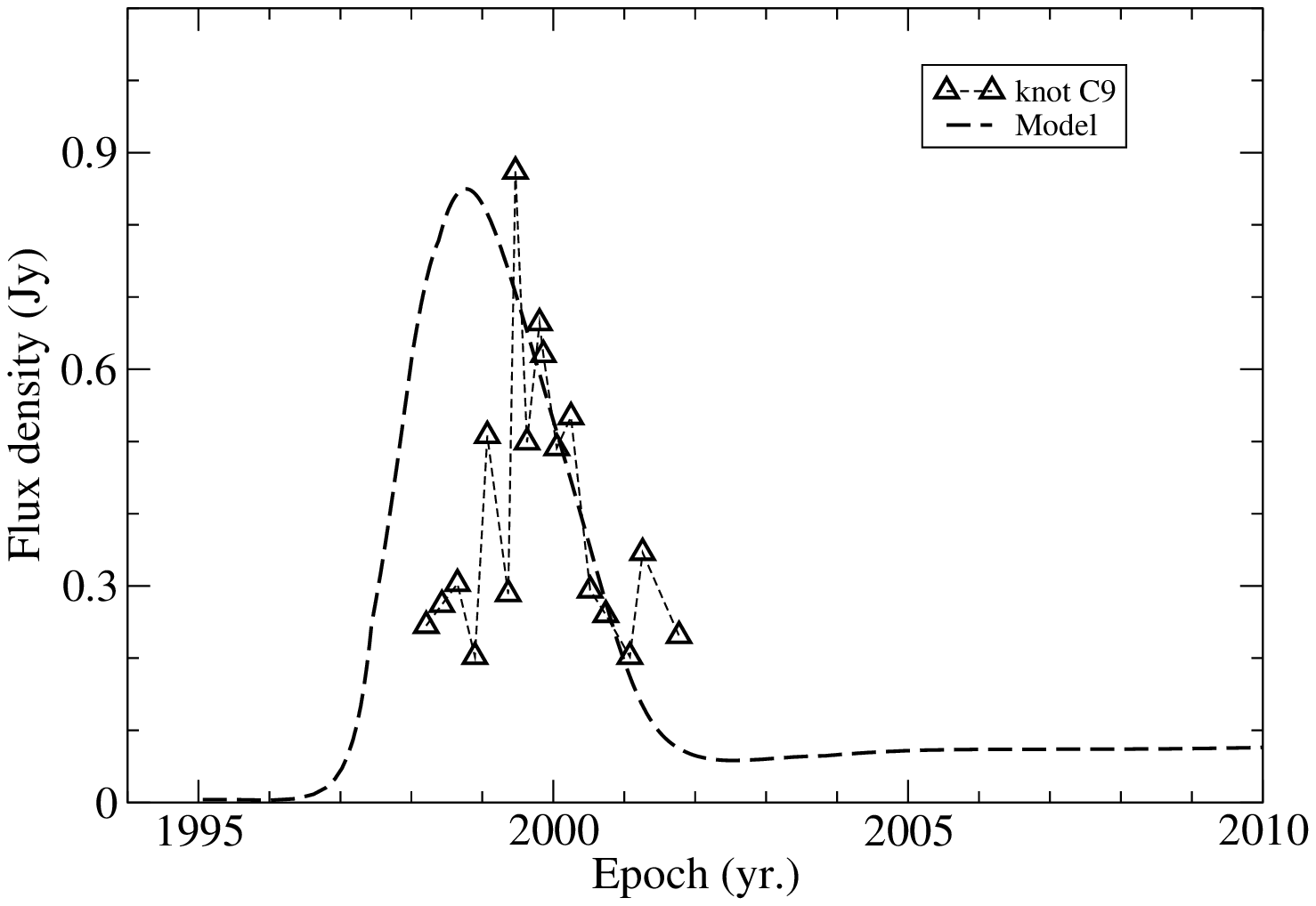}
    \caption{Knot C9: The 43\,GHz light curve $S_{obs}(t)$ (shown by the 
    triangles) as a function of core  distance (left panel) and as a function
    of time (right panel) [the observational data 
    are adopted from Jorstad et al. \cite{Jo05}]. The decaying part of
    the light curve is well fitted by the Doppler boosting profile predicted 
    in our model-simulation of its kinematics (Qian \cite{Qi22b}).}
    \end{figure*}
    \begin{figure*}
    \centering
    \includegraphics[width=6cm,angle=-90]{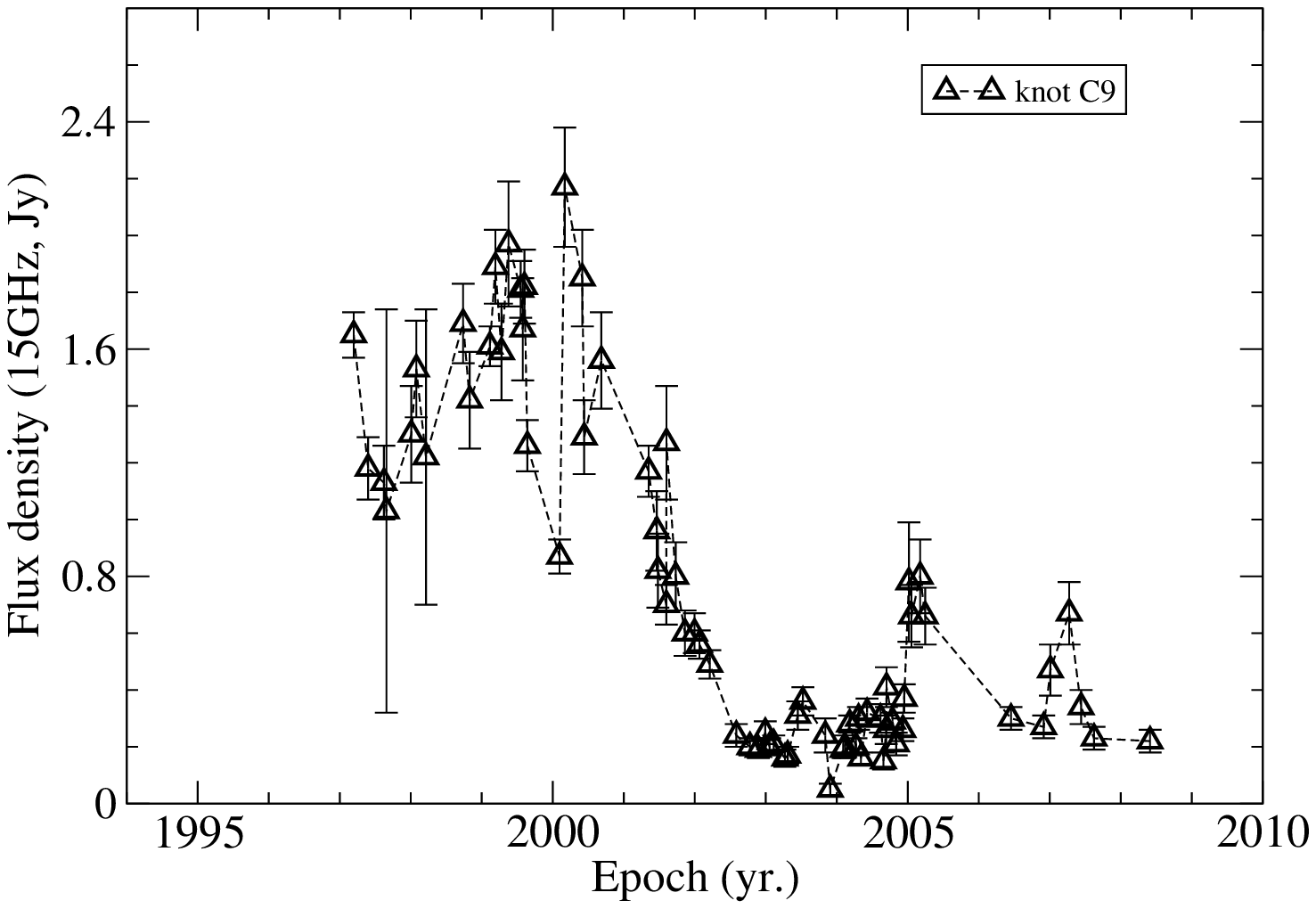}
    \includegraphics[width=6cm,angle=-90]{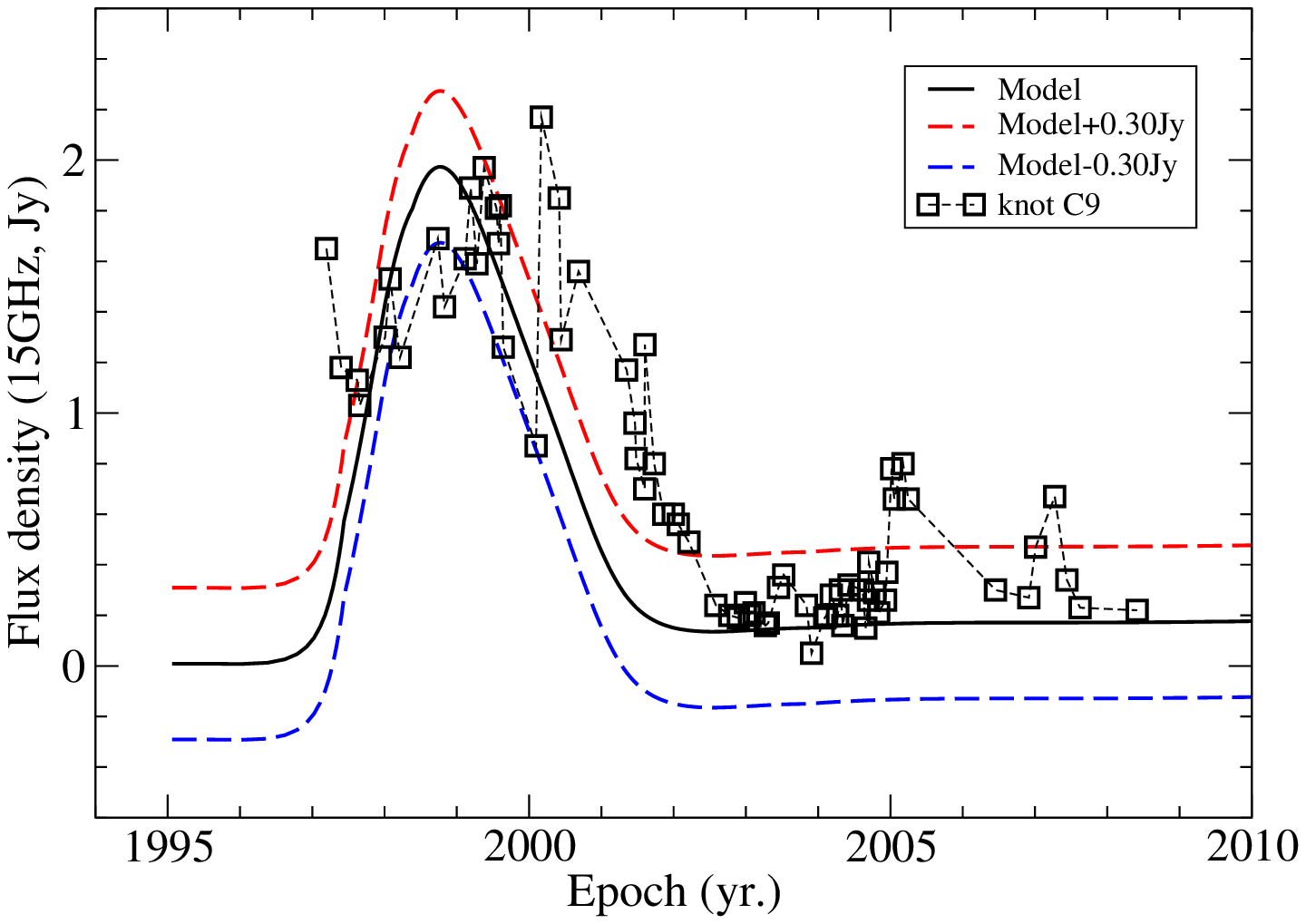}
    \includegraphics[width=6cm,angle=-90]{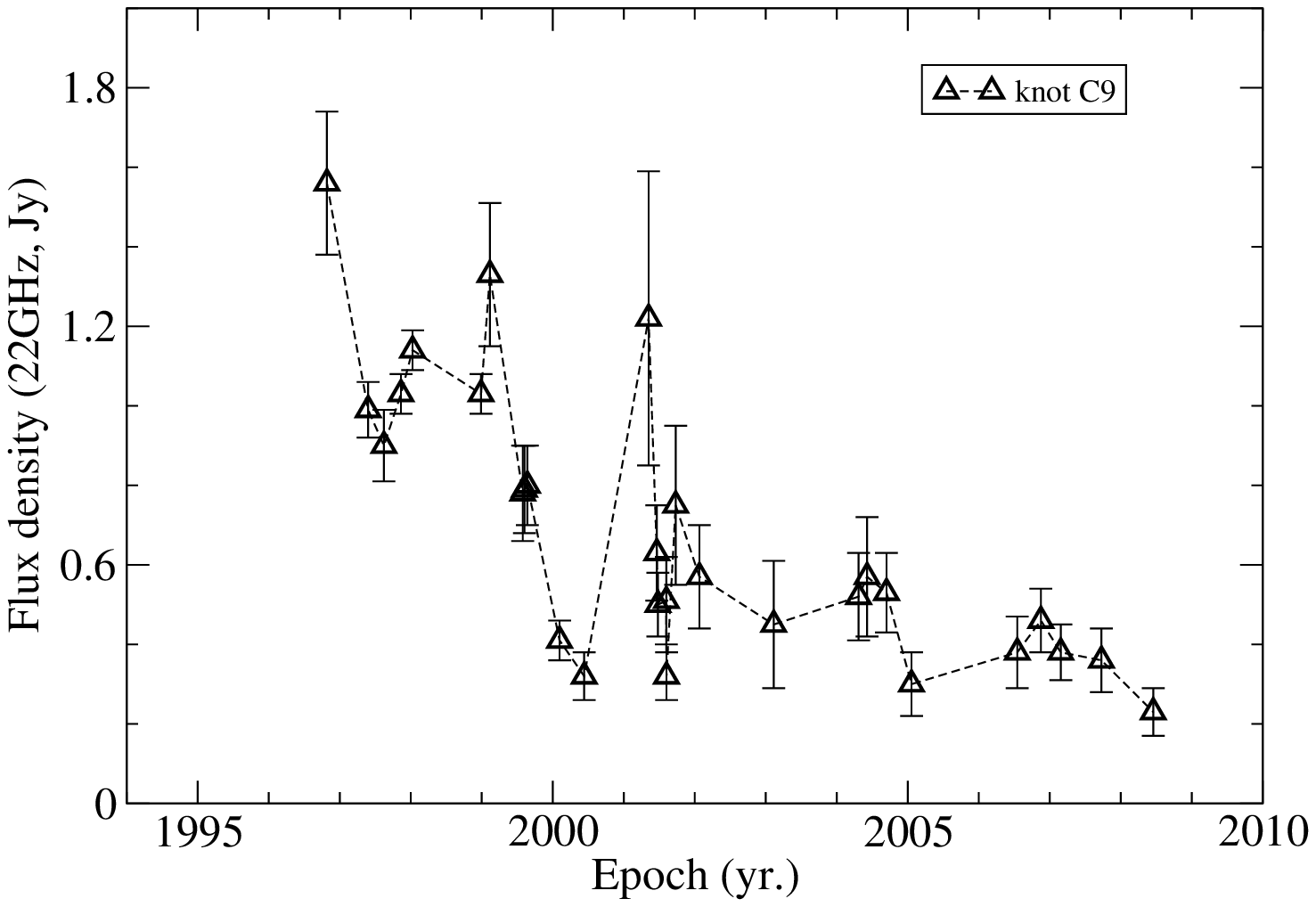}
    \includegraphics[width=6cm,angle=-90]{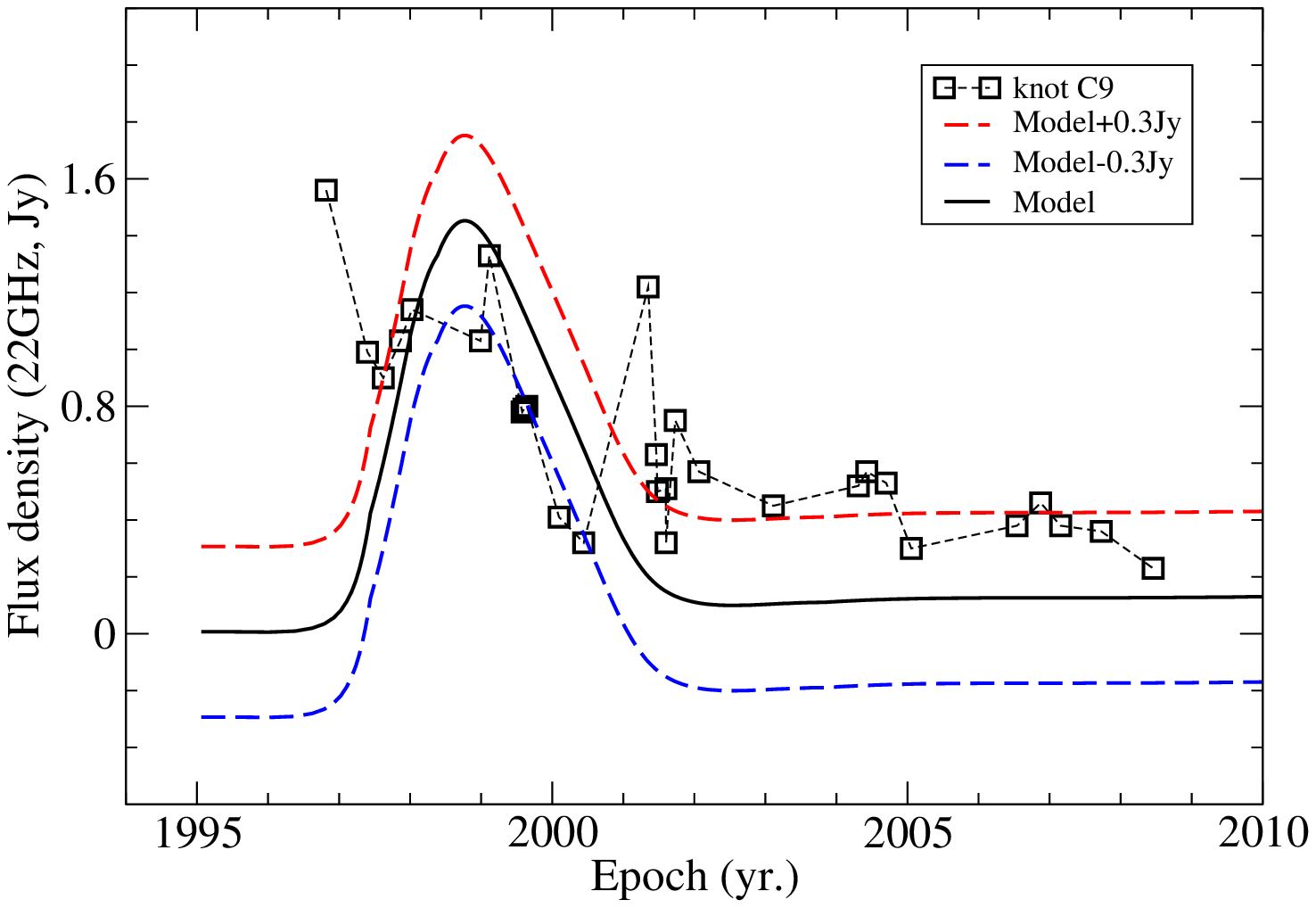}
    \includegraphics[width=6cm,angle=-90]{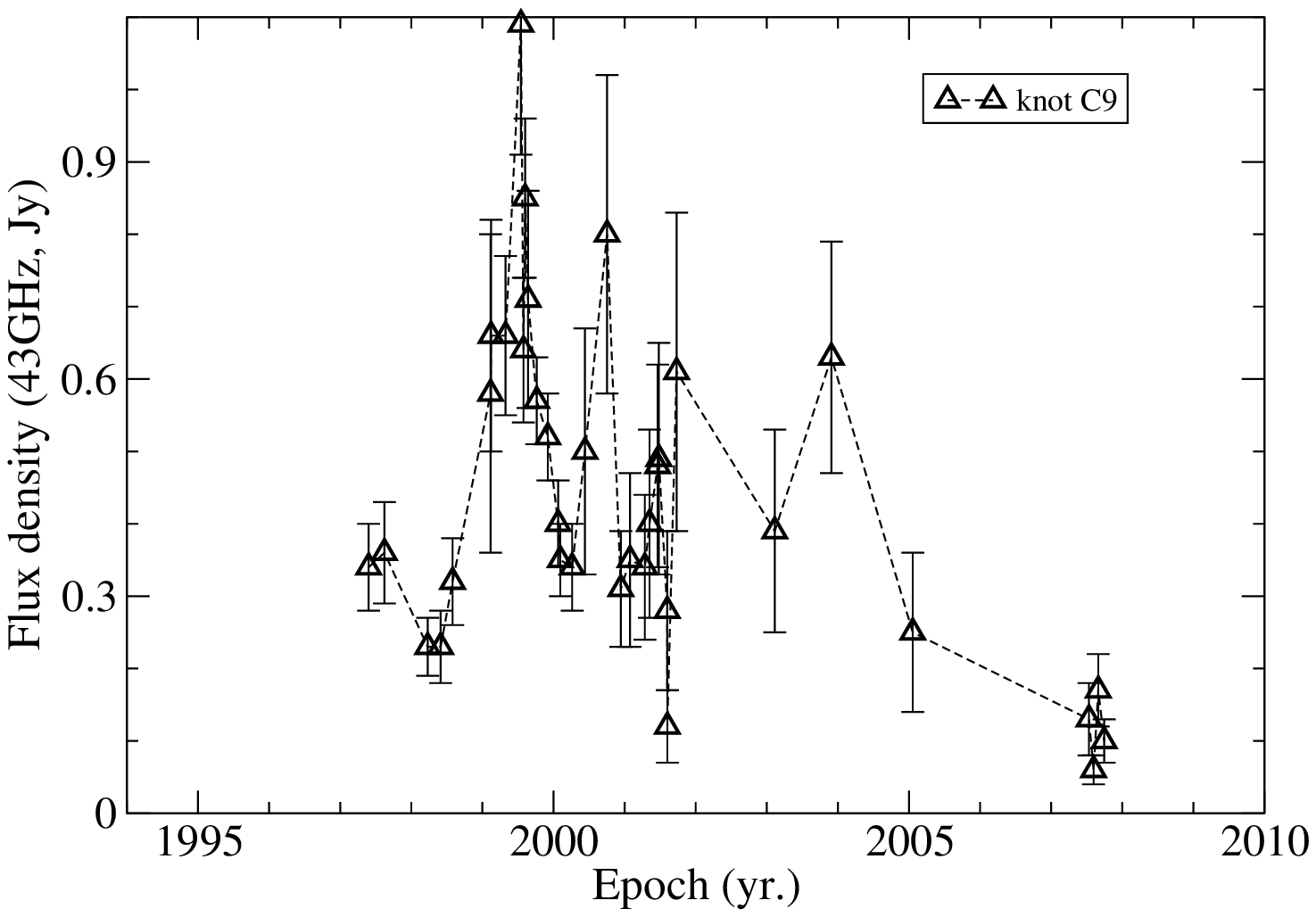}
    \includegraphics[width=6cm,angle=-90]{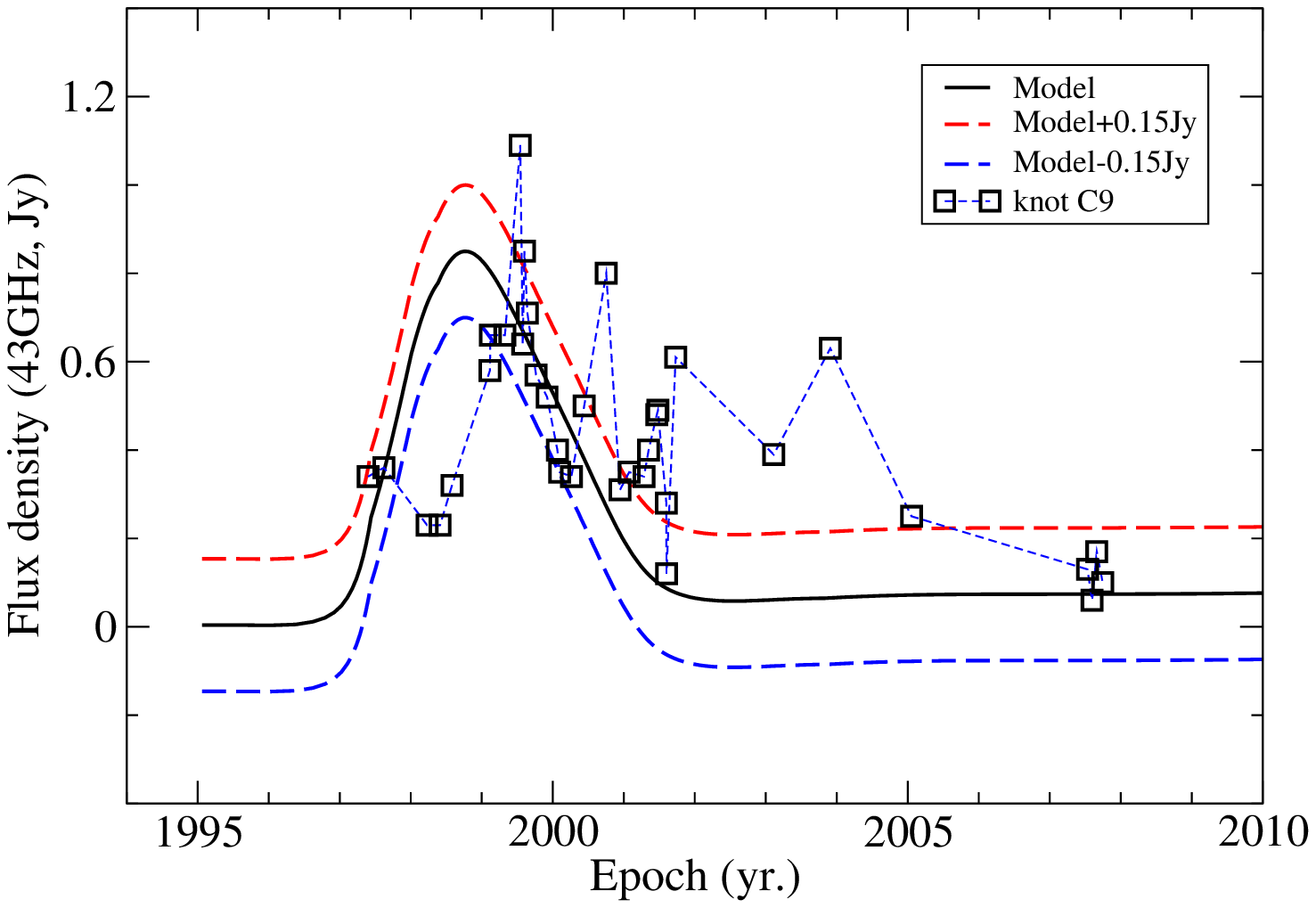}
    \caption{Knot C9: The light curves measured at 15, 22
    and 43\,GHz (left panels) and the model fits  (right panels)
    in terms of their Doppler boosting effect 
    ($S_{obs}(t)$=$S_{int}$$[\delta(t)]^{3+\alpha}$ with $\alpha$(15-22-43GHz)=0.80;
    $S_{\nu}$$\propto{{\nu}^{-\alpha}}$).
    All the three light curves are reasonably well fitted by the model-predicted
    Doppler boosting profiles (solid black lines), except some isolated 
    spikes of short-time scales ($\leq$0.5\,yr) which may be caused by 
   the  variations in the intrinsic flux density of knot C9 itself. The intrinsic 
   flux densities were adopted as $S_{int}$=3.83, 2.58 and 1.65$\mu$Jy
    at 15, 22 and 43\,GHz, respectively. The dashed red and blue lines 
   (right panels) represent the profiles which deviate from the 
    Doppler boosting  profiles by $\pm$0.30\,Jy (for the 22 and 15\,GHz curves) 
    and $\pm$0.15\,Jy (for the 43\,GHz curve), demonstrating that  most of the
    measured data-points are fitted between the two lines and the observed
    light curves are reasonably well fitted. In addition, there are several
    isolated spikes on short time-scales during 2000--2008  superimposed on
     the regular variations. These spikes may be explained  in terms of 
     the variations in the intrinsic flux density and spectral index of knot C9
     (a traveling shock or plasmoid).}
    \end{figure*}
    %\begin{figure*}
    %\centering
    %\includegraphics[width=6cm,angle=-90]{F15GC9uv.ps}
    %\includegraphics[width=6cm,angle=-90]{F43GC9uv.ps}
   % \includegraphics[width=6cm,angle=-90]{Dif43GC9a.ps}
   % \caption{Residual flux density at 15\,GHz for knot C9, showing the 
   % validity of the fit to the measured light curve. Squares denote the 
   % model-fit and red/blue lines represent the model-fit by Doppler profile
   % $\pm$0.30\,Jy (15GHz) and 0.15\,Jy (43GHz).}
   % \end{figure*}
    \begin{itemize}
    \item Based on an assumed helical pattern of precessing common trajectory 
   for jet-A the trajectory distribution of the superluminal components of
   jet-A is shown in Figure 1 (left panel) for different knots ejected at 
   different precession phseses. It clearly demonstates both the swing of 
   their ejection position angles and different curved tracks cooresponding to
   different ejection times, which have been successfully applied to modelfit
     the observed kinematics of the 
    superluminal components of jet-A (Qian \cite{Qi22a}, \cite{Qi22b}).
    \item Generally, the superluminal knots moved along their precessing common
    trajectories in the inner-jet regions, while they moved along their own 
    individual tracks in the outer jet regions. This transition of trajectory
    pattern occurred at different core distances for different knots. The 
    apparent trajectory of knot C9 was measured extending to core distance 
    of $\sim$4.16\,mas, equivalent to a traveled distance
     Z$\sim$106\,mas=705\,pc. Its entire trajectory is well fitted in terms of 
     our precessing nozzle scenario as shown in Figure 1 (right). Particularly,
     its inner trajectory can be very well fitted 
    by the precessing common trajectory pattern (Figure 2),
     corresponding to a precession
    phase $\phi_0$=5.54+2$\pi$ (or ejection time 1995.06) and extending to a
    core distance $\sim$1.25\,mas, which is equivalent to a traveled 
    distance Z$\sim$44.8\,mas=298\,pc. The precessing common trajectory 
    pattern assumed in the scenario for jet-A is also applicable to interpret
    the trajectories measured for other superluminal knots in jet-A (e.g. C4,
    C5, C7, C10--C13, C22 and C23; Qian \cite{Qi22a}, \cite{Qi22b}).   
    \item The curved jet-axis and the helical pattern assumed for the 
    precessing common trajectory of jet-A in our scenario result in prominent
    curvatures in the apparent trajectories of superluminal components
    (Fig.1, left panel). The remarkable trajectory curvature measured for 
     knot C9 is very well model-fitted in our precessing nozzle scenario 
    (Figure 2). Within $X_n$=1.22\,mas knot C9 moved along the precessing 
    common track. Most of the observed data-points are located extremely well 
    within the area delimited by the red and blue lines which indicate the 
    trajectories calculated for the precession phases 
    $\phi_0$=5.54${\pm}$0.31\,rad (or $\pm$5\% of the precession period). This
    distinctively demonstrates that the helical pattern assumed for the 
    precessing  common trajectory is appropriate and valid to describe the
    kinematic behavior of knot C9 and other knots (e.g., C4, C5 and C22;
     for knot C4, its common track extends to $X_n{\sim}$1.14\,mas, equivalent
     to a traveled distance of $\sim$267\,pc). Moreover, the model-fits to the
     apparent trajectories of multi-knots imply the existence of both a 
    precessing nozzle and a precessing common track of helical pattern.
    \item For knot C9, its core distance $r_n(t)$, coordinates $X_n(t)$ and $Z_n(t)$
     are well fitted and  shown in Figure 3 (upper two and bottom left panels).
    Both the model-derived apparent velocity $\beta_{app}(t)$ and viewing angle
    $\theta(t)$ as continuous functions of time shown in Figure 3 
    (bottom right panel) reveal prominent structural details. 
    During the period of $\sim$1997--2005 $\beta_{app}(t)$ exhibits a structure of 
    acceleration/deceleration/re-acceleration/deceleration: (a) at 1997.44 (core 
    separation $r_n$=0.33\,mas) $\beta_{app}$=17.91 (a maximum) and 
    $\theta$=$2.41^{\circ}$; (b) at 1998.77 ($r_n$=0.88\,mas)
     $\beta_{app}$=11.44 (a minimum) and $\theta$=$1.14^{\circ}$ (a minimum,
     corresponding to the maximal Doppler factor $\delta_{max}$=31.86, see
    Figure 4 below); (c) at 2001.28 ($r_n$=1.77\,mas) $\beta_{app}$=17.91
    (a maximum) and $\theta$=$2.94^{\circ}$.
    \item The bulk Lorentz factor $\Gamma(t)$ and Doppler factor $\delta(t)$
     derived for knot C9 are shown in Figure 4. During the period of 
    $\sim$1997.5-2001.4 the bulk Lorentz factor $\Gamma$$\simeq$18.5--18.0, 
    while the  Doppler factor $\delta$ has a smooth bump structure, which 
    was completely caused by the change in the viewing angle, as shown in 
    Figure 3 (bottom right panel). The maximal Doppler
    factor is 31.86 at 1998.77 (at $r_n$=0.88\,mas), coincident with the 
    minimal viewing angle $\theta$=$1.14^{\circ}$ and minimal apparent speed
    $\beta_{app}$=11.44.
    \item In order to compare our results with those obtained by Jorstad et al.
   (\cite{Jo05}) we show the apparent speed $\beta_{app}$ as a function  of 
   core distance $r_n$ for knot C9 in Figure 5. It exhibits a
    pronounced oscillating structure with two maxima (both $\beta_{app,max}$=17.91)
    at core distances $r_n$=0.33\,mas and 1.77\,mas, and a minimum 
   $\beta_{app,min}$=11.4 at $r_n$=0.88\,mas, corresponding to a minimal
   viewing angle $\theta_{min}$=$1.14^{\circ}$. In addition, the decreasing part
   (in the range of $r_n$$\sim$1.2--1.8\,mas)  of the 43\,GHz light curve
    measured by Jorstad et al. (\cite{Jo05}) for knot C9 can be well fitted by 
   the Doppler boosting profile ${S_{int}}[\delta(t)/\delta_{max}]^{3+\alpha}$
    derived as a function of core distance in this paper as shown
    in Figure 6. Thus, we unexpectedly find that the 
    model-simulated results obtained within the framework of our precessing
     nozzle scenario are fully consistent with the phenomena observed and 
    analyzed by Jorstad et al. (\cite{Jo05}) as follows:\\
      \\
    {\bf{ The change in apparent speed and the twisted trajectory are also
    pronounced for C9. Although the historically extrapolated position of C9
    (using the average apparent speed of 15.6c) agrees well with the the
    measured position at epoch 1996.81 (Ros et al. 2000). It decelerates 
    significantly from  $\sim$20c near the core to $\sim$10c at 1\,mas and 
    then accelerates beyond 1.5\,mas. The deceleration is accompanied by 
    brightening of the knot and rotation of the EVPA by $\sim$$50^{\circ}$,
    making it almost  perpendicular to the jet direction. }} \\
    \\ 
     It is worth noticing that the rapid variations  in its intrinsic flux 
    density (or the steepening of the local spectrum at 43\,GHz) resulted 
    in the peaking stage of the flare not observed at 43GHz. Detailed 
    model-fits to the light curves measured by Schinzel (\cite{Sc11a}) at 15, 22
     and 43\,GHz will be presented below in Figure 7.
    \item It should be emphasized that the bulk Lorentz factor
     $\Gamma(t)$, viewing angle $\theta(t)$ and Doppler factor $\delta(t)$
    as functions of time were directly derived from the model fitting of 
    the kinematics only, and the Doppler boosting effect 
    ($[\delta(t)]^{3+\alpha}$) is purely predicted by the precesssing nozzle 
    scenario. Thus, whether the predicted Doppler boosting  
    profile ${S_{int}}[\delta(t)/\delta_{max}]^{3+\alpha}$ can explain
    the flux evolution of superluminal knots in 3C345 would be a significant
     test for our precessing nozzle scenario. In the model-simulation the 
    intrinsic flux density $S_{int}$ were adopted to be 3.83, 2.58
   and 1.65\,$\mu$Jy for 15, 22 and 43\,GHz, respectively. The spectral index was 
   adopted as $\alpha$(15-22-43GHz)=0.80 (${S_{\nu}}\,{\propto}\,{{\nu}^{-\alpha}}$).
   \item The light curves observed at 15, 22 and 43\,GHz for knot C9
   \footnote{Data are adopted from Schinzel (\cite{Sc11a}) only.} are shown in 
   Figure 7 (left panels). They exhibit very complicated structures with flux
   density fluctuations  on short time-scales of $\sim$0.5\,years, but the 
   Doppler boosting effect having a time scale of $\sim$5 years (1996.5--2001.5)
    as predicted by our model-simulation can be explicitly discerned: 
   (1) During both the rising 
   ($\sim$1997.0--1998.5) and decaying ($\sim$1999.5--2001.0) stages the flux
    densities measured at 15\,GHz are well fitted by the Doppler boosting 
    profiles (solid black lines in Fig.7, right panels). Thus the 
    model-predicted Doppler-boosting profiles can be regarded
   as correctly determined with its maximum at 1998.77, where the apparent 
   speed and viewing angle are at minimum; (2) At both 22 and 43\,GHz there are 
    a few data-points are also
    fitted by the profiles during the rising satge ($\sim$1997--1998); (3)
    The low flux densities  measured at 43\,GHz during 1998--1999 and
     measured at 22\,GHz during the decaying stage
    (1999.5--2000.5) are obviously due to the intrinsic variations of knot C9.
    For example, during 1998--1999 the measured spectral index 
   $\alpha$(15-43GHz)$\sim$1.8 much larger than that 
   ($\alpha$(15--43GHz)$\sim$0.80) during both rising and decaying stages; 
   (4) The fluctuations on time-scales of 0.5--1 years in flux density
    measured at 15, 22 and 43\,GHz during  2001--2008 can  also be interpreted 
   in terms of the intrinsic variations of knot C9, because the Doppler 
    boosting effect has diminished  during that period. Thus in order to fully
    explain the flux evolution of superluminal components in 3C345 both 
    intrinsic variations and Doppler boosting effect should be taken 
    into account; (5) In Figure 7 (right panels)
    the dashed red and blue lines show the profiles deviating from the 
    model-derived Doppler boosting profiles (solid lines in black) by $\pm$0.15\,Jy
    (for 43\,GHz) and 0.30\,Jy (for 22 and 15\,GHz). It can be seen that most
    of the measured data-points fall between the two lines, indicating that the 
    model-fits to the measured light curves are successful.
    \end{itemize}
   The brief summary described above for the midel-fitting results of the
    kinematic behavior and flux evolution for the superluminal components of
   jet-A in 3C345 explicitly shows that our processing nozzle scenario is 
   most appropriate and  valid to explain the kinematic, dynamic and emission
    properties observed in the superluminal components of blazar 3C345.\\
     In the following we shall investigate the flux evolution associated 
    with the Doppler boosting effect for five superluminal knots (C19, C20,
    C21, B5 and B7) belonging to the jet-B of 3C345. As in
    the previous work (Qian \cite{Qi22b}) the observational data are adopted 
    from Schinzel (\cite{Sc11a}) and Jorstad et al. (\cite{Jo05}).
   \section{Geometry and model parameters}
    According to the precessing nozzle scenario proposed for 3C345 (Qian 
    \cite{Qi22a}, \cite{Qi22b}) the superluminal knots could be separated into
    two groups (group-A and group-B) which were hypothetically assumed to 
    be related to a double-jet structure (jet-A plus jet-B), because the two 
    jets were found to be precessing with the same period of 7.3\,yr, but 
    having different patterns of precessing common trajectory.\\
    The geometry and model-parameters describing the precessing nozzle 
    scenario proposed for 3C345 have been discussed in detail in the previous
    works (Qian \cite{Qi22a}, \cite{Qi22b}). Here we only list the 
     main points of the scenario related to the investigation of the
    kinematics of the superluminal knots of group-B as follows:\\
    \\
    (1) Jet-axis\\
    \\
   The direction of the axis of jet-B in space, around which superluminal knots 
    of group-B move, is defined by parameters $\epsilon$ and $\psi$. The 
    jet-axis is assumed as:\\
      \begin{equation}
       {x_0}=p({z_0}){{z_0}^{\zeta}}
      \end{equation}
     where $\epsilon$=$1.5^{\circ}$, $\psi$=$12.0^{\circ}$, $\zeta$=1.0 and
     p=1.34$\times$$10^{-4}$ are adopted.  \\
    \\
    (2) Amplitude and phase of trajectory\\
    \\
    The amplitude and phase of the trajectory of superluminal 
    knots are defined as:
    \begin{equation}
    {A({Z})}={{A_0}[\sin({\pi}{Z}/{Z_1})]} 
    \end{equation}
    \begin{equation}
    {\phi}({Z})={{\phi}_0}
     \end{equation}
     where $A_0$ represents the amplitude coefficient of the common trajectory
     pattern. $A_0$=1.09\,mas and $Z_1$=396\,mas are adopted. 
      ${\phi}_0$ is the precession phase of an individual knot, which 
     is related to its ejection time $t_0$:
      \begin{equation}
       {\phi_0(rad)}=5.70+{\frac{2\pi}{T_0}}({t_0}-2002.12)
      \end{equation}
     where $T_0$ is the precession period of the jet nozzle.\\
      The distribution of the precessing trajectories of the knots 
     is shown in Figure 8, displaying a bunch of straight-line trajectories, 
     which is completely different from that of the curved trajectories of the 
      knots in jet-A. \\
      \begin{figure*}
      \centering
      \includegraphics[width=7cm,angle=-90]{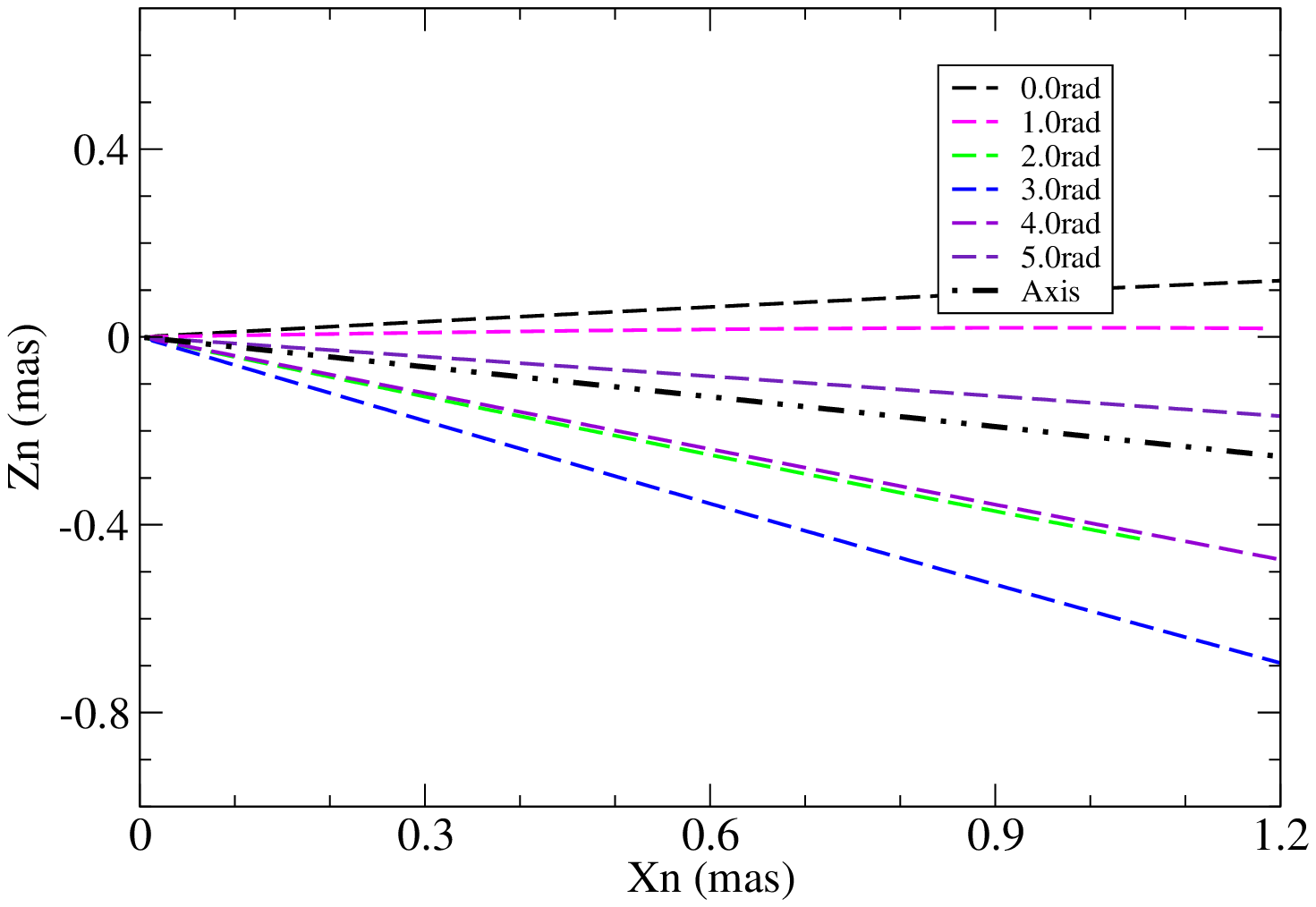}
      \caption{Jet B: Distribution of the precessing trajectories for
      precession phases $\phi_0$=0.0, 1.0, 2.0, 3.0, 4.0 and 5.0\,rad.  
       The jet axis is at position angle --$102^{\circ}$. This distribution  
      modeled for jet-B is like a bunch of straight-line tracks, which is 
     completely different from that of the curved 
     trajectories caused by the helical pattern of precessing common 
     trajectory assumed for jet-A. This distribution illustrates both the 
     swing of the ejection  position angle and the apparent tracks
     of superluminal knots in jet-B ejected at different precession phases. 
     Jet-B consists of 14 components: C15, C15a, C16 to C21,  B5 to B8, B11 
    and B12 (Qian \cite{Qi22a}).}
      \end{figure*}
      \\
       (3) Doppler boosting effect\\
     \\
      In order to investigate the relation between  the flux evolution of
      superluminal components and their Doppler boosting effect during their
      accelerated/decelerated motion and model-fitting of the measured 
      light curves, we define the Doppler boosting profile  as
      ${S_{int}}[\delta(t)]^{3+\alpha}$ or 
      $[\delta(t)/{\delta_{max}}]^{3+\alpha}$, where $S_{int}$--the intrinsic
      flux density of the knot, $\delta(t)$--Doppler factor and
       $\alpha$--spectral index (in the frequency range of 10.7--43\,GHz). 
      The observed flux density $S_{obs}(\nu,t)$ of the superluminal components
     is Doppler-boosted:
     \begin{equation}
   {S_{obs}}(\nu,t)={S_{int}}(\nu){\times}{\delta(t)}^{3+\alpha}
     \end{equation}
     $S_{int}(\nu)$=$S_0$\,${({\nu/{\nu_0})}^{-\alpha}}$,
     $S_0$--the flux density at the fiducial frequency $\nu_0$. \footnote{Here
     in this paper we assume $S_{int}$ and $\alpha$ are 
    constants, not depending on time and frequency. In general case both 
    $S_{int}$ and $\alpha$ could be defined as functions of frequency and time.
     Thus flux evolution of superluminal knots would be associated with
     more complex Doppler-boosting effect.}\\
    \\ 
   (4) Cosmological model\\
    \\
    We will apply the concordant cosmological model (Spergel et al. 
    \cite{Sp03}, Hogg \cite{Ho99}) with $\Omega_{\lambda}$=0.73 and 
    $\Omega_{m}$=0.27, and $H_0$=71km$s^{-1}$${Mpc}^{-1}$. Thus the luminosity
    distance of 3C345 $D_L$=3.49\,Gpc, the angular-diameter distance
    $D_a$=1.37\,Gpc, 1\,mas=6.65\,pc and 1\,mas/yr=34.6\,c. 1\,c is equivalent
    to an angular speed 0.046\,mas/yr in the rest frame.\\
    \begin{figure*}
    \centering
    \includegraphics[width=6cm,angle=-90]{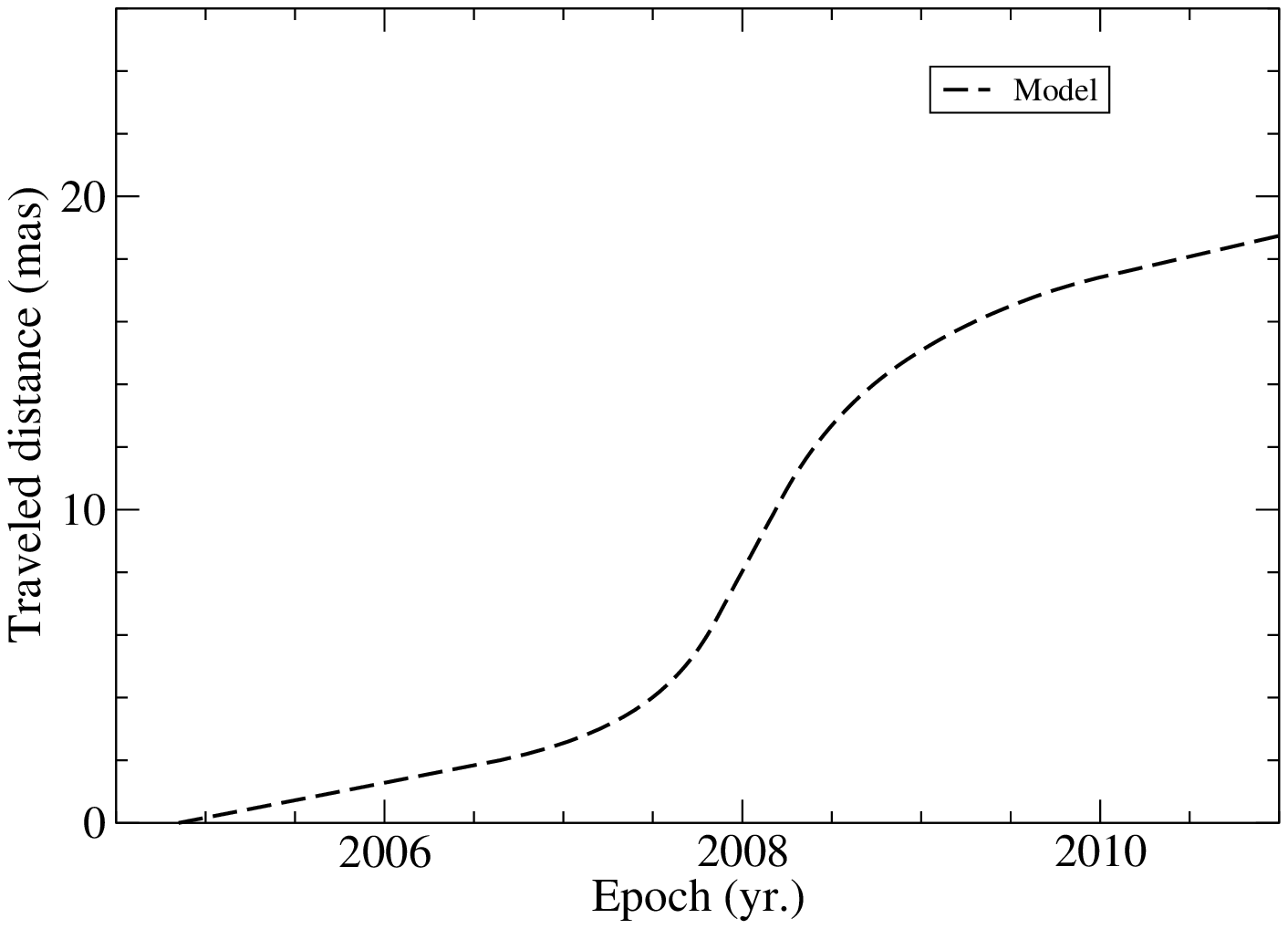}
    \includegraphics[width=6cm,angle=-90]{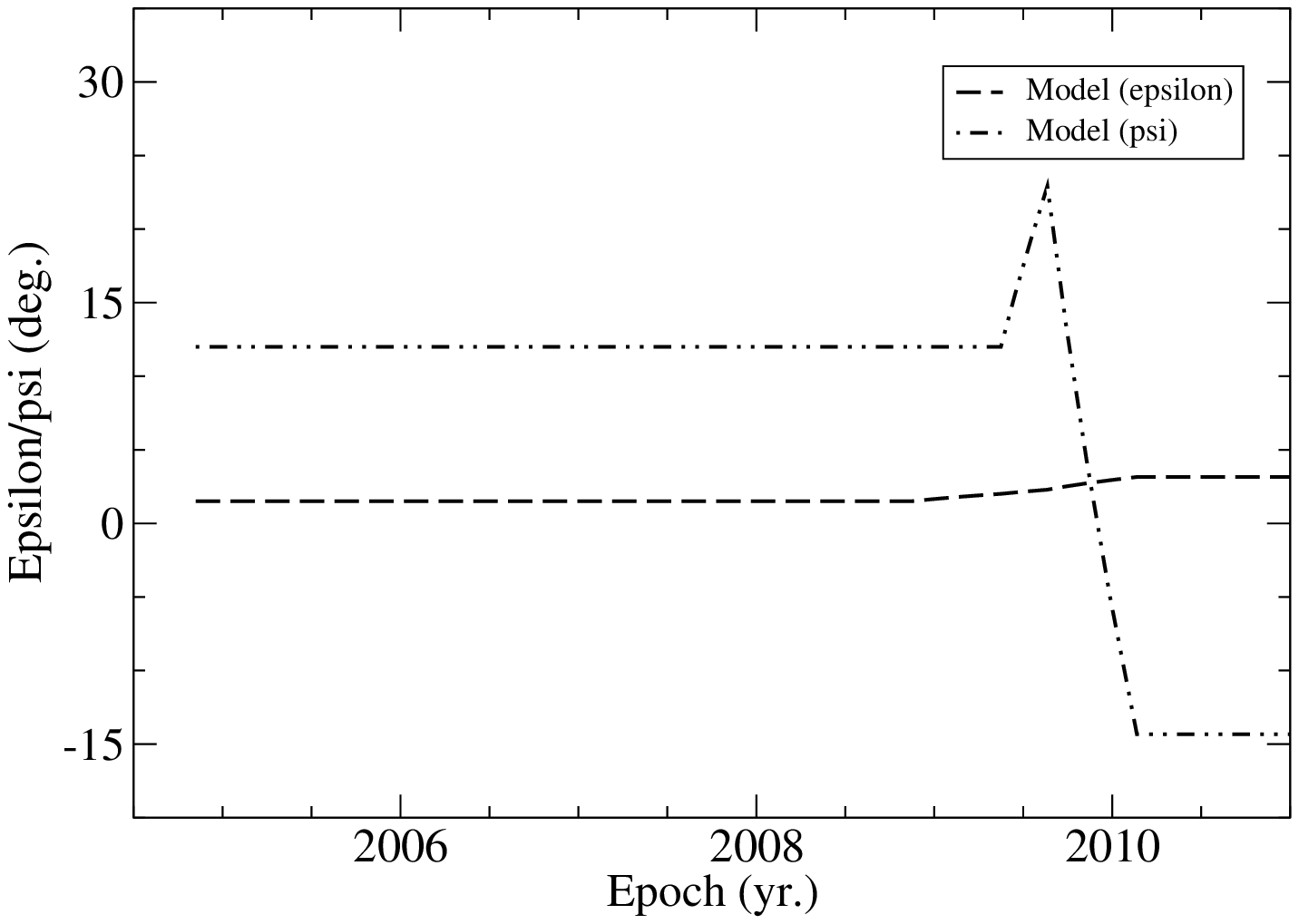}
    \caption{Knot C19: Model-derived traveled distance Z(t), parameters
    $\epsilon(t)$ and $\psi(t)$. Before 2008.87 ($X_n{\leq}$0.25\,mas)
     $\epsilon$=$1.5^{\circ}$ and $\psi$=$12.0^{\circ}$, knot C19 moved along
     the precessing common trajectory, while after 2008.87 ($X_n{>}$0.25\,mas)
    $\epsilon$ and $\psi$ started to change, and
    knot C19 started to move along its own individual track, deviating from
    the precessing common track. }
    \end{figure*}
    \begin{figure*}
    \centering
    \includegraphics[width=8cm,angle=-90]{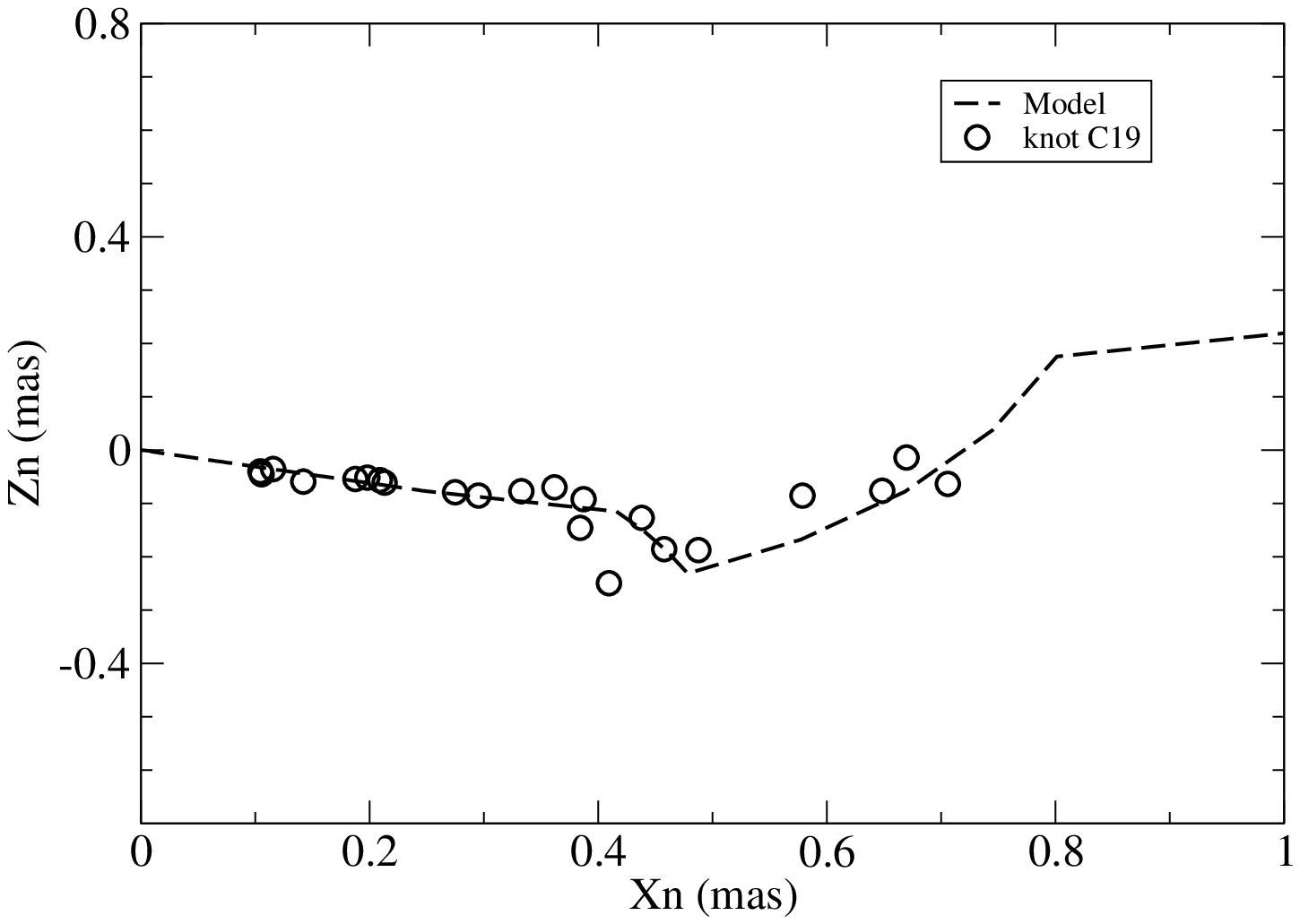}
    \caption{Knot C19: Model fitting of the entire trajectory $Z_n(X_n)$.
     Within $X_n$=0.25\,mas knot C19 moved along the precessing common 
    trajectory, while beyond that it started to move along its own individual
    track.}
    \end{figure*}
    \begin{figure*}
    \centering
    \includegraphics[width=6cm,angle=-90]{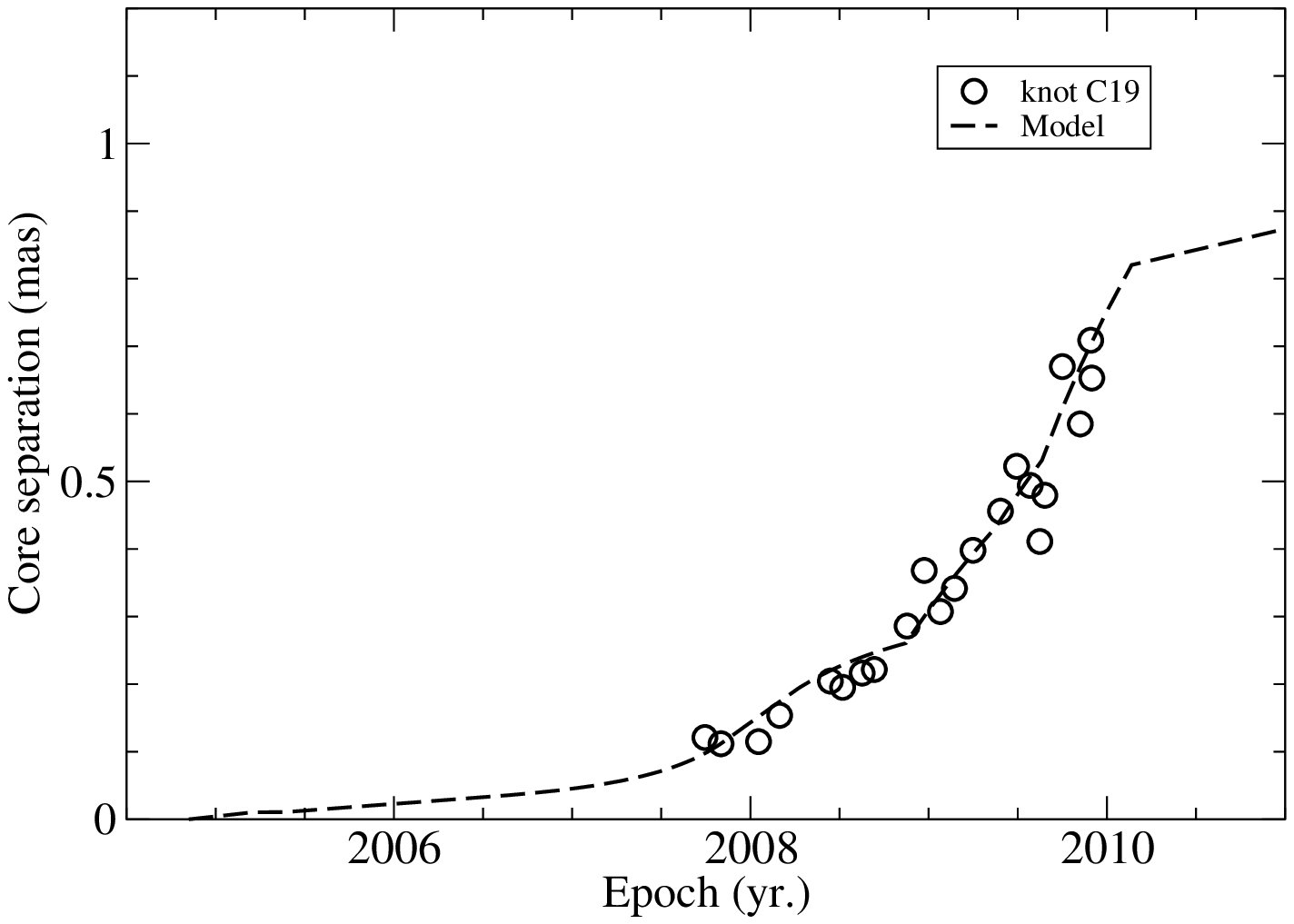}
    \includegraphics[width=6cm,angle=-90]{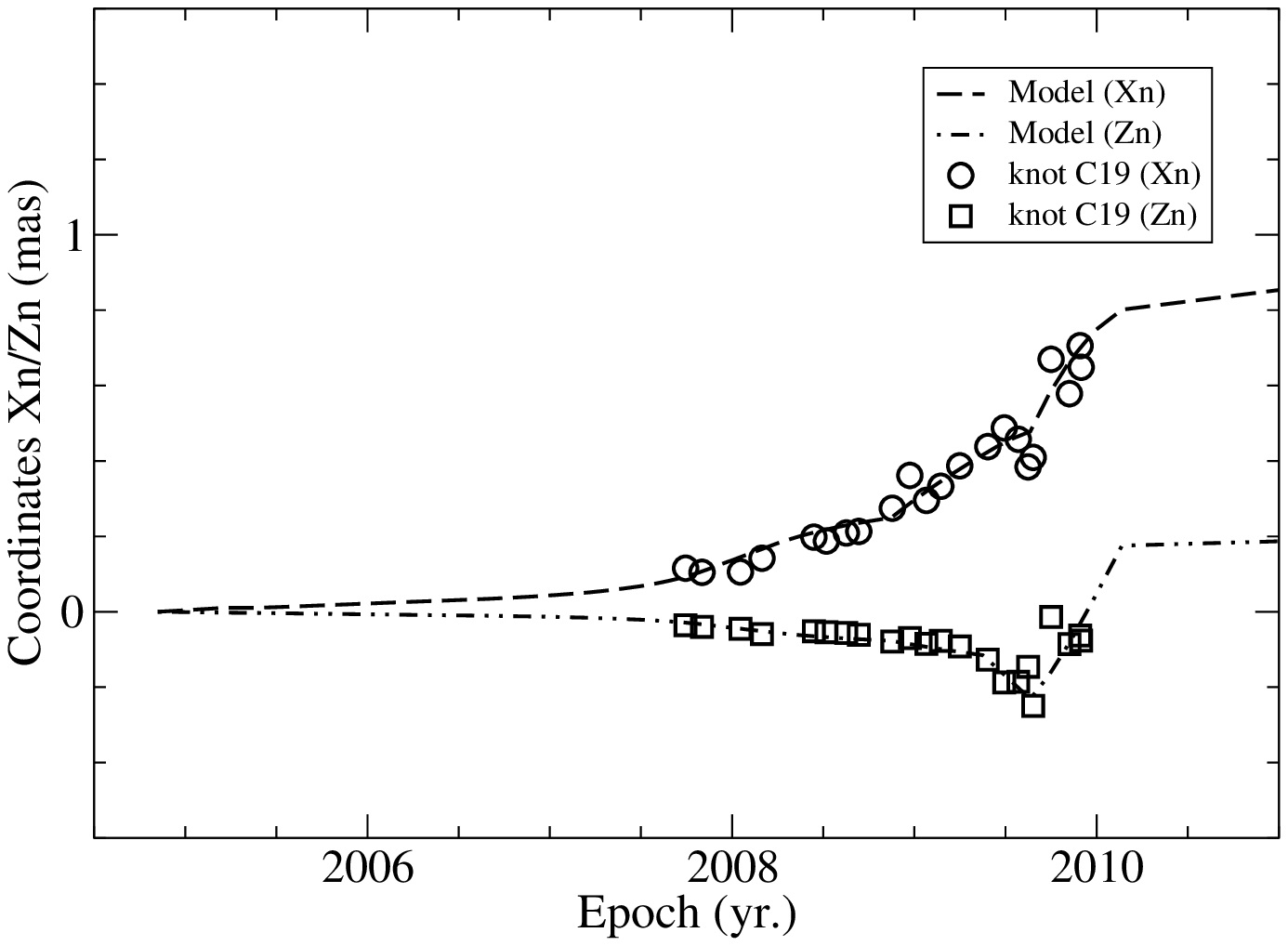}
    \caption{Knot C19: Model fitting of the core separation $r_n(t)$,
     coordinates $X_n(t)$ and $Z_n(t)$. Both the precessing common trajectory
     (inner track; before 2008.66, $X_n{\leq}$0.25\,mas) and the individual
     trajectory (outer track; after 2008.66, $X_n{>}$0.25\,mas) are well
     fitted.}
    \end{figure*}
    \begin{figure*}
    \centering
    \includegraphics[width=6cm,angle=-90]{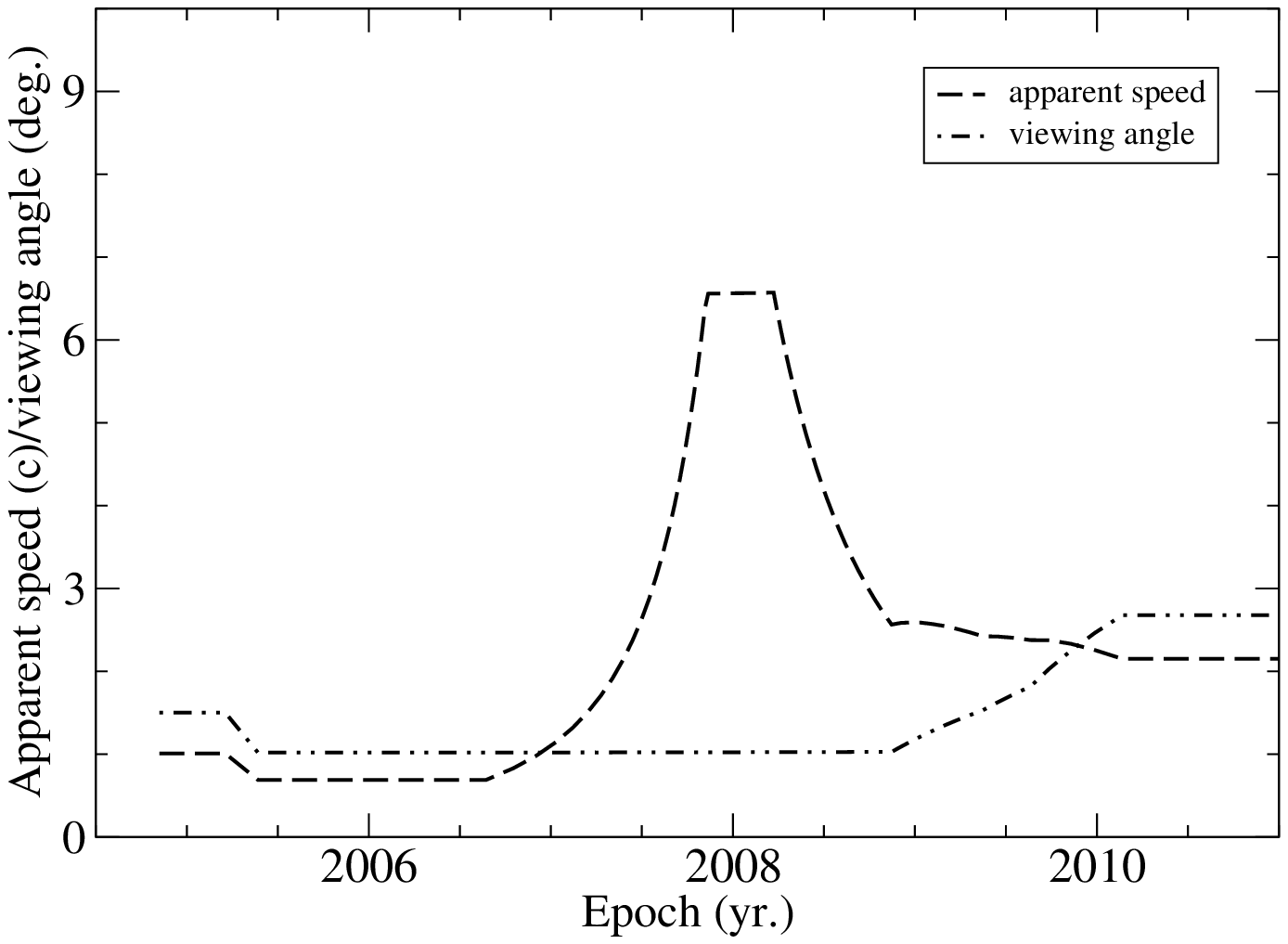}
    \includegraphics[width=6cm,angle=-90]{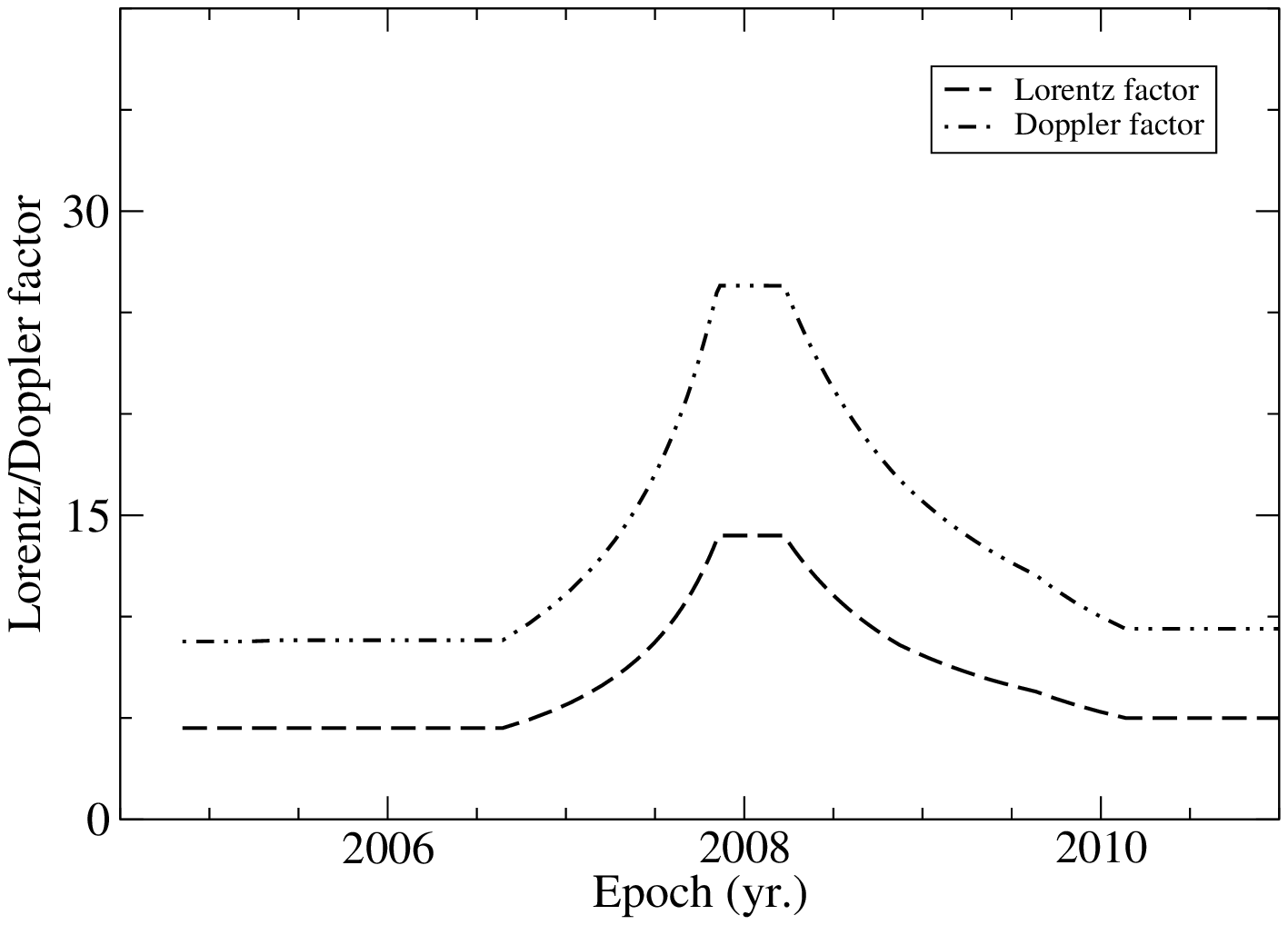}
    \caption{Knot C19. Left panel: the model-derived apparent speed 
    $\beta_{app}(t)$ and viewing angle $\theta(t)$ as  functions 
    of time. The apparent speed has a maximum $\beta_{app}(t)$=6.57 at 2008.22
     and the viewing angle $\theta(t){\simeq}$$1.02^{\circ}$ during 
    2007.5--2008.8. Right panel: the model-derived bulk Lorentz factor
    $\Gamma(t)$ and Doppler factor $\delta(t)$ as  functions of time.
    During 2007.86--2008.22  $\Gamma(t)$=18.40 and 
    $\delta(t){\simeq}$26.32 (a maximum). }
    \end{figure*}
    \begin{figure*}
    \centering
    \includegraphics[width=6cm,angle=-90]{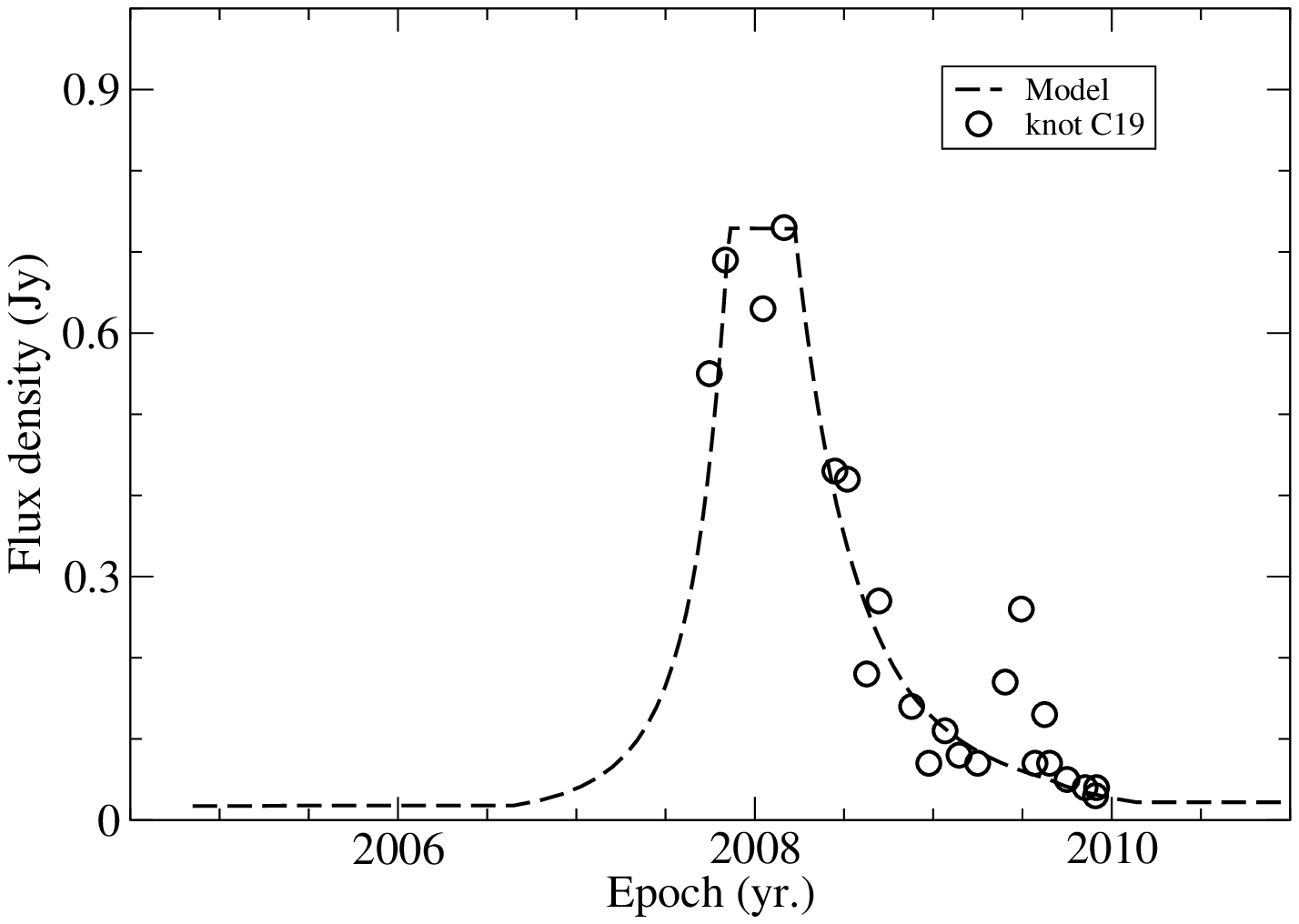}
    \includegraphics[width=6cm,angle=-90]{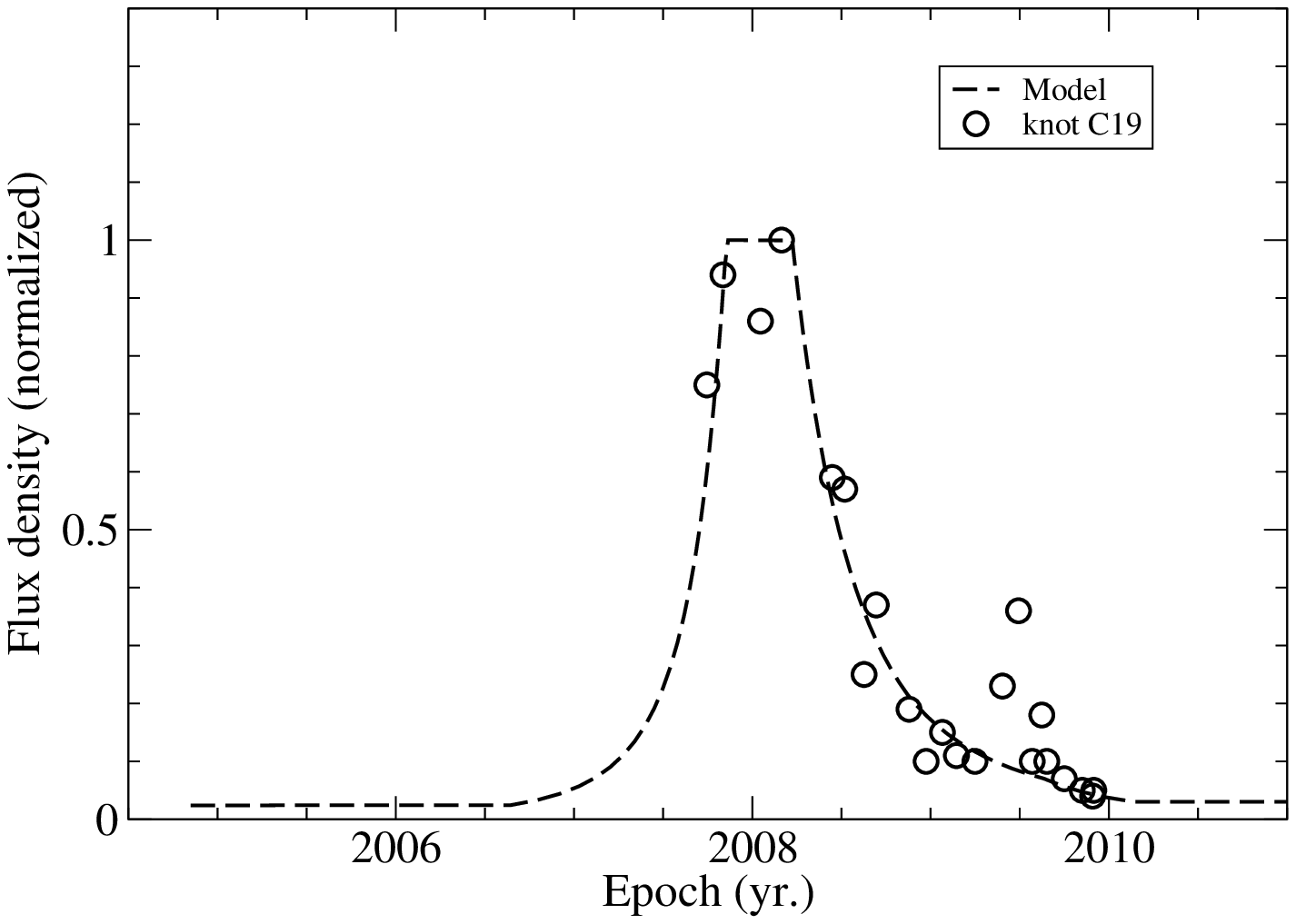}
    \caption{Knot C19: Model fitting of the 43GHz light curve (left
    panel) and its corresponding normalized light curve (right panel). 
     Both  are very well fitted by the  Doppler boosting profile
     $S_{int}[\delta(t)]^{3+\alpha}$ and 
    $[\delta(t)/\delta_{max}]^{3+\alpha}$, respectively. An intirnsic 
    flux density $S_{int}$=10.8\,$\mu$Jy and 
    a spectral index $\alpha$=0.40 are adopted.}
    \end{figure*}
    \section{Knot C19: Interpretation of kinematics and flux evolution} 
    According to the precessing nozzle scenario for jet-B, knot C19 has its 
    ejection time $t_0$=2004.85 and corresponding precession phase 
    $\phi_0$=1.77+2$\pi$. Its traveled distance Z(t) and parameters 
    $\epsilon(t)$ and $\psi(t)$ are model-fitted as shown in Figure 9.
    Before 2008.87 $\epsilon$=$1.50^{\circ}$ and $\psi$=$12.0^{\circ}$, knot C9
    moved along the precessing common trajectory. The cooresponding traveled
    distance Z=14.6\,mas=97.1\,pc.    After 2008.87 
    (core separation $r_n{>}$0.26\,mas or $X_n{>}$0.25\,mas) $\epsilon$
    and $\psi$ started to change and knot C19 started to move along  its own 
    individual track, deviating from the precessing common track. 
    \subsection{Knot C19: Model simulation of kinematics}
    The model-fitting results of the entire trajectory $Z_n(X_n)$, core 
    separation $r_n(t)$ and coordinates $X_n(t)$ and $Z_n(t)$ for knot C19 
    are shown in Figures 10 and 11 They are all well fitted
    because the transition between the common track in the inner jet region 
    and the individual track in the outer jet region has been 
   taken into account.\\
    The model-derived apparent speed $\beta_{app}(t)$ and viewing angle 
    $\theta(t)$ as continuous functions of time are shown in Figure 12 (right 
   panel): $\beta_{app}(t)$ has a prominent bump during 2007.5--2008.8
   with its peak of  6.57 (at 2008.22), while 
   $\theta(t)$$\simeq{1.0^{\circ}}$. The modeled bulk Lorentz factor
   $\Gamma(t)$ and Doppler factor $\delta(t)$ as continuous functions of time
   are shown in Figure 12 (left panel), also demonstrating a distinct bump 
    structure with $\Gamma_{max}$=14.8 and $\delta_{max}$=26.3.
      \subsection{Knot C19: Doppler boosting effect and flux evolution}
   The light curve observed at 43\,GHz and its corresponding normalized light 
   curve are well fitted (Figure 13) in terms of the Doppler boosting profile 
   ${S_{int}}[\delta(t)]^{3+\alpha}$ and $[\delta(t)/\delta_{max}]^{3+\alpha}$,
   respectively. A spectral index $\alpha$=0.40 and an intrinsic flux density 
    $S_{int}$=10.8$\mu$Jy are adopted.
     \begin{figure*}
    \centering
    \includegraphics[width=6cm,angle=-90]{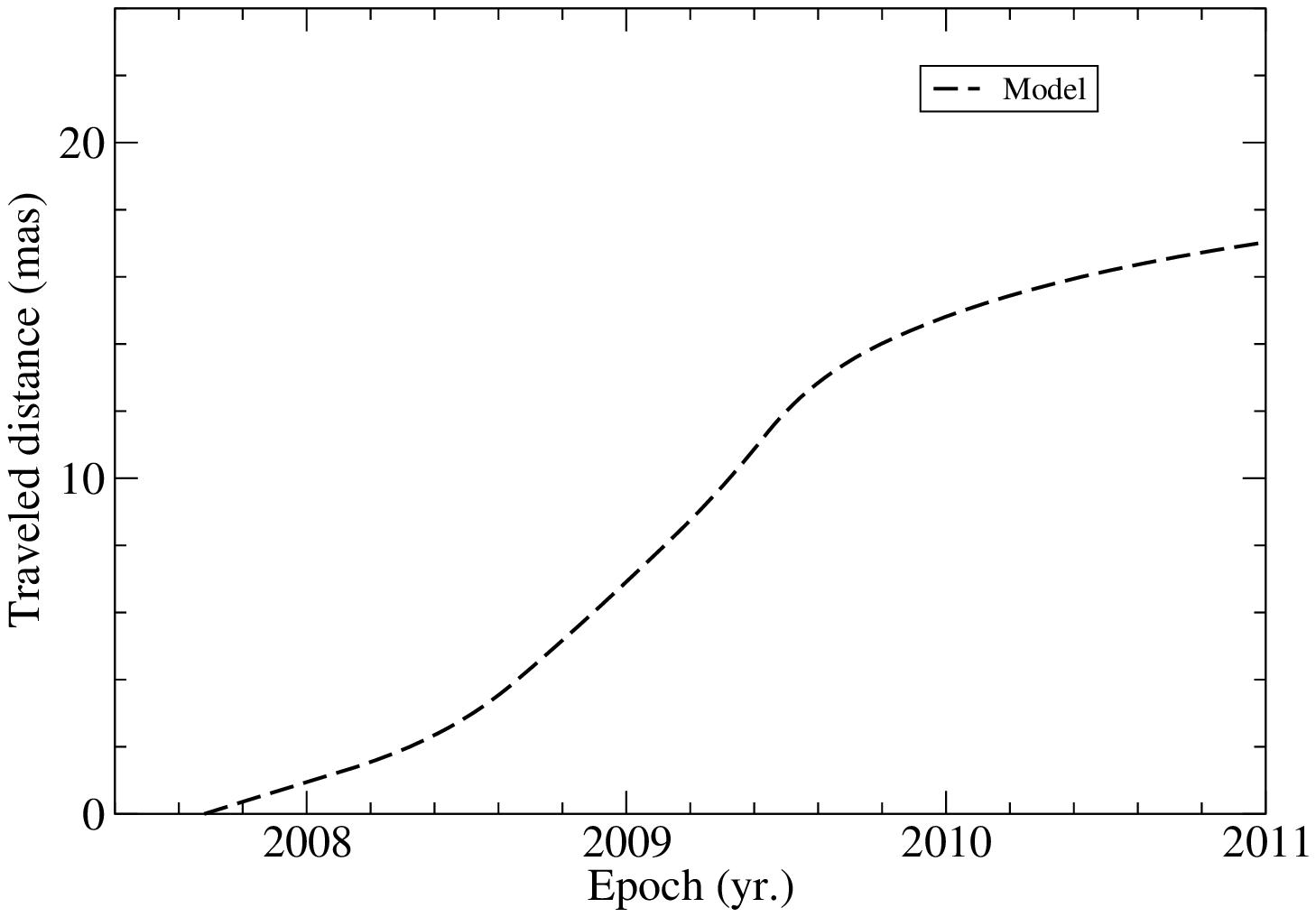}
    \includegraphics[width=6cm,angle=-90]{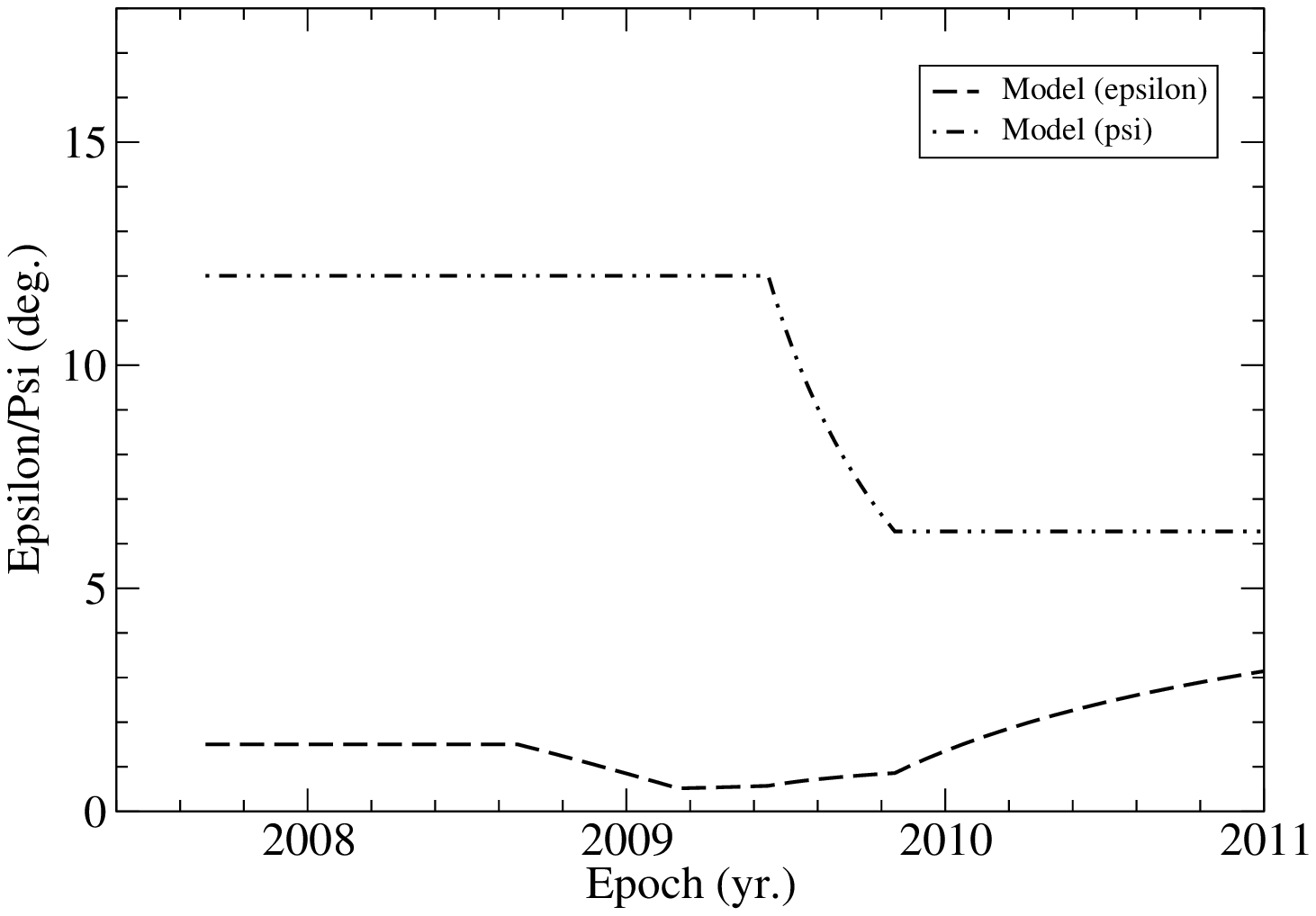}
    \caption{Knot C20: The model-derived  traveled distance Z(t) (left panel),
    parameters ${\epsilon}$(t) and ${\psi}$(t) (right panel). Before 
    2008.66 ($X_n{\leq}$0.13\,mas) $\epsilon$=$1.50^{\circ}$ and
     $\psi$=$12.0^{\circ}$ knot C20 moved
    along the precessing common trajectory. The corresponding traveled distance
    extends to Z=4.0\,mas=26.6\,pc. After 2008.66 $\epsilon$ started to change 
    ($\psi$ started to change after 2009.44) and  knot C20 started to move
    along its own individual track. That is, the  transition from the
    precessing common track to its individual track occurred at 2008.66.}
    \end{figure*}
    \begin{figure*}
   \centering
   \includegraphics[width=8cm,angle=-90]{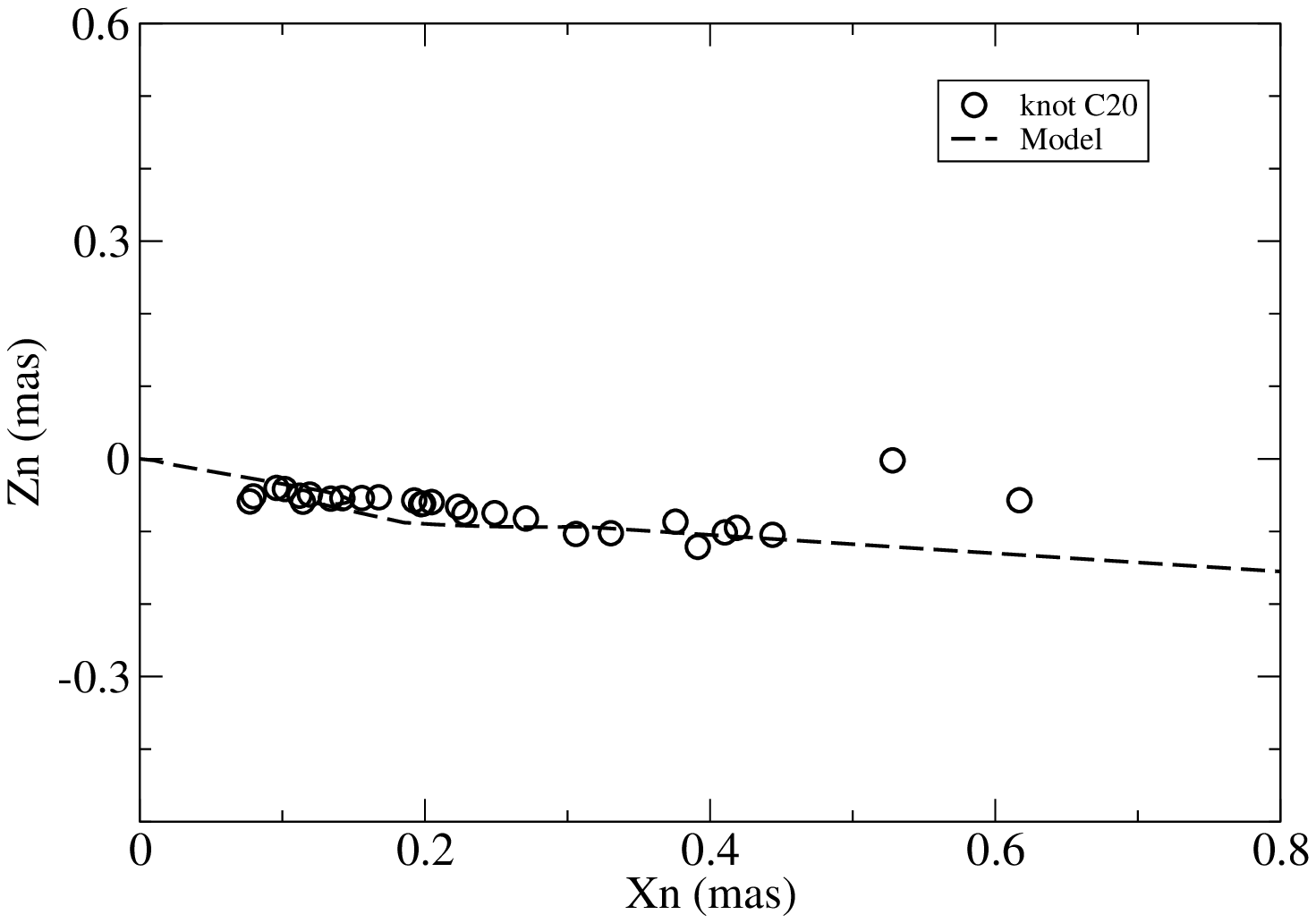}
   \caption{Knot C20: Model-fitting of the entire trajectory $Z_n(X_n)$. Within 
    $X_n$=0.13\,mas, knot C20 moved along the precessing common track and 
    beyond $X_n$=0.13\,mas it started to move along its own individual track.
    Thus transition  from the precessing common track to its own individual 
    track occurred at $X_n$=0.13\,mas. }
    \end{figure*}
   \begin{figure*}
   \centering
   \includegraphics[width=6cm,angle=-90]{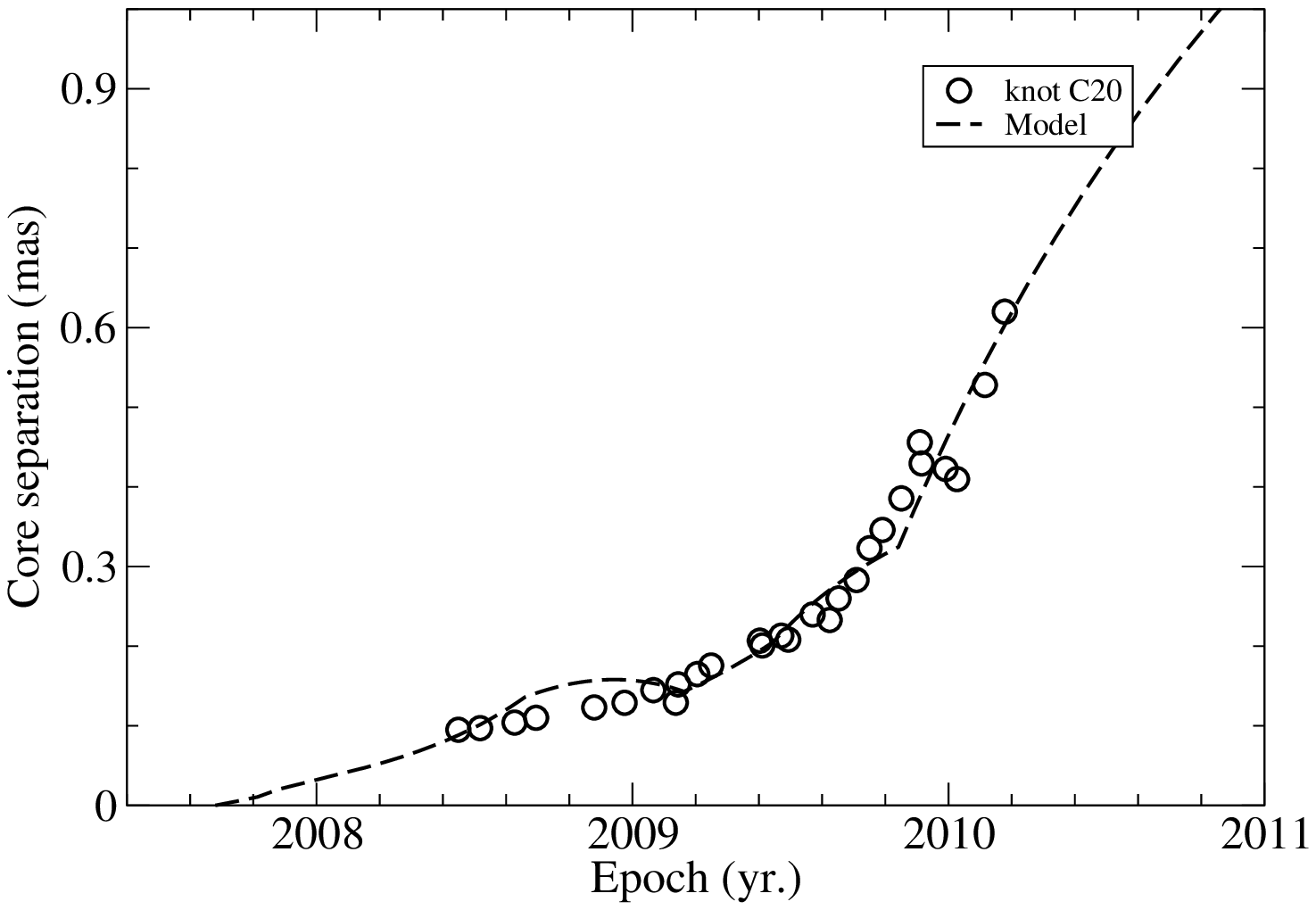}
   \includegraphics[width=6cm,angle=-90]{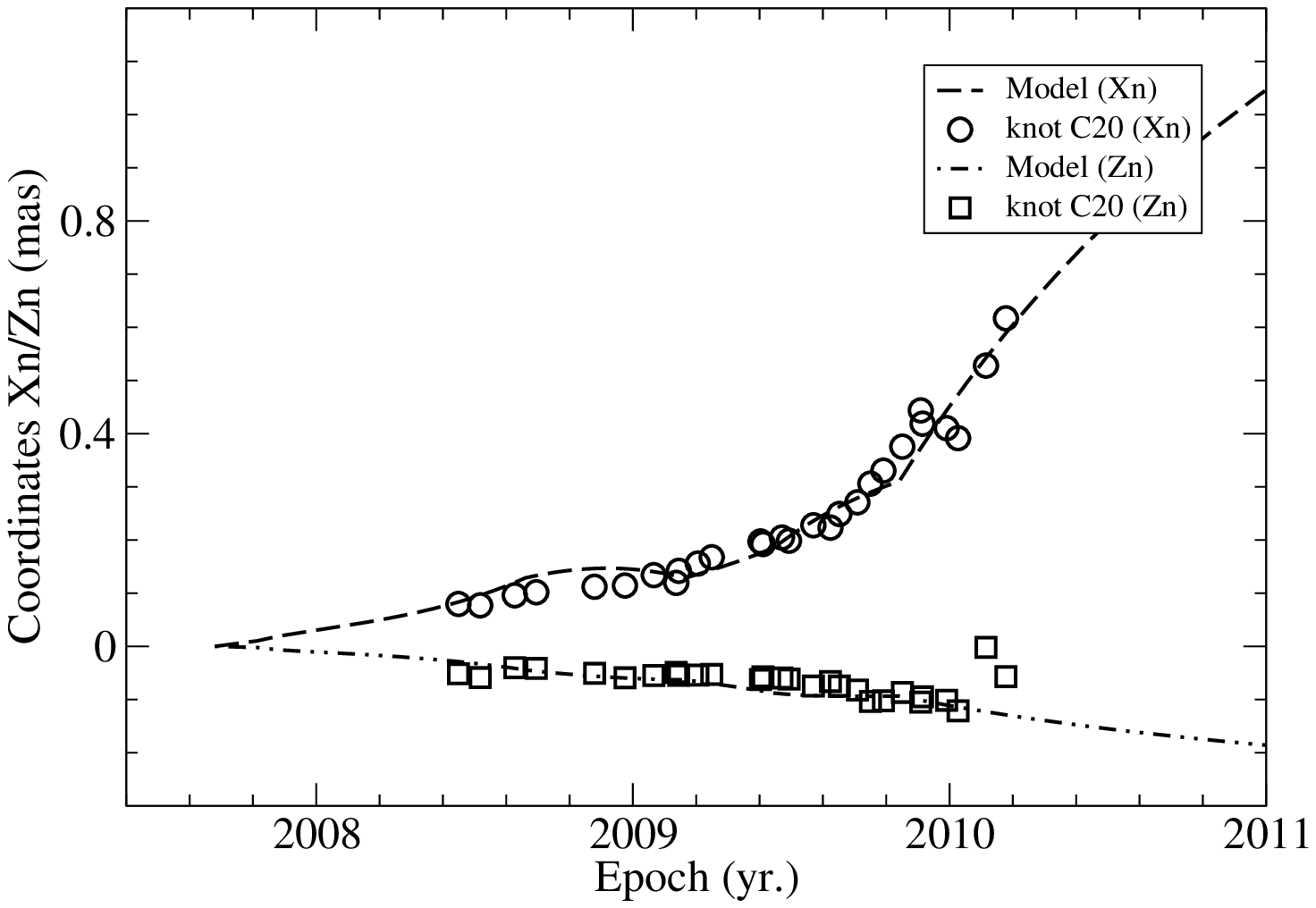}
   \caption{Knot C20: Model fitting of the core separation $r_n(t)$, 
   coordinates $X_n(t)$ and $Z_n(t)$. Both the inner precessing common track 
   ($X_n{\leq}$0.13\,mas) and the outer individual track ($X_n{>}$0.13\,mas)
     are well fitted.}
   \end{figure*}
   \begin{figure*}
   \centering
   \includegraphics[width=6cm,angle=-90]{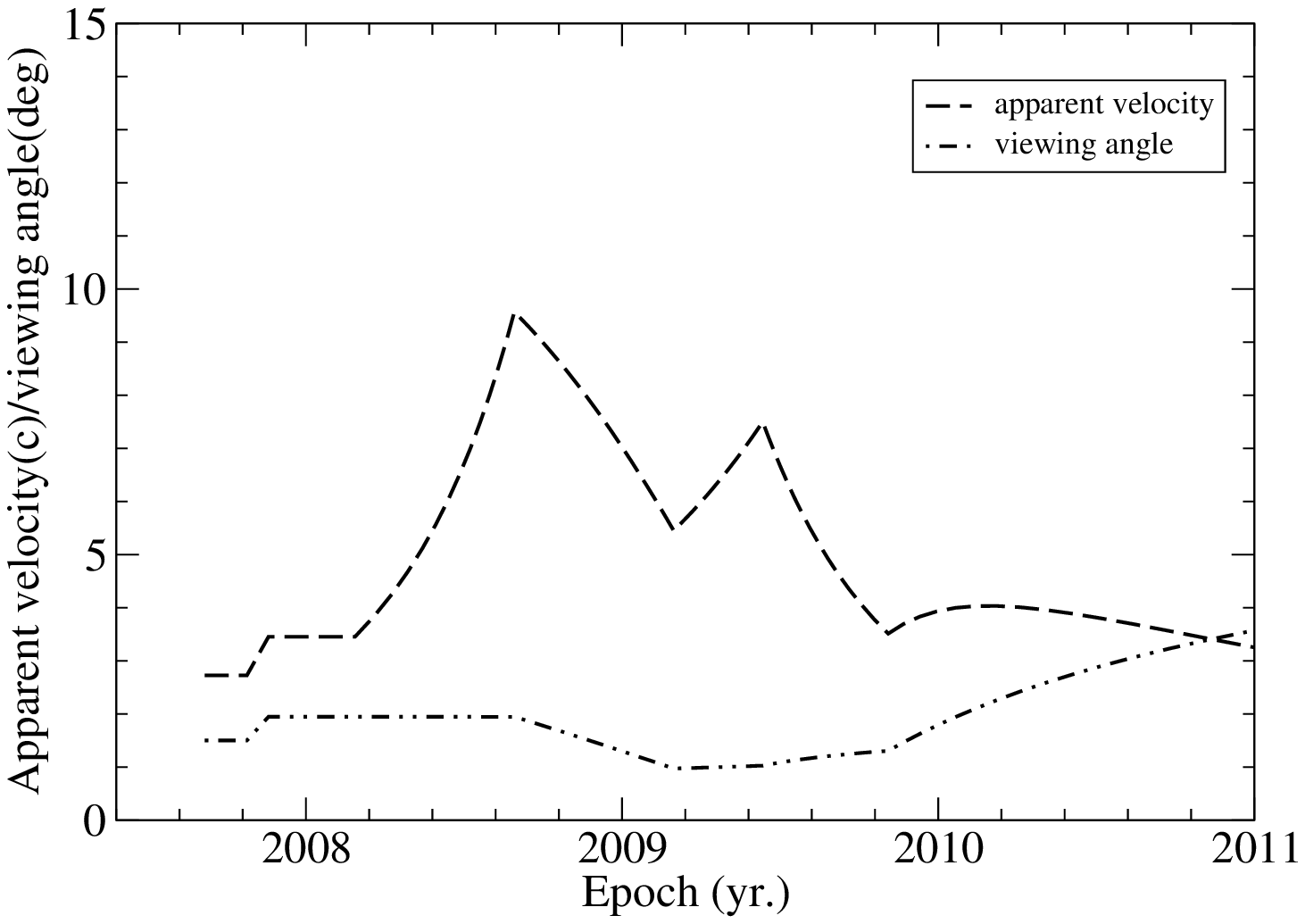}
   \includegraphics[width=6cm,angle=-90]{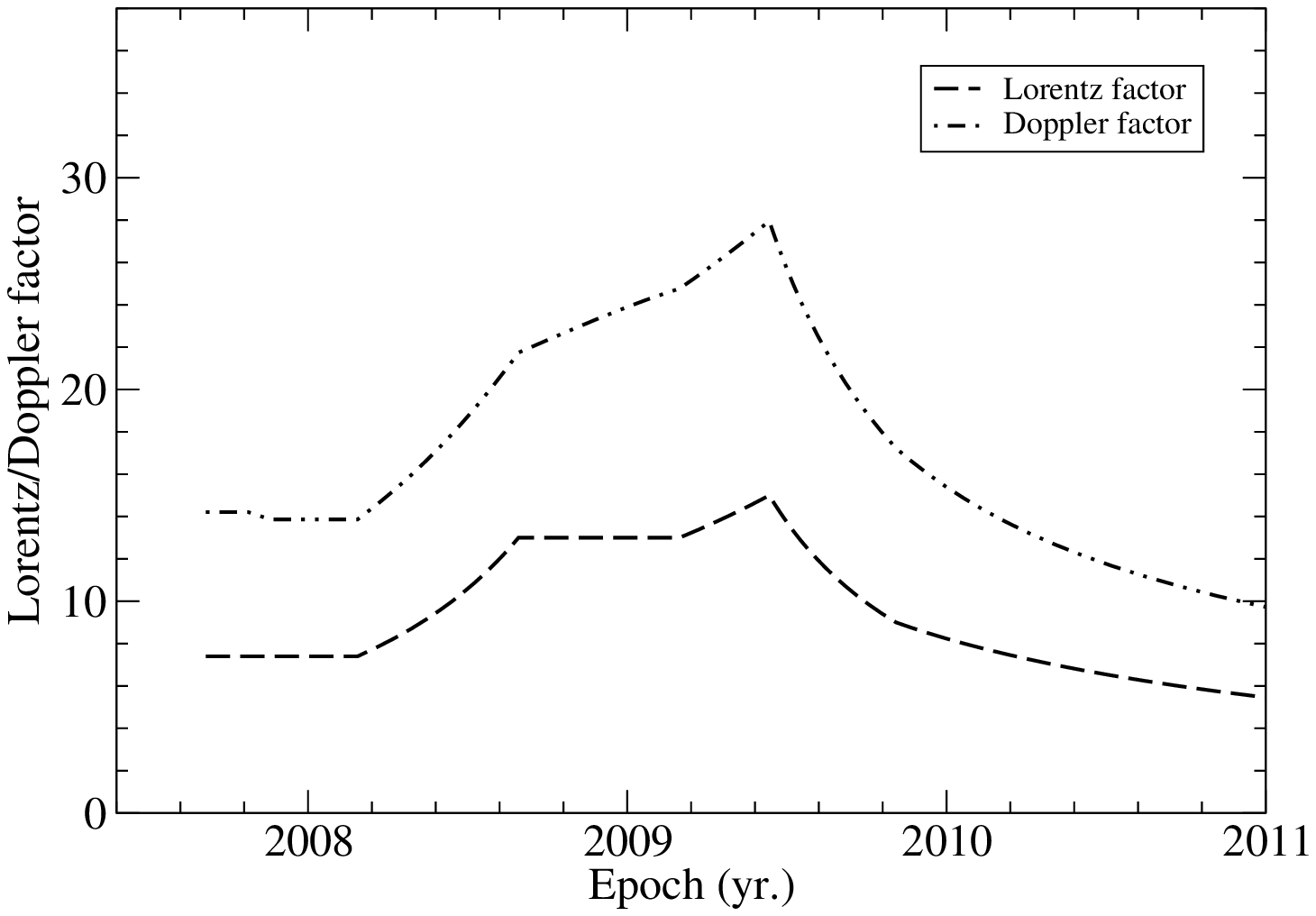}
   \caption{Knot C20. Left panel: the model-derived apparent speed
    $\beta_{app}(t)$ and viewing angle $\theta(t)$ as functions of
    time. The apparent speed has two peaks: $\beta_{app}$=9.57 at 2008.66
    and 7.50 at 2009.44. The corresponding viewing angles are
    $\theta$=$1.95^{\circ}$ and =$1.03^{\circ}$, respectively. Right panel: the 
    model-derived bulk Lorentz factor $\Gamma(t)$ and Doppler factor
    $\delta(t)$. Both have a peak $\Gamma_{max}$=15.00 and $\delta_{max}$=27.95
    at 2009.44, coincident with the second peak in the apparent speed
     $\beta_{app}$.}
  \end{figure*}
    \begin{figure*}
    \centering
    \includegraphics[width=6cm,angle=-90]{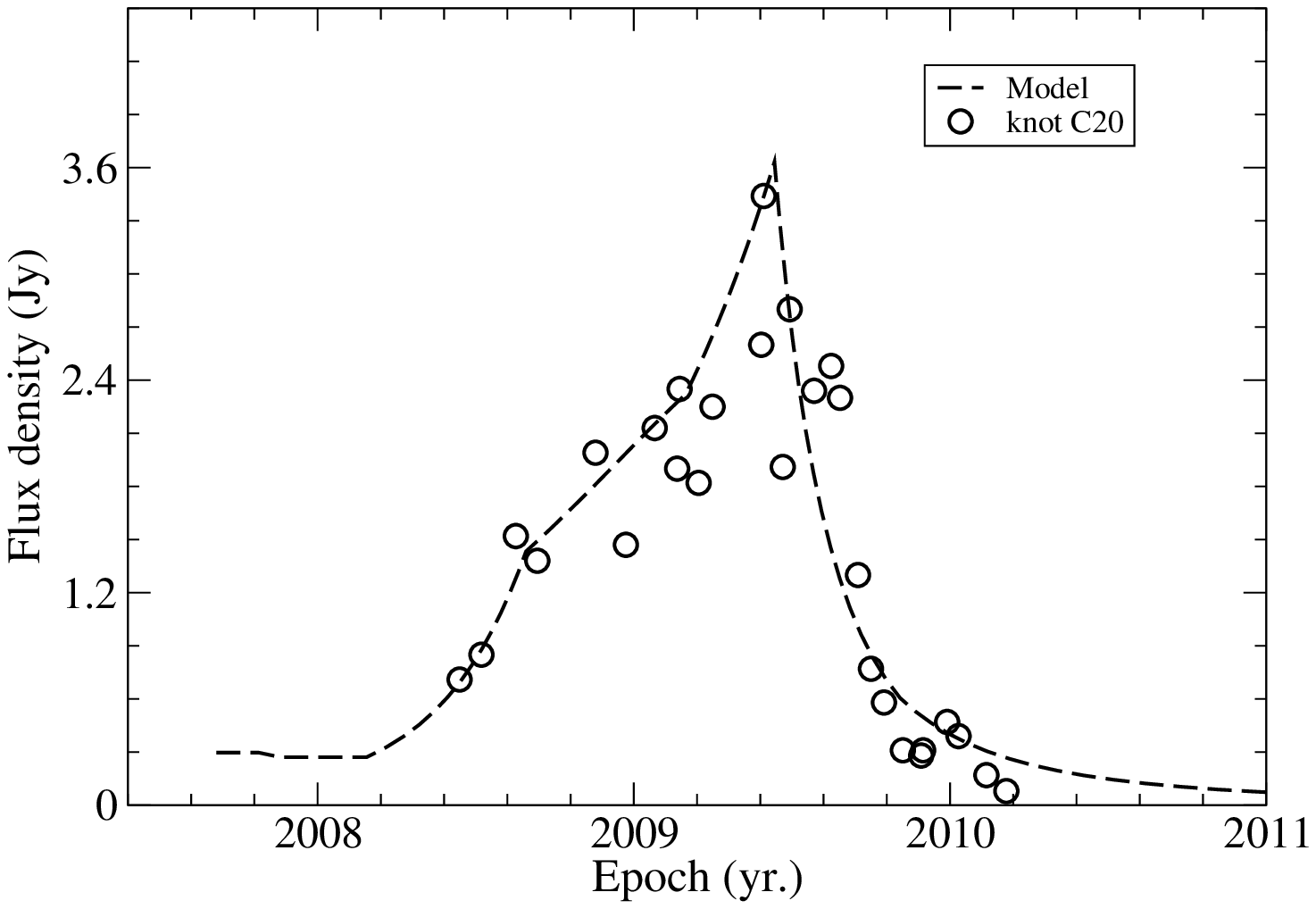}
    \includegraphics[width=6cm,angle=-90]{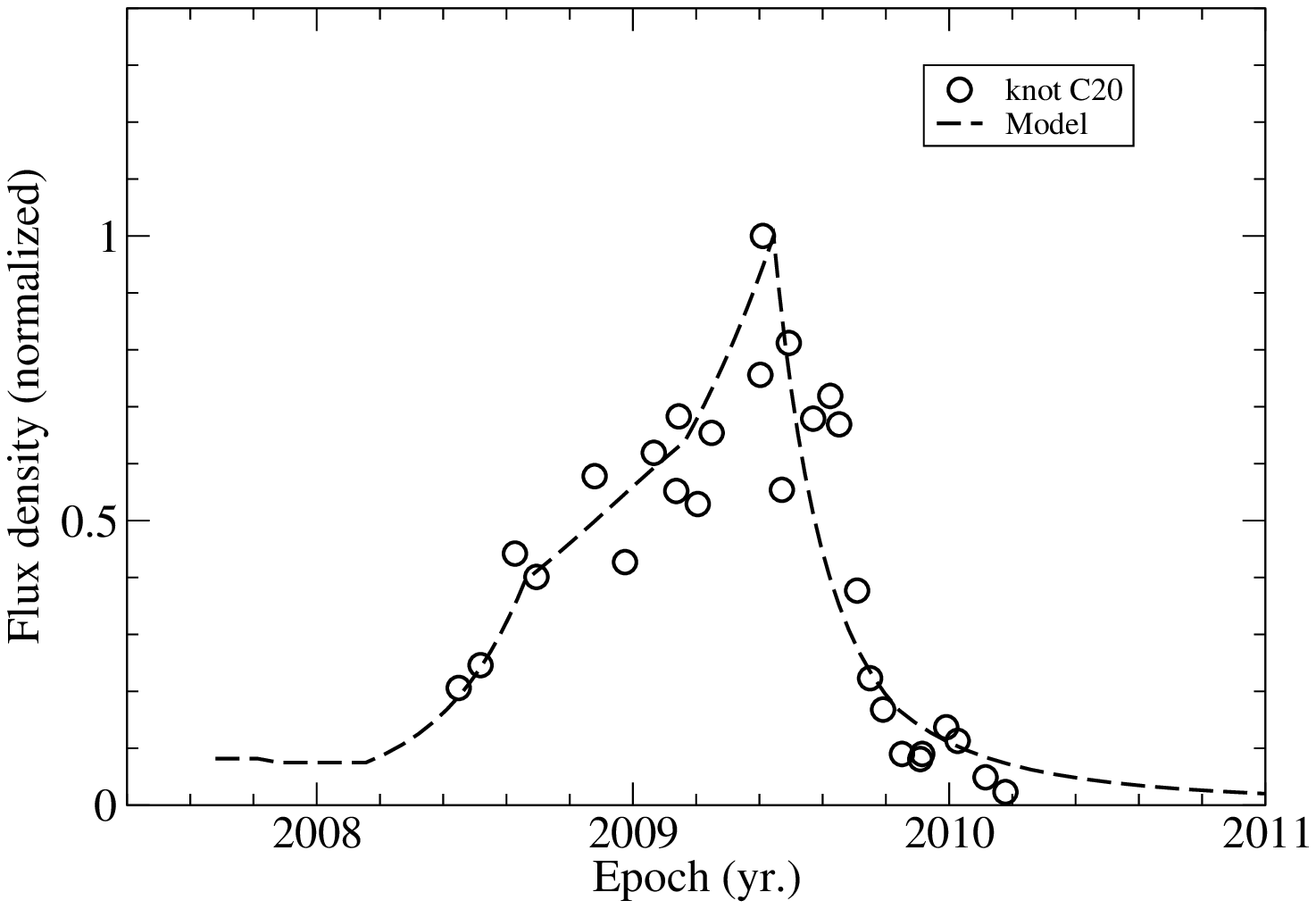}
    \caption{Knot C20: model fitting of the 43\,GHz light curve (left
    panel) and its coresponding normalized light curve (right panel). 
    They are well fitted
    by the Doppler boosting profile  $S_{int}[\delta(t)]^{3+\alpha}$  and
     $[\delta(t)/\delta_{max}]^{3+\alpha}$, repectively. An intrinsic flux 
    density  $S_{int}$=16.2\,$\mu$Jy and a spectral index $\alpha$=0.70
    are adopted.} 
    \end{figure*}
   \section{Knot C20: Interpretation of kinematics and flux evolution}
     According to the precessing nozzzle scenario for jet-B, the ejection time
    of knot C20 is $t_0$=2007.68 and corresponding precession phase 
     $\phi_0$=4.20+2$\pi$. Its traveled distance Z(t) and parameters
    $\epsilon(t)$ and $\psi(t)$ were model-fitted as functions of time as showm
    in Figure 14. Before 2008.66 $\epsilon$=$1.50^{\circ}$ and 
    $\psi$=$12.0^{\circ}$, knot C20 moved along the precessing common 
    trajectory. The corresponding traveled distance Z=4.0\,mas=26.6\,pc.
    After 2008.66 (core separation $r_n{>}$0.14\,mas or coordinate 
    $X_n{>}0.13$\,mas) parameter $\epsilon$ started to change (parameter
    $\psi$ started to change after 2009.4) and its motion
     started to follow its own individual track, deviating from the precessing
     common track. 
   \subsection{Knot C20: Model simulation of kinematics}
    The model-fitting results of the entire trajectory $Z_n(X_n)$, core 
    distance $r_n(t)$, coordinates $X_n(t)$ and $Z_n(t)$ for knot C20 are 
    presented in Figure 15 and Figure 16. They are all well fitted
    because the transition between the common track in the inner jet region and
    the individual track in the outer jet region has been considered.
    The model-derived apparent speed $\beta_{app}(t)$ and viewing angle 
    $\theta(t)$ as continuous functions of time are shown in Figure 17 (left 
    panel).  The apparent speed has two peaks at 2008.66 and 2009.44:
    $\beta_{app}$=9.57 and 7.50, respectively. The corresponding viewing angles
    are $\theta$=$1.95^{\circ}$ and $1.03^{\circ}$.\\
     The model-derived  bulk Lorentz factor $\Gamma(t)$ and Doppler
    factor $\delta(t)$ as continuous functions of time are shown in Figure 17
     (right panel). Both have a peak at 2009.44, coincident with the second
    peak in the apparent speed: $\Gamma_{max}$=15.0 and $\delta_{max}$=27.9.
      \subsection{Knot C20: Doppler boosting effect and flux evolution}
     The model fitting of the measured  43\,GHz light curve and its 
    corresponding normalized light  curve are shown in Figure 18.
    Both are well model-fitted in terms of the Doppler boosting effect with
    an assumed spectral index at 43\,GHz of $\alpha$=0.70 and an intrinsic
    flux density $S_{int}$=16.2\,$\mu$Jy. The maximal flux density 
    $S_{max}$=3.63\,Jy at 2009.44.
    \begin{figure*}
    \centering
    \includegraphics[width=6cm,angle=-90]{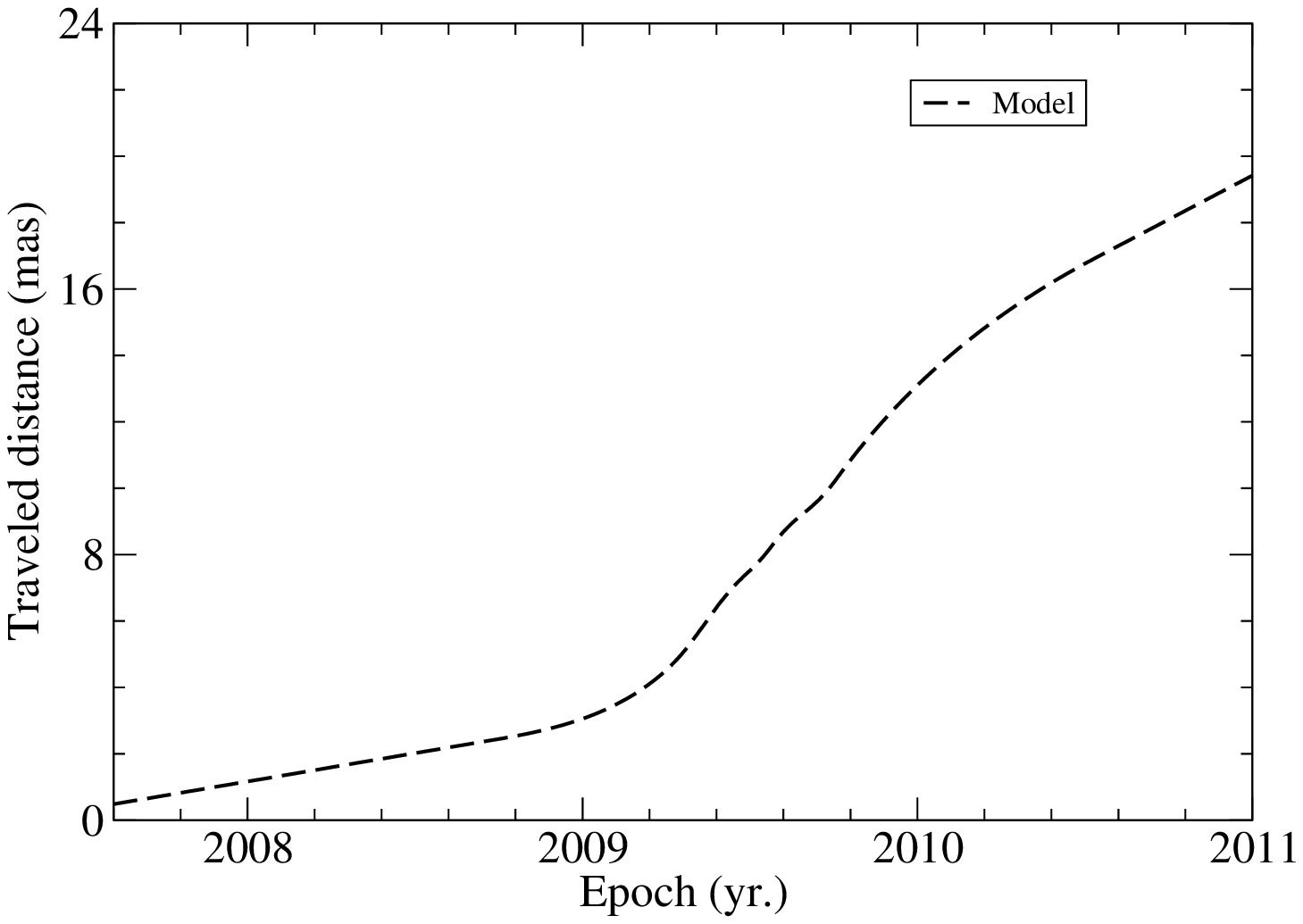}
    \includegraphics[width=6cm,angle=-90]{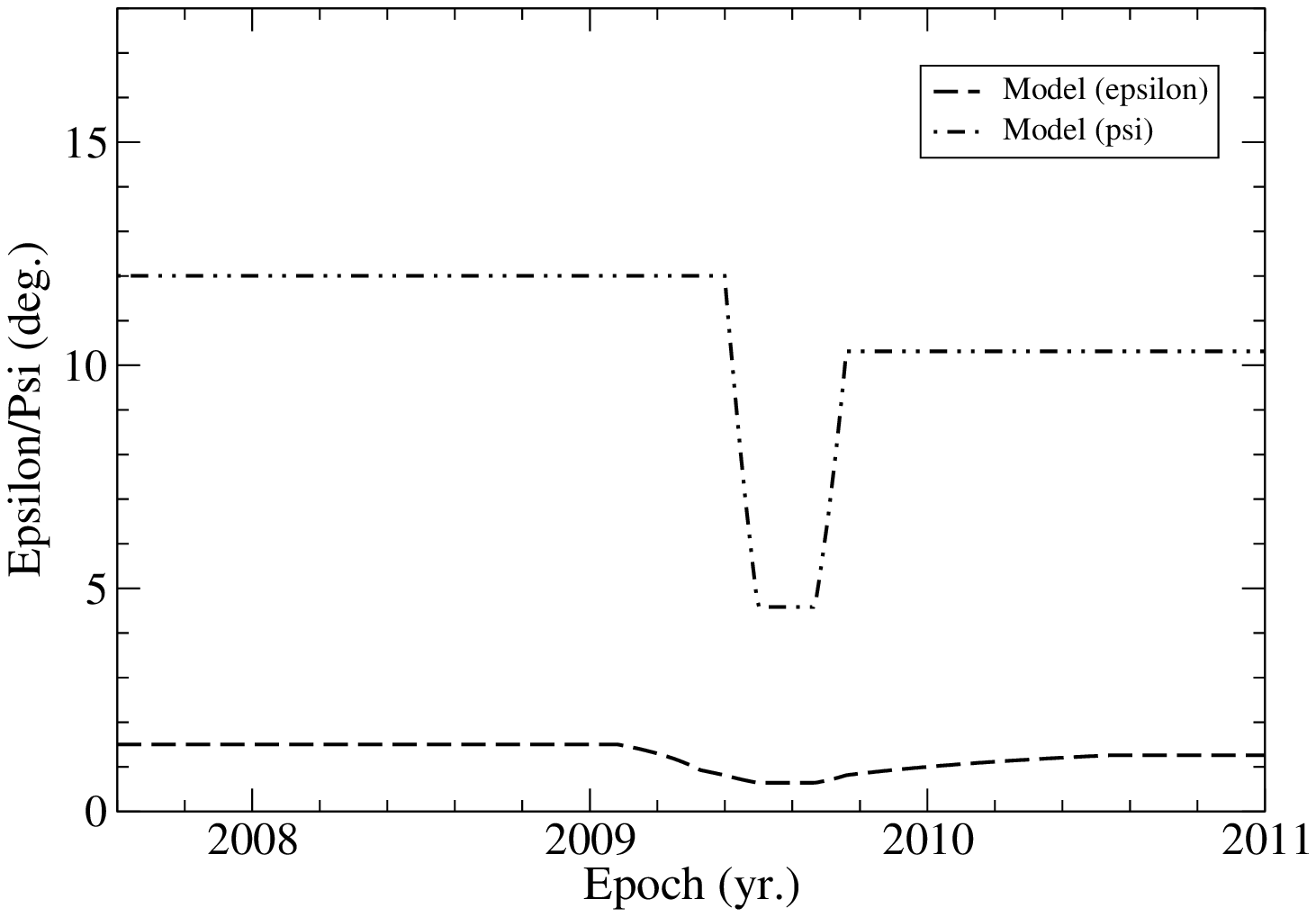}
    \caption{Knot C21: The model-derived traveled distance Z(t), parameters
    $\epsilon(t)$ and $\psi(t)$ as functions of time. Before 2009.08 
    ($X_n{\leq}$0.10\,mas) $\epsilon$=$1.50^{\circ}$ and $\psi$=$12.0^{\circ}$,
    knot C21 moved along the precessing common trajectory, while after 2009.08
    ($X_n{>}$0.10\,mas) $\epsilon$ started 
    to change ($\psi$ started to change after 2009.40), knot C21 started to
    move along its own individual track, deviating from the precessing 
    common track. Thus the transition from the precessing 
    common track to its own individual track occurred at $X_n$=0.10\,mas.
    The corrresponding traveled distance is Z=3.40\,mas=22.6\,pc.}
    \end{figure*}
      \begin{figure*}
    \centering
    \includegraphics[width=8cm,angle=-90]{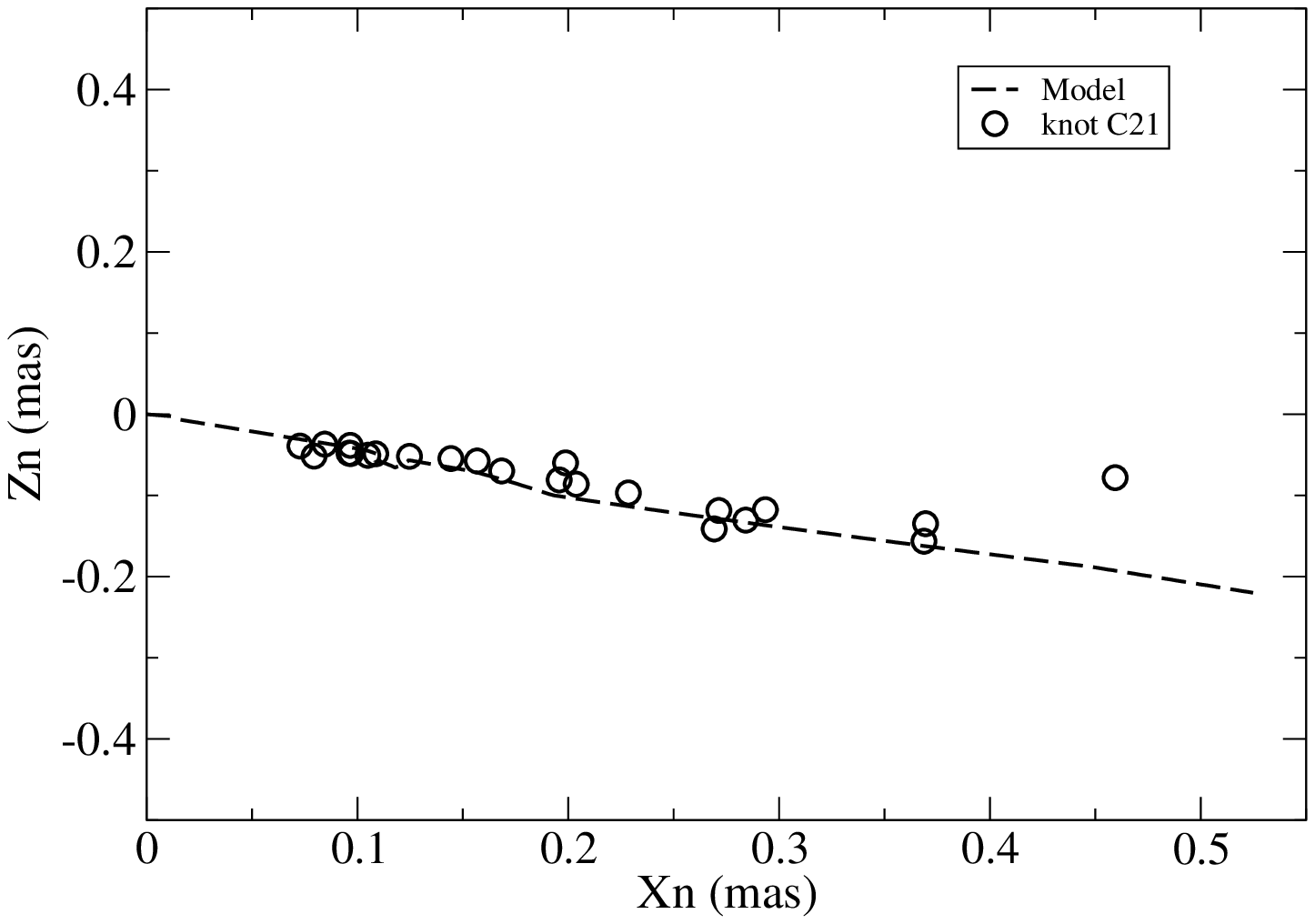}
    \caption{Knot C21: Model fitting of the entire trajectory $Z_n(X_n)$.
     Within $X_n$=0.10\,mas knot C21 moved along the precessing common 
    trajectory, while beyond $X_n$=0.10\,mas it started to move along its own
    individual track. Both the inner precessing common track and the outer 
    individual track are well fitted.}
    \end{figure*}
    \begin{figure*}
    \centering
    \includegraphics[width=6cm,angle=-90]{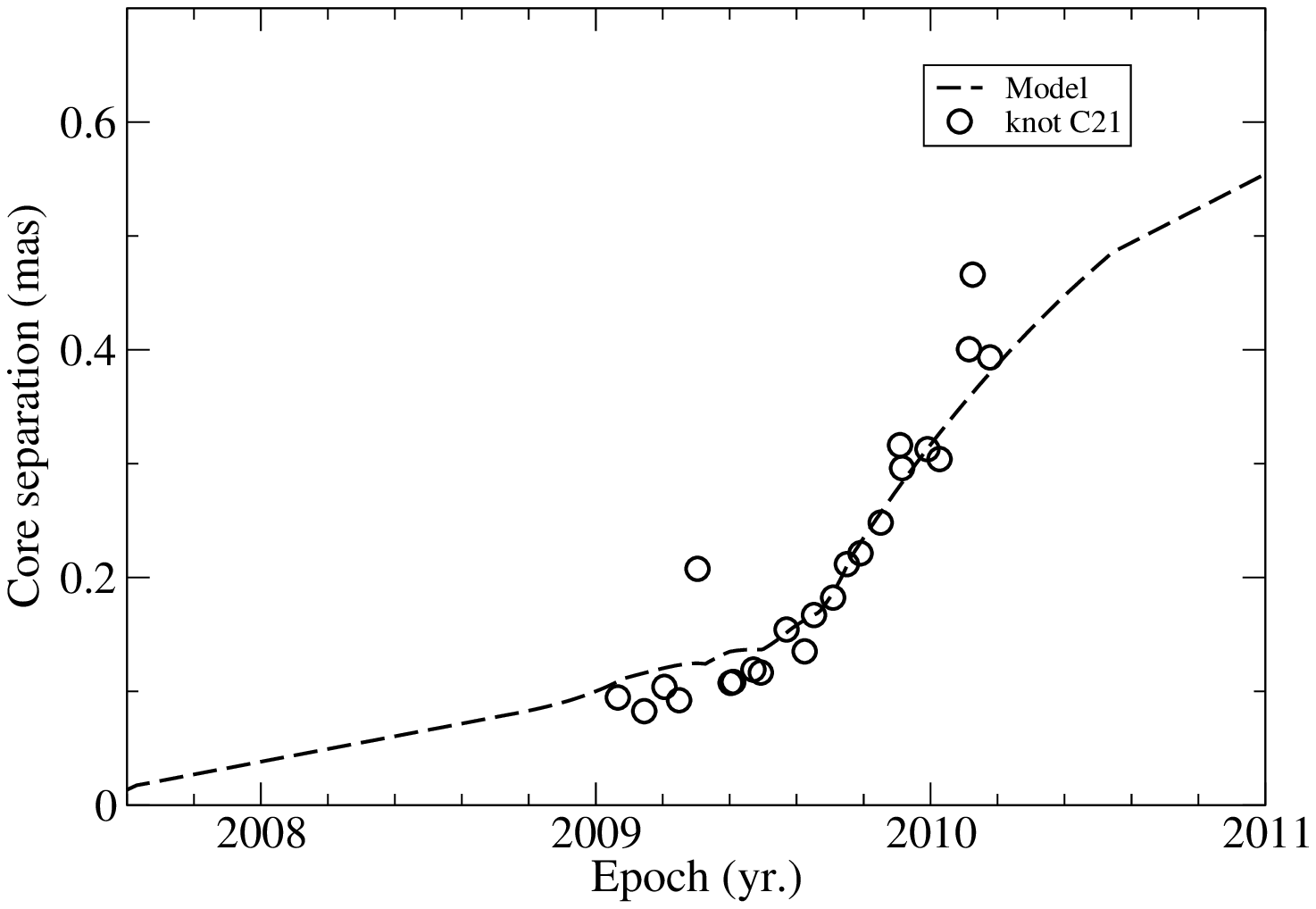}
    \includegraphics[width=6cm,angle=-90]{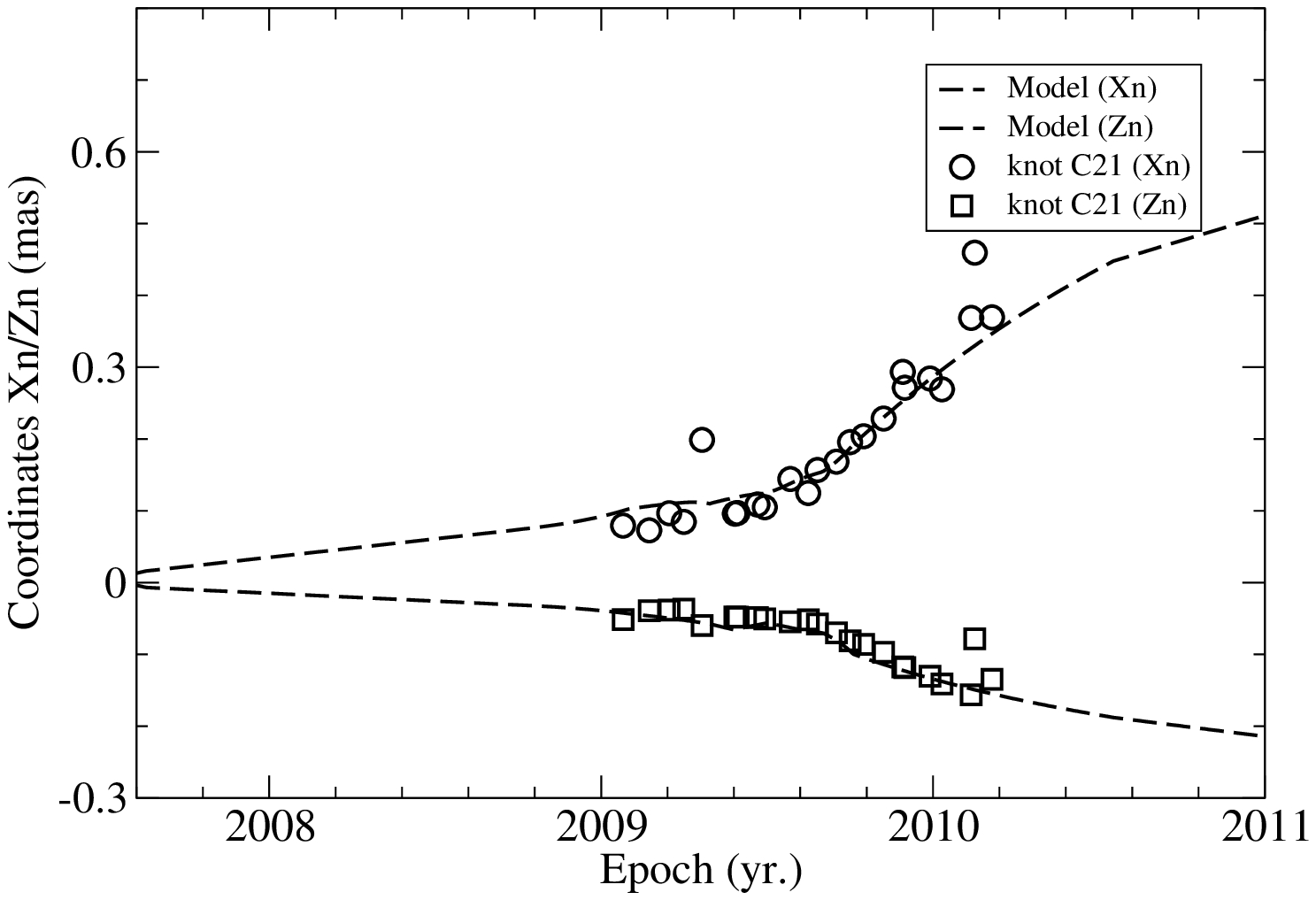}
    \caption{Knot C21: Model fitting of the core separation $r_n(t)$, 
    coordinates $X_n(t)$ and $Z_n(t)$ as functions of time. They are all 
    well fitted.}
    \end{figure*}
    \begin{figure*}
    \centering
    \includegraphics[width=6cm,angle=-90]{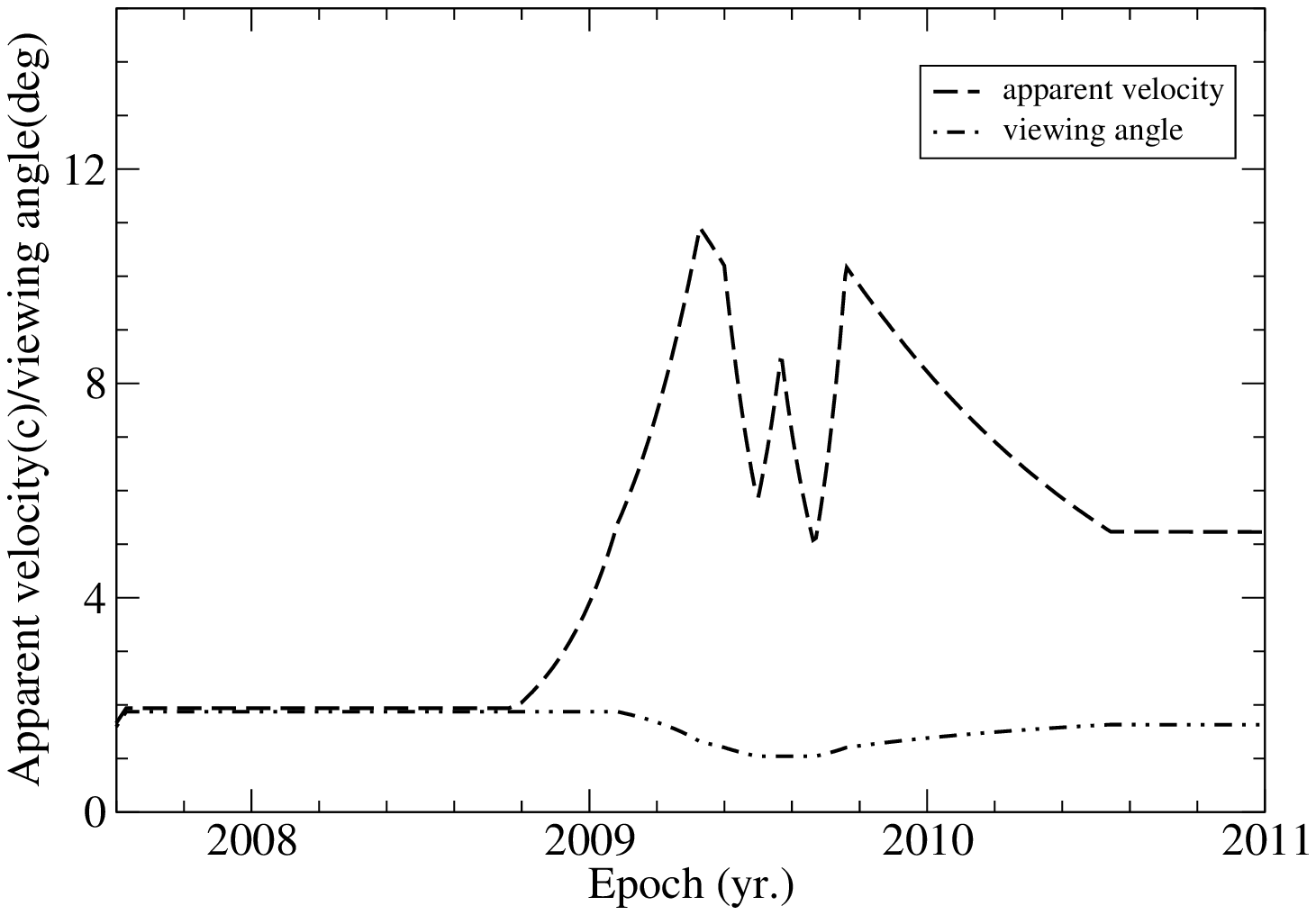}
    \includegraphics[width=6cm,angle=-90]{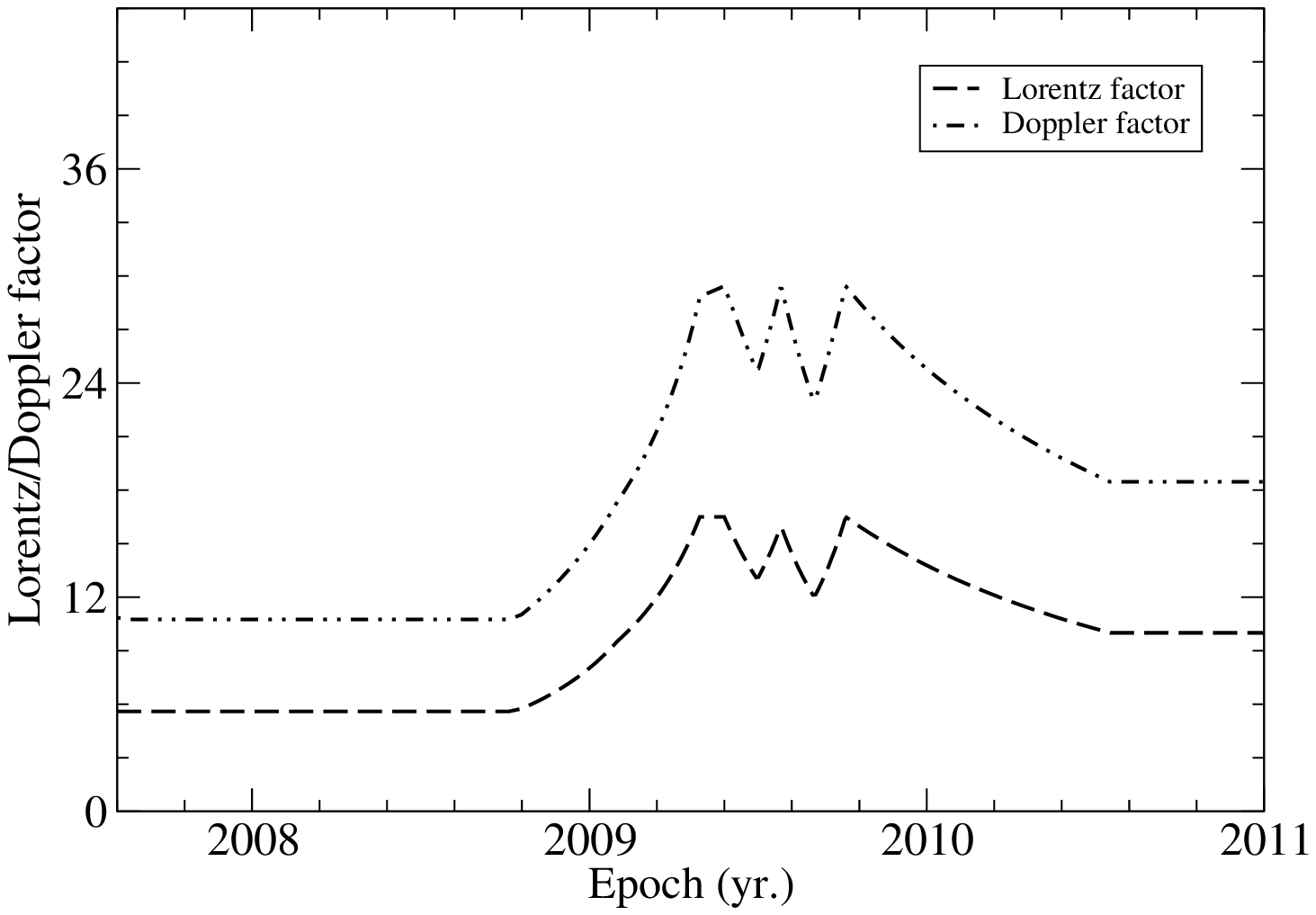}
    \caption{Knot C21. Left panel: the model-derived apparent speed 
    $\beta_{app}(t)$ and viewing angle $\theta(t)$ as functions 
    of time. The apparent speed  has three peaks $\beta_{app}=$10.91, 8.43
    and  10.16 at 2009.33, 2009.56 and 2009.76, respectively. The corresponding
   viewing angles are $\theta$=$1.32^{\circ}$, $1.04^{\circ}$ (a minimum) and
    $1.21^{\circ}$, respectively. Right panel:  the model-derived bulk Lorentz 
   factor $\Gamma(t)$ and Doppler factor $\delta(t)$ as continuous 
   functions of time.
   The Doppler factor has three peaks $\delta$=29.43 (at 2009.40), 29.29 (at
    2009.56) and 29.38 (at 2009.76). The corresponding bulk Lorentz
    factor $\Gamma$=16.50, 15.88 and  15.87, respectively.}
    \end{figure*}
    \begin{figure*}
    \centering
    \includegraphics[width=6cm,angle=-90]{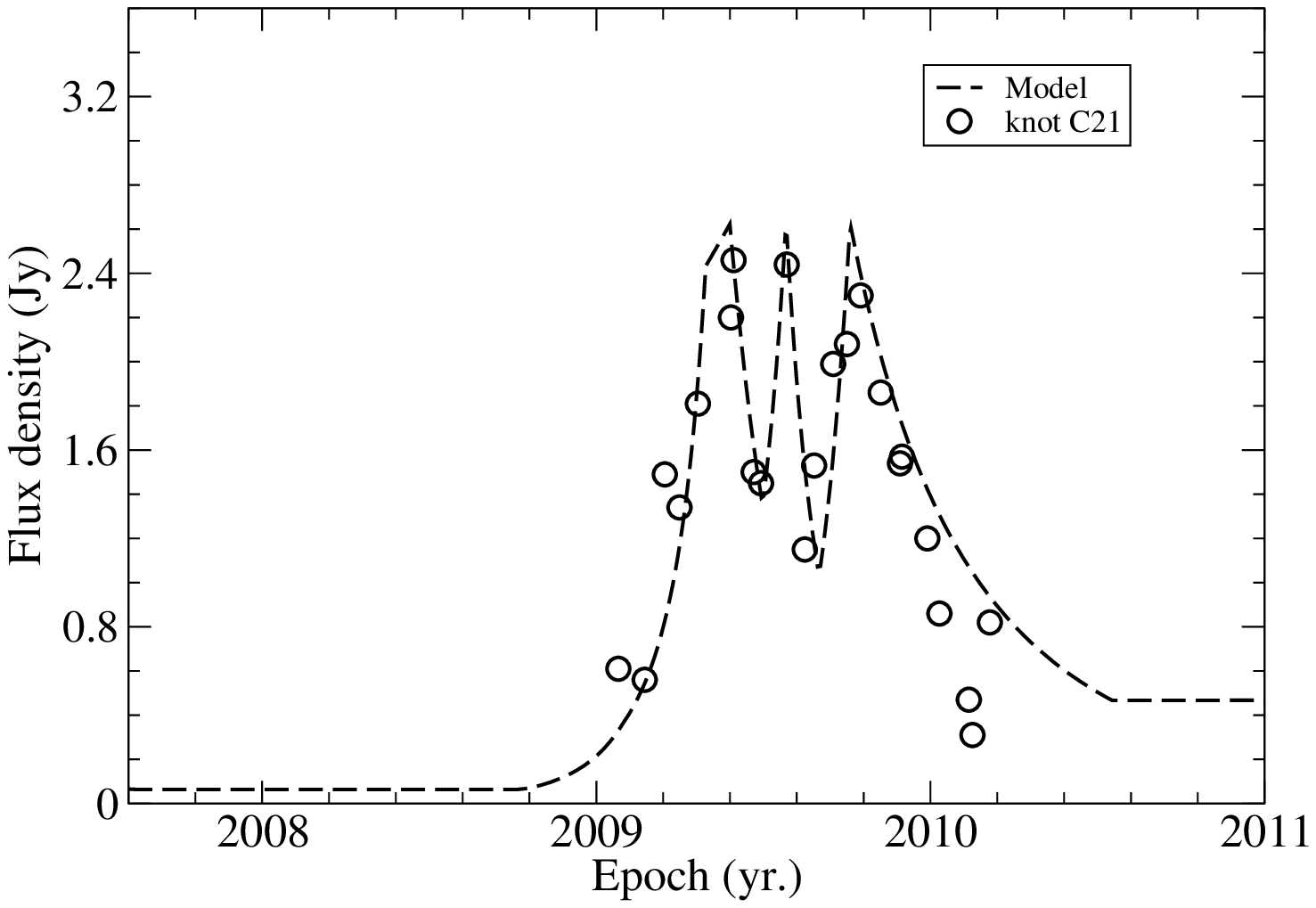}
    \includegraphics[width=6cm,angle=-90]{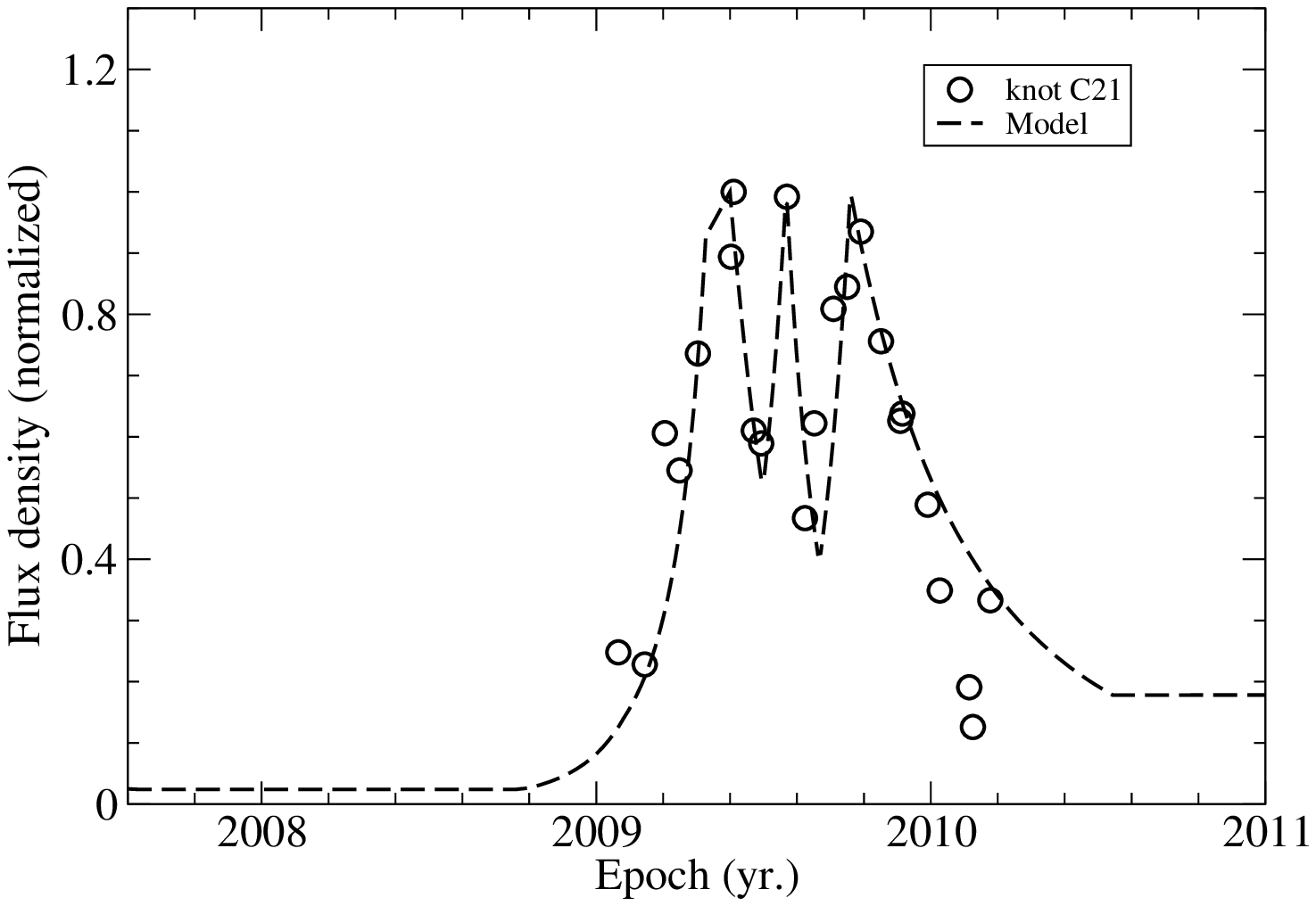}
    \caption{Knot C21: Model fitting of the 43\,GHz light curve (left 
    panel) and its corresponding normalized light curve (right panel). Both
     are well fitted  by the model-derived Doppler boosting profile
    $S_{int}[\delta(t)^{3+\alpha}]$ and $[\delta(t)/\delta_{max}]^{3+\alpha}$,
     respectively. An intrinsic flux density $S_{int}$=9.64\,$\mu$Jy and a 
    spectral index $\alpha$=0.70 are adopted. The modeled peak flux 
    densities are 2.62\,Jy (at 2009.40), 2.57\,Jy (at 2009.56) and 2.61\,Jy
     (at 2009.76), respectively.}
    \end{figure*}
    \section{Knot C21: Interpretation of kinematics and flux evolution}
   According to the precessing nozzle secnario for jet-B the ejection time
   of knot C21 is $t_0$=2007.32 and the corresponding precession phase
   $\phi_0$=3.90+2$\pi$. Its traveled distance Z(t) and modeled parameters
   $\epsilon(t)$ and $\psi(t)$ as continuous functions of time are shown in
   Figure 19. Before 2009.08 $\epsilon$=$1.50^{\circ}$ and 
   $\psi$=$12.0^{\circ}$, knot C21 moved along the precessing common track,
    which extends to the traveled distance Z=3.40\,mas=22.6\,pc. After 2009.08
   ($r_n{\geq}$0.11\,mas or $X_n{\geq}$0.10\,mas) parameter $\epsilon$ started 
   to change (parameter $\psi$ started to change after 2009.40) and knot C21
   started to follow its own individual track, deviating from the precessing 
    common track. That is, the 
   transition between the precessing common track and its individual track
   occurred at $X_n{\sim}$0.10\,mas.
    \subsection{Knot C21: model simulation of kinematics}
    The model-fitting results of the entire trajectory $Z_n(X_n)$, and the core 
   seperation $r_n(t)$ and coordinates $X_n(t)$ and $Z_n(t)$ as 
    functions of time are shown in Figures 20 and 21. They are all
   well fitted  in both inner and outer jet regions.\\
    The model-derived apparent speed $\beta_{app}(t)$ and viewing angle
    $\theta(t)$ as functions of time  are shown in Figure 22 (right panel),
    Both exhibit three peaks at 2009.33, 2009.56 and 2009.76 with 
    $\beta_{app}$=10.9, 8.4 and 10.2, respectively. The corresponding
    viewing angles are $\theta$=$1.32^{\circ}$, $1.04^{\circ}$ and 
    $1.20^{\circ}$, respectively.\\
    The model-derived bulk Lorentz factor $\Gamma(t)$ and Doppler factor
    $\delta(t)$ as continuous functions of time are shown in Figure 22 (left
    panel). Both also show three peaks at 2009.40, 2009.56 and 2009.76 with
    $\Gamma$=16.50, 15.87 and 16.46, respectively. The corresponding Doppler
    factor $\delta$=29.43, 29.29 and 29.38, respectively. These peaks are 
    well coincident with the peaks in the measured light curve.
    \subsection{Knot C21: Doppler boosting effect and flx evolution}
    The model fitting results of the measured 43\,GHz light curve and its 
    corresponding normalized light curve are shown in Figure 23. It can be
    seen that the entire light curve with three peaks (2.62\,Jy at 2009.40,
    2.57\,Jy at 2009.56 and 2.61 at 2009.76) is very well fitted 
    in terms of the model-derived Doppler boosting profile with an assumed 
    spectral index $\alpha$=0.70 and an intrinsic flux density 
    $S_{int}$=9.64$\mu$Jy, indicating that the modeled  Doppler boosting 
    effect may dominate the complex flaring event and the intrinsic emission 
    of knot C21 is quite stable. 
    \begin{figure*}
    \centering
    \includegraphics[width=5cm,angle=-90]{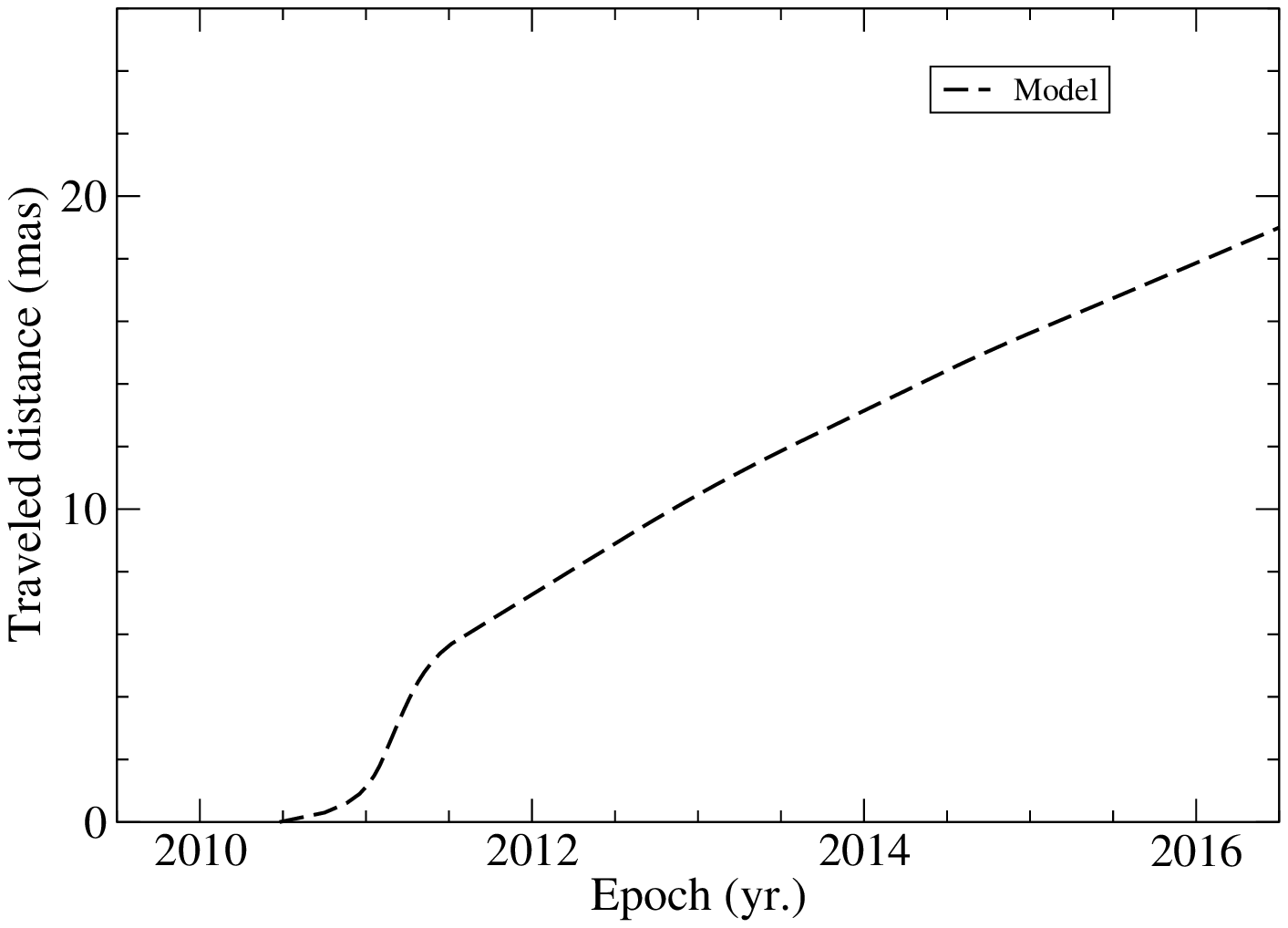}
    \includegraphics[width=5cm,angle=-90]{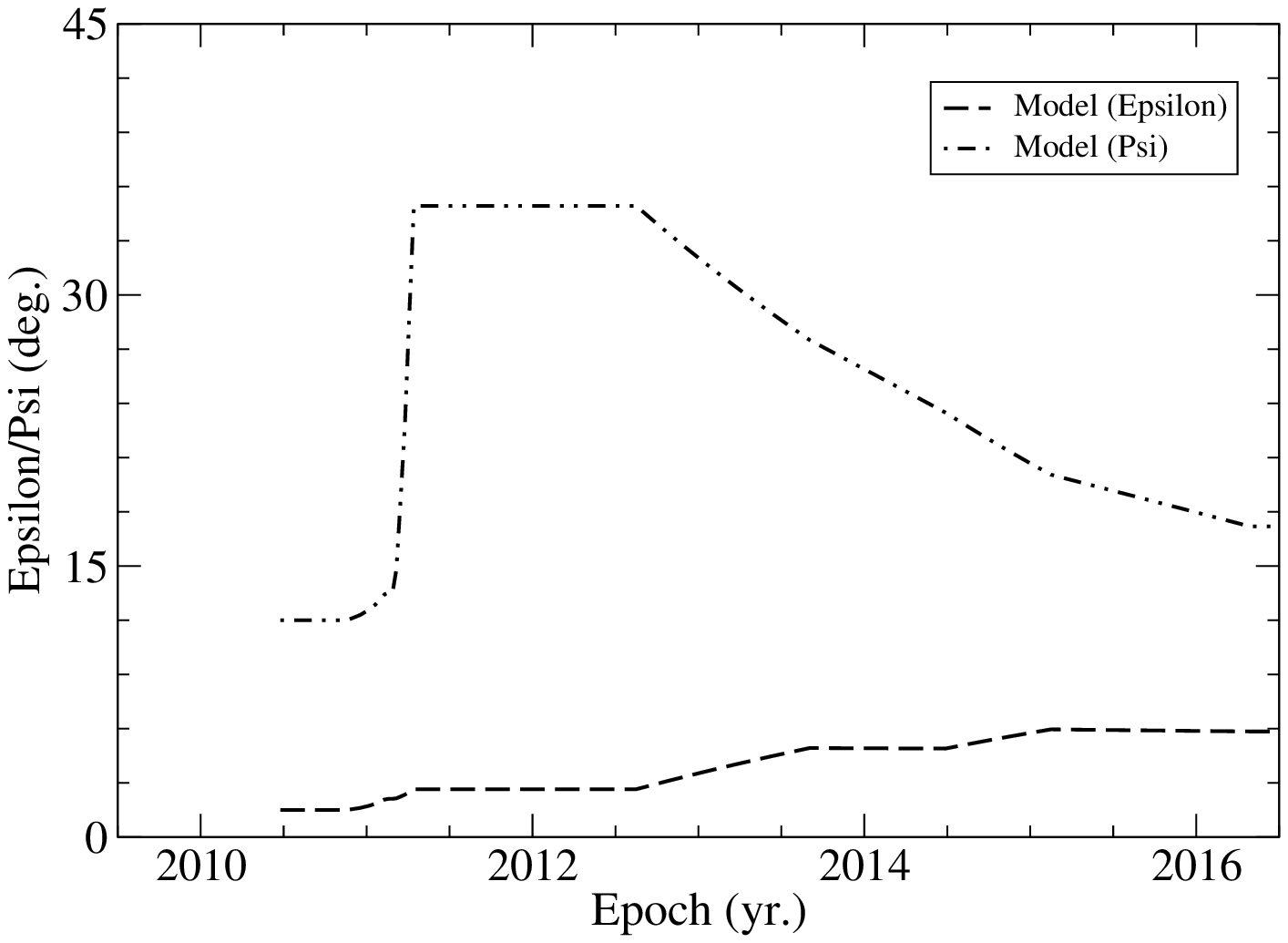}
    \caption{Knot B5: The modeled traveled distance Z(t),  parameters
    $\epsilon(t)$ and $\psi(t)$ as functions of time. Before
    2010.88 $\epsilon$=$1.50^{\circ}$ and $\psi$=$12.0^{\circ}$, knot B5 moved
    along the precessing common trajectory, while after that epoch 
    both $\epsilon$ and $\psi$ started to change and knot B5 started 
    to move along its own 
    individual track, deviating from the common track. Thus the transition 
     from the precessing common track to its own individual track occurred 
    at $\sim$2010.88 ($X_n$=0.015\,mas), corresponding to a 
    traveled distance Z=0.6\,mas=4.0\,pc. Only its individual track in the 
    outer jet region was observed.}
        \end{figure*}
     \begin{figure*}
    \centering 
    \includegraphics[width=6cm,angle=-90]{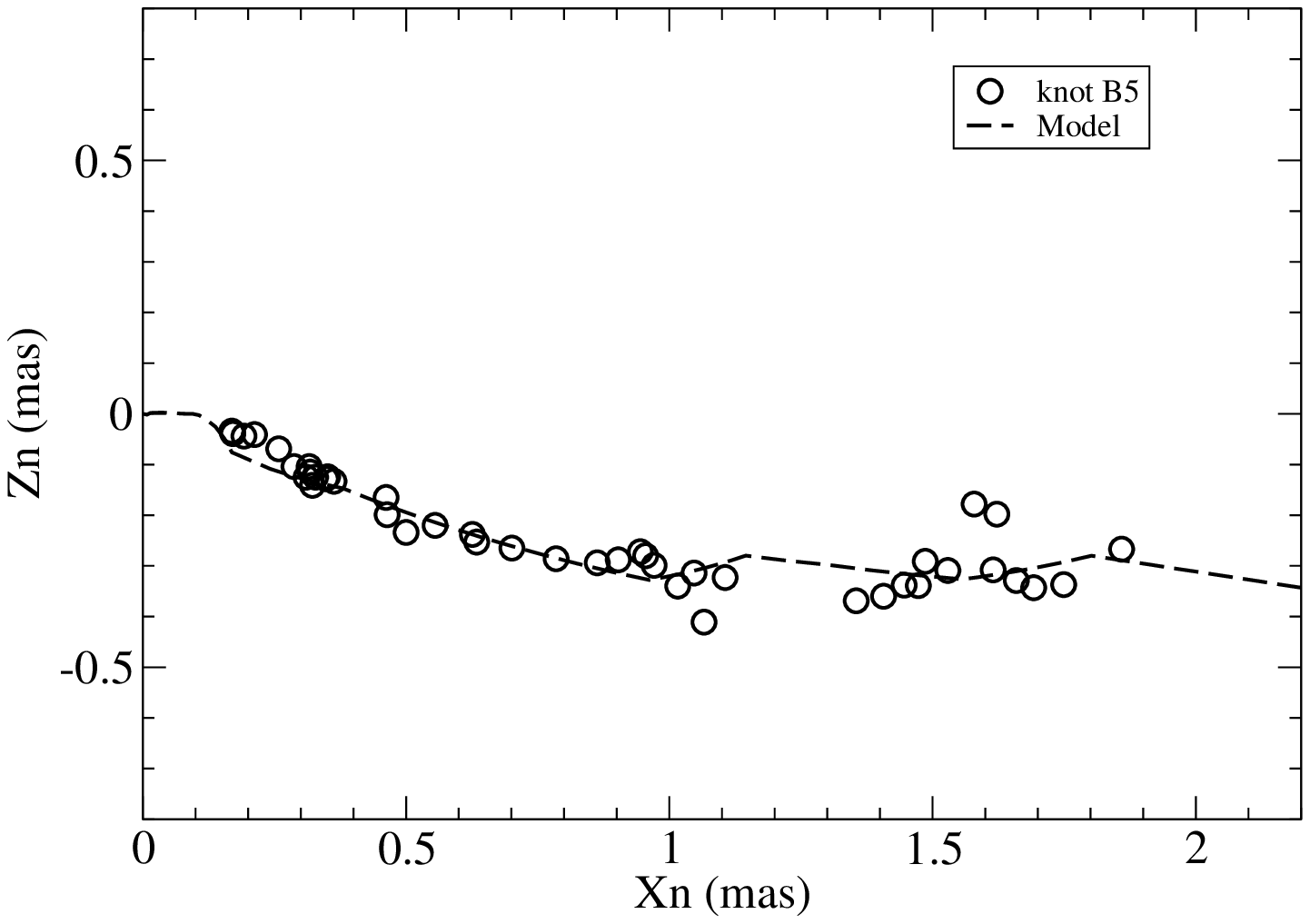}
    \caption{Knot B5: Model fitting of the entire trajectory $Z_n(X_n)$.
    Within $X_n$=0.015\,mas knot B5 moved along the precessing common 
    trajectory (with a precession phase $\phi_0$=0.33+4$\pi$), while beyond 
    $X_n$=0.015\,mas it started to move along its own individual track.
    The entire trajectory is very well fitted, but only its individual track
    in the outer jet region was observed.}
    \end{figure*}
    \begin{figure*}
     \centering
    \includegraphics[width=5cm,angle=-90]{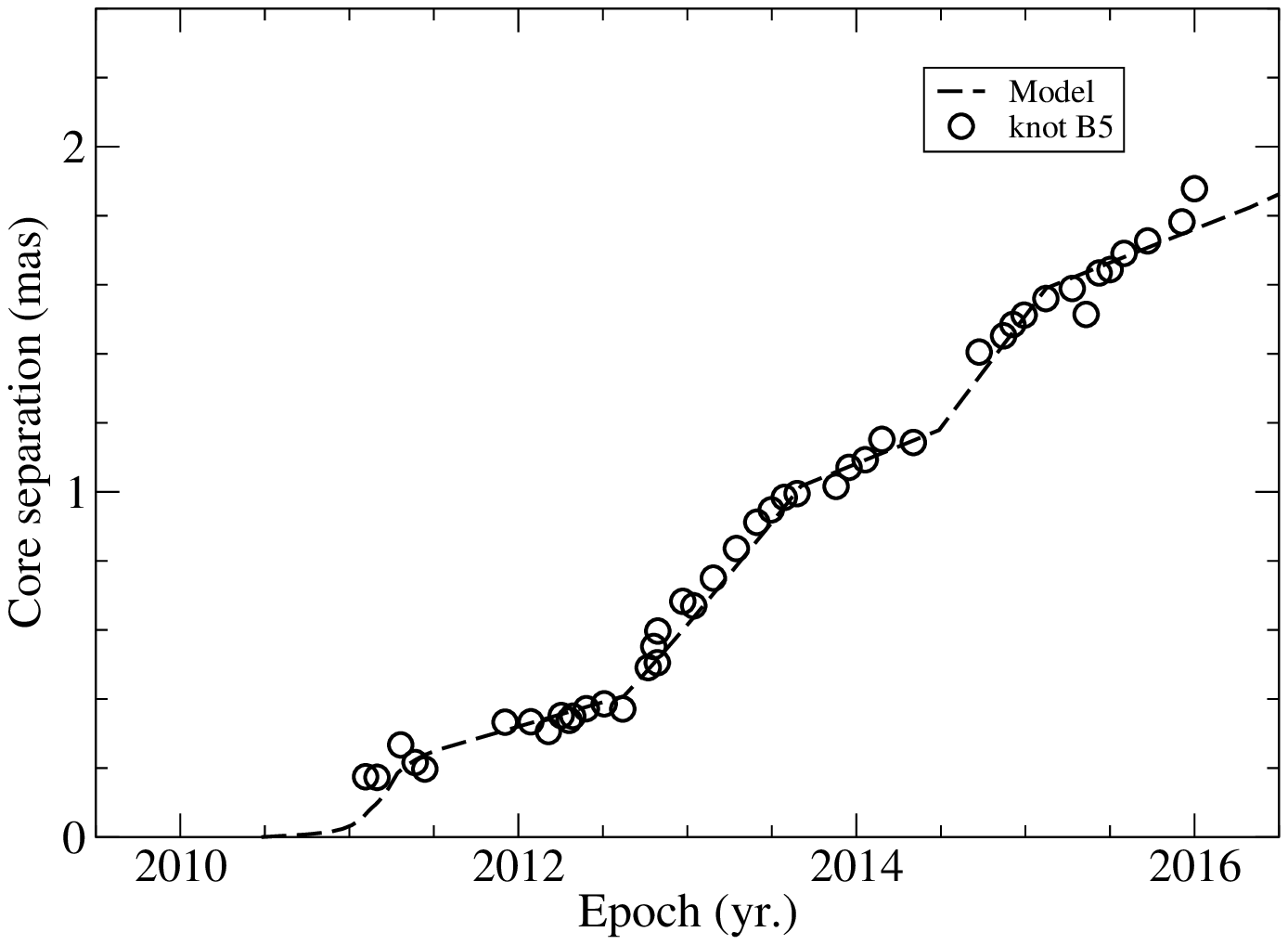}
     \includegraphics[width=5cm,angle=-90]{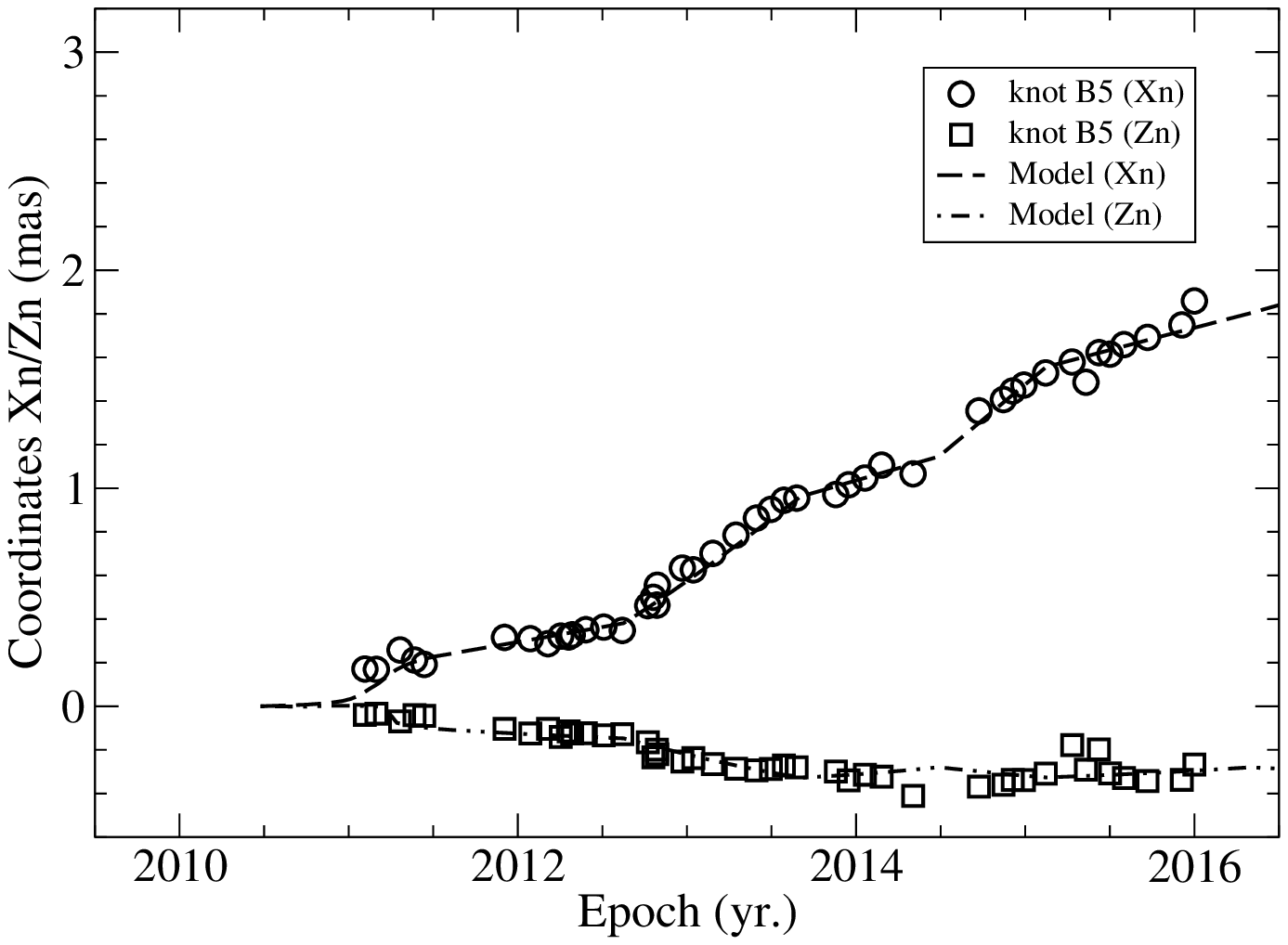}
    \caption{Knot B5: Model fitting of the core separation $r_n(t)$, 
    coordinates $X_n(t)$ and $Z_n(t)$ as  functions of time. They are all
    well fitted. Only its individual track in the outer jet region 
   (after 2010.88, $X_n{\geq}$0.015\,mas) was observed. }
    \end{figure*}
    \begin{figure*}
    \centering
     \includegraphics[width=5cm,angle=-90]{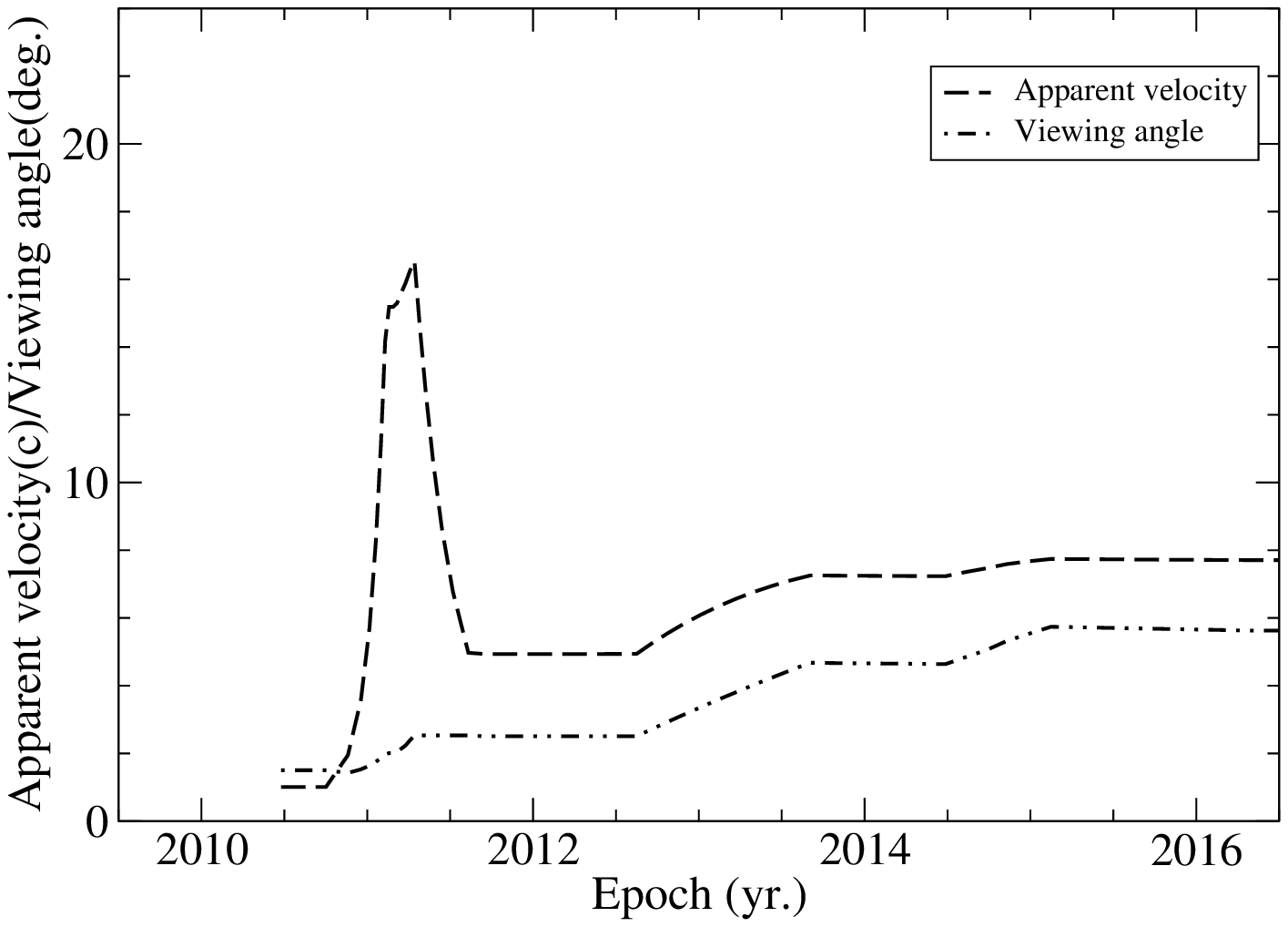}
     \includegraphics[width=5cm,angle=-90]{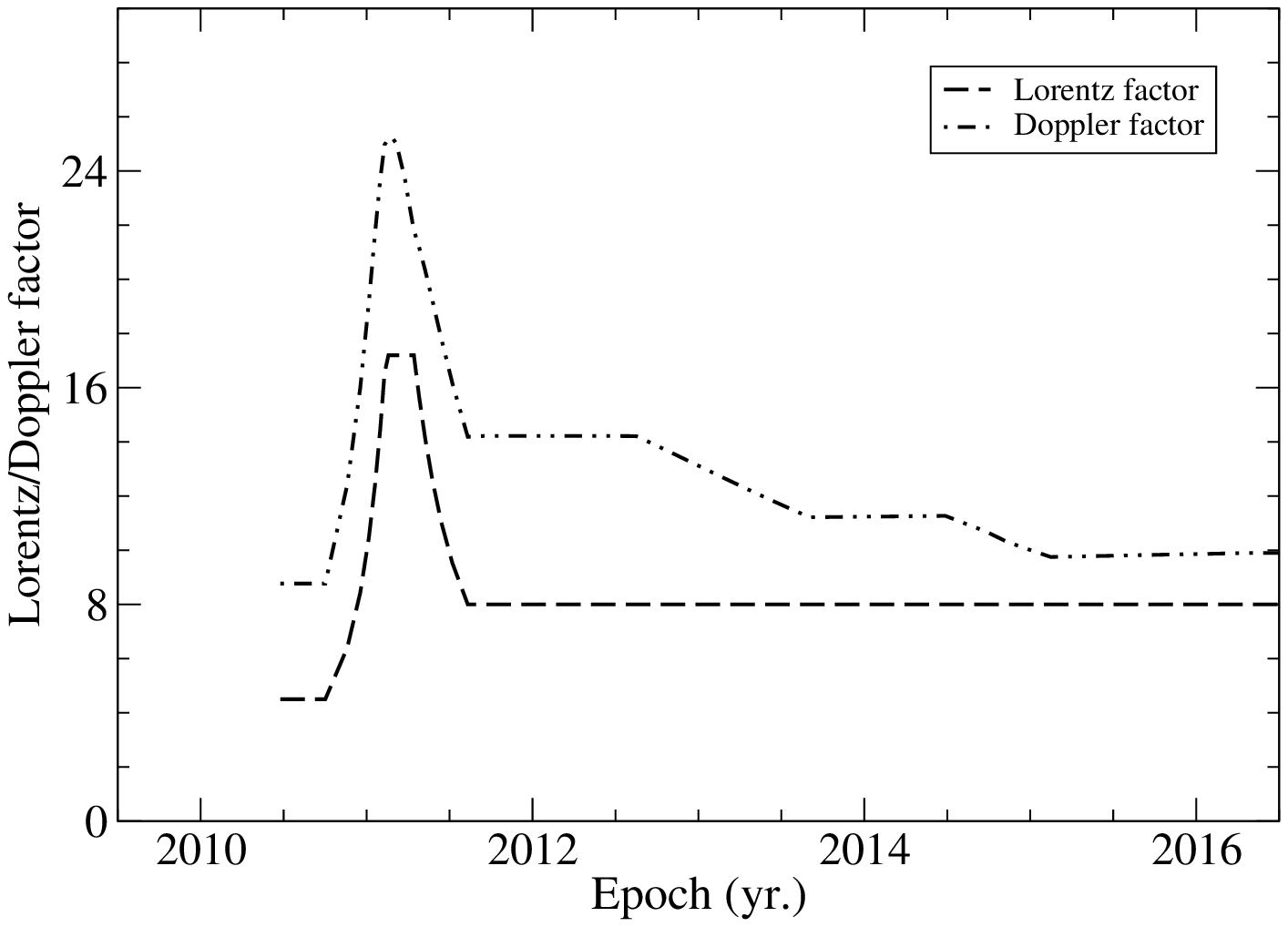}
    \caption{Knot B5. Left panel: the model-derived apparent speed 
    $\beta_{app}(t)$ and viewing angle $\theta(t)$ as functions of 
    time. The apparent velocity has a maximum $\beta_{app,max}$=16.54 at 
    2011.28 and the corresponding viewing angle $\theta$=$2.53^{\circ}$,
    $\Gamma$=17.20 and $\delta$=21.81.
       Right panel: the model-derived bulk Lorentz factor $\Gamma(t)$  and
    Doppler factor $\delta(t)$ as  continuous functions of time. 
    The Doppler factor has a maximum $\delta_{max}$=25.21 at 2011.13 and 
    the corresponding Lorentz factor $\Gamma$=17.20 and
     viewing angle $\theta$=$2.01^{\circ}$ (a local minimum). The 
    Lorentz factor $\Gamma$=17.20 (during 2011.13--2011.28, at maximum),
     coincident with the maximal Doppler factor.}
    \end{figure*}
    \begin{figure*}
    \centering
     \includegraphics[width=5cm,angle=-90]{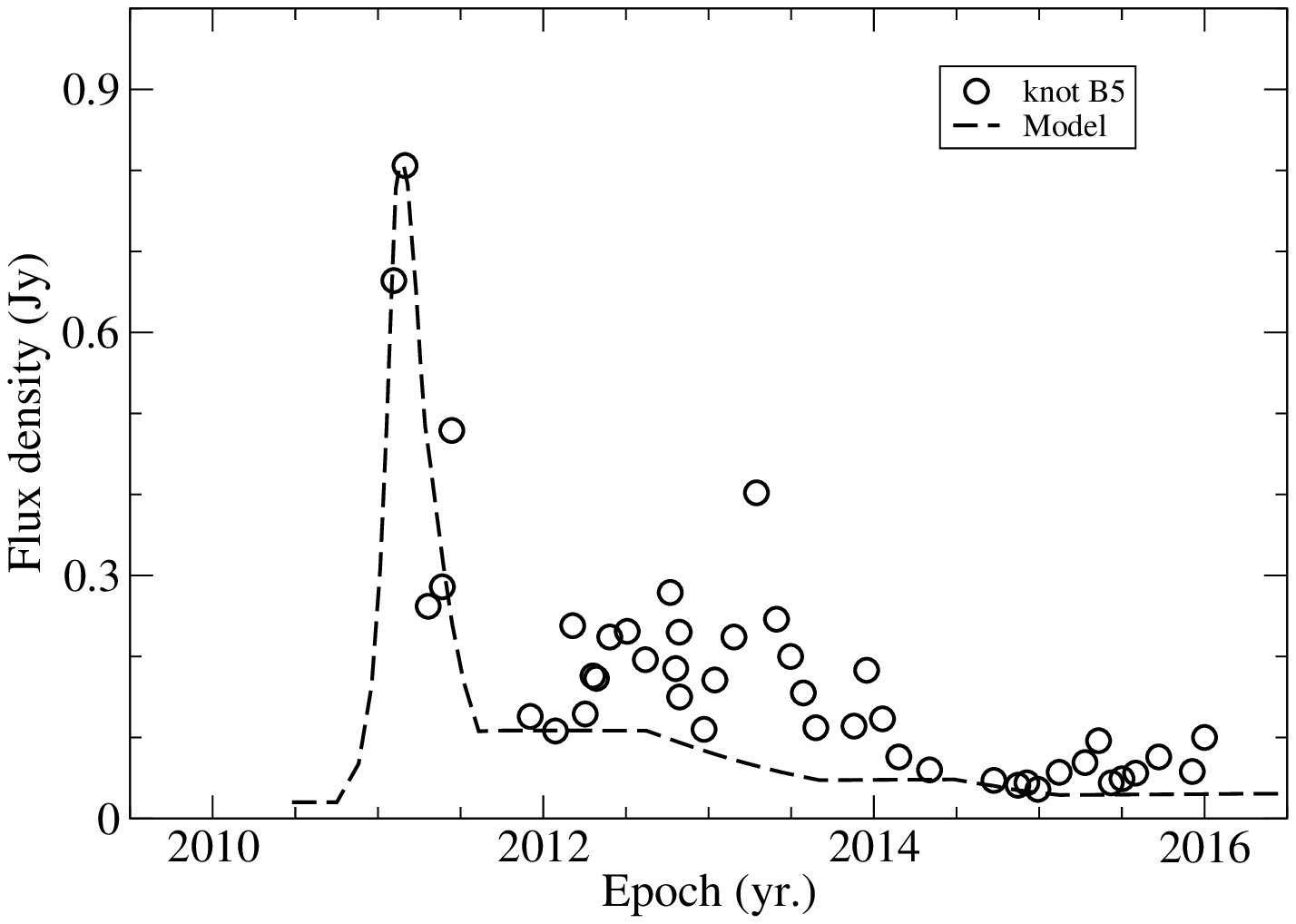}
     \includegraphics[width=5cm,angle=-90]{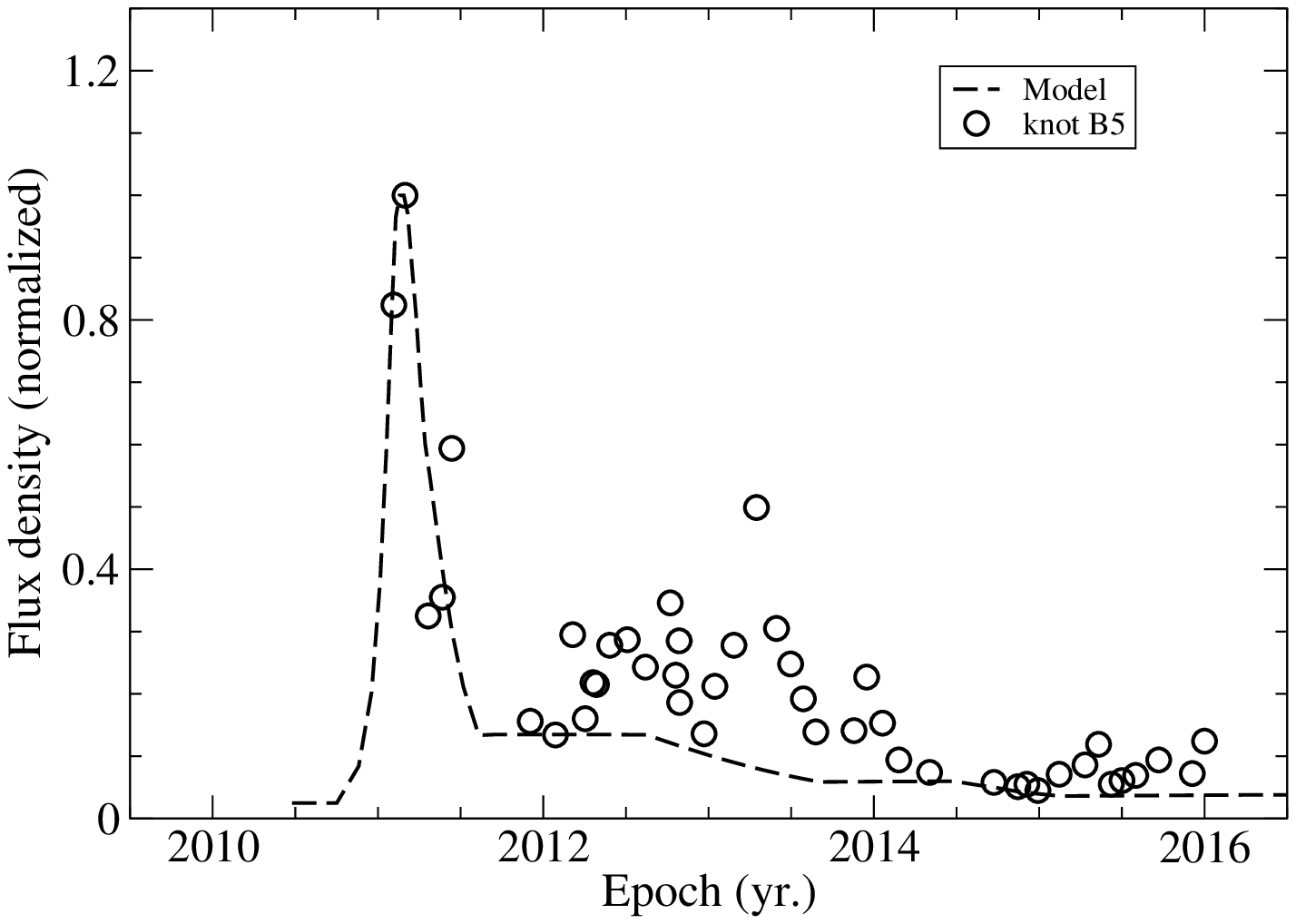}
     \caption{Knot B5: Model fitting of the 43\,GHz light curve (left
     panel) and its corresponding normalized light curve (right panel).
     Both are well fitted by the model-derived Doppler boosting profile
     $S_{int}[\delta(t)]^{3+\alpha}$ and $[\delta(t)/\delta_{max}]^{3+\alpha}$,
     respectively. An intrinsic flux density $S_{int}$=12.9\,$\mu$Jy and
     a  spectral index $\alpha$=0.50 were adopted. The rapid
     flux density fluctuations during 2012--2014 may be due to the
      variations in its intrinsic flux density and spectral index.}
    \end{figure*}
   \section{Knot B5: Interpretation of kinematics and flux evolution}
   According to the precessing nozzle scenario for jet-B the ejection time of
   knot B5 is $t_0$=2010.48 and the corresponding precession phase 
   $\phi_0$=0.33+4$\pi$. Its traveled distance Z(t) and modeled parameters
   $\epsilon(t)$ and $\psi(t)$  as functions of time are shown in Figure 24.
   Before 2010.88 ($X_n{\leq}$0.015\,mas) $\epsilon$=$1.50^{\circ}$ and
    $\psi$=$12.0^{\circ}$ knot B5 moved along the precessing common trajectory
    which extends to the traveled distance Z=0.60\,mas=4.0\,pc. After that epoch
   both $\epsilon$ and $\psi$ started to change and knot B5 started to 
  follow its own individual track, deviating from the precessing common track.
    That is, the transition between the  precessing common track and its
    individual track occurred very near to the core (at 
   $r_n{\sim}{X_n}$=0.015\,mas) and the Doppler-boosting effect occurred 
     completely in the outer jet regions. 
   \subsection{Knot B5: Model simulation of kinematics}
   The model-fitting results of the entire trajectory $Z_n(X_n)$ and the core
    separation $r_n(t)$, coordinates $X_n(t)$ and $Z_n(t)$ as 
    functions of time are shwon in Figures 25 and 26. They are all
    well fitted during the five years ($\sim$2011.0--2016.0). The rapid
    increase in core distance $r_n$ during $\sim$2012.8--2014.0 was mostly 
    induced by the increase in the viewing angle $\theta$.\\
   The model-derived  apparent velocity $\beta_{app}(t)$ and viewing angle 
   $\theta(t)$  as continuous functions of time are shown in Figure 27 (left
   panel). The maximal apparent speed $\beta_{app,max}$=16.54 occurred 
   at 2011.28 and the corresponding viewing angle $\theta$=$2.53^{\circ}$,
   $\Gamma$=17.2 and $\delta$=21.81.\\
    The model-derived bulk Lorentz factor $\Gamma(t)$ and Doppler factor
   $\delta(t)$ as continuous functions of time are shown in Figure 27 (right 
   panel). the maximal Lorentz factor $\Gamma_{max}$=17.20 (during 
   2011.13--2011.28). 
    The maximal Doppler factor $\delta_{max}$=25.21 at 2011.13 and the 
   corresponding viewing angle $\theta$=$2.01^{\circ}$ ( a local minimum).
   \subsection{Knot B5: Doppler boosting effect and flux evolution}
   The model fitting results of the measured 43\,GHz light curve and its 
   corresponding normalized light curve are shown in Figure 28. The light
   curve of the radio burst measured during 2011--2011.4 is well
   fitted  in terms of the Doppler boosting effect with an assumed spectral
   index $\alpha$=0.50 and an intrinsic flux density $S_{int}$=12.9\,$\mu$Jy.
   The flux fluctuations observed during $\sim$2012.2--2014.2 may be due to
   the variations in the intrinsic flux density of knot B5.
      \begin{figure*}
     \centering
     \includegraphics[width=5cm,angle=-90]{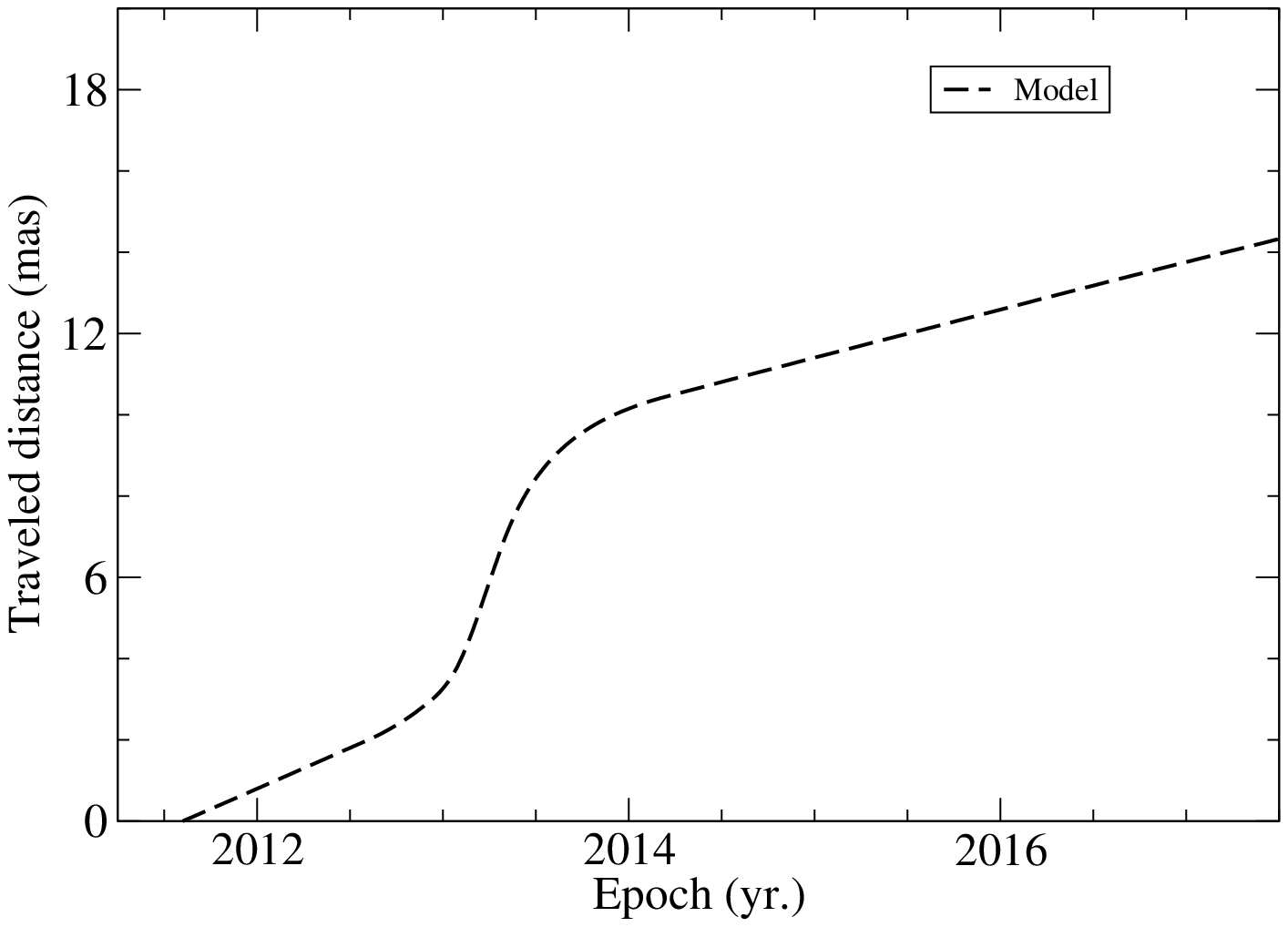}
     \includegraphics[width=5cm,angle=-90]{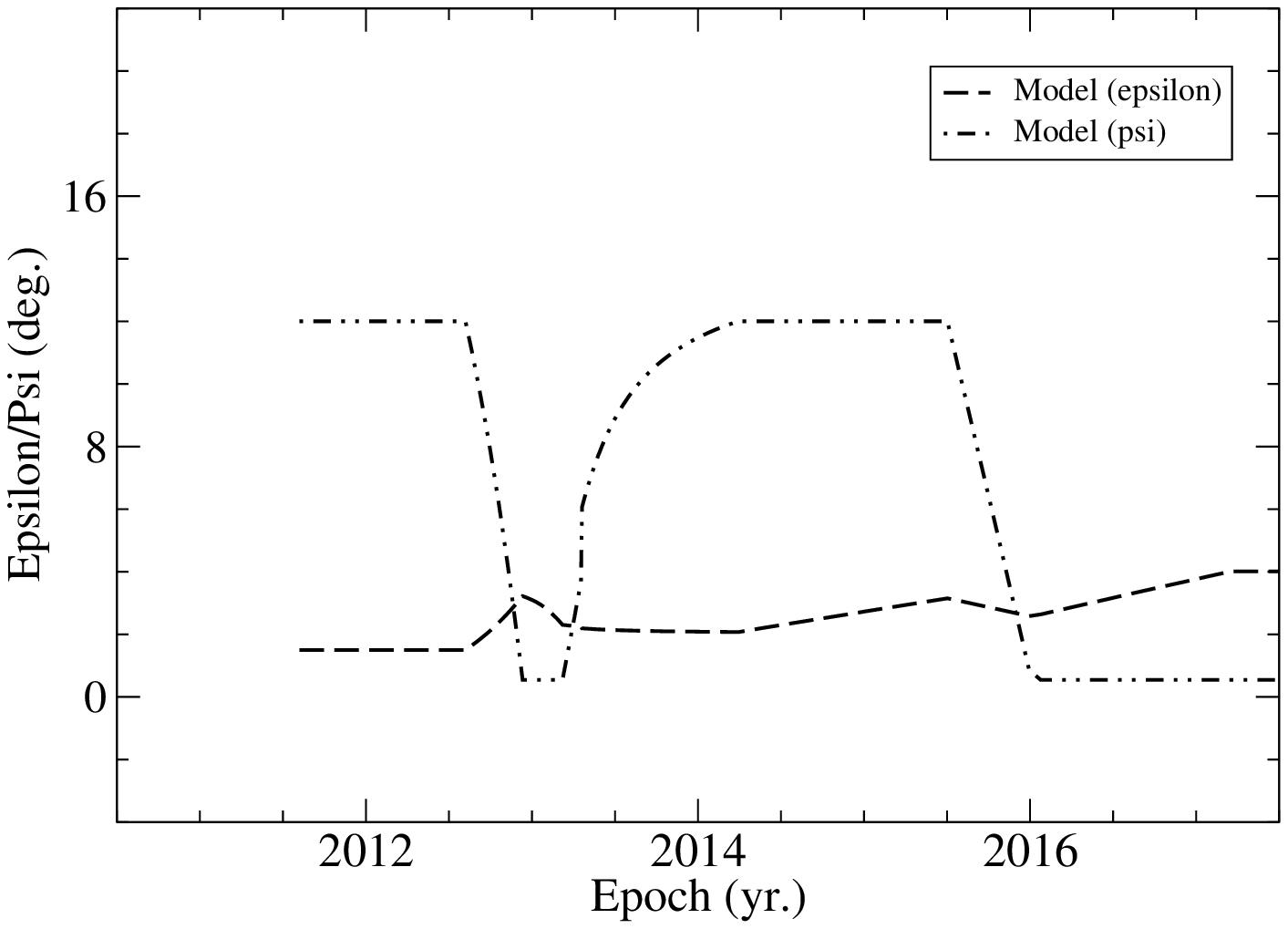}
     \caption{Knot B7: the modeled traveled distance Z(t), parameters
     $\epsilon(t)$ and $\psi(t)$ as functions of time. Before 2012.60 
     $\epsilon$=$1.50^{\circ}$ and $\psi$=$12.0^{\circ}$, knot B7 moved along 
    the precessing common trajectory, while
    after 2012.60 both parameters started to change and knot B7 started to move
    along its own individual track, deviating from the precessing common track. Thus
    its transition from the precessing common trajectory to its individual
   track occurred at $X_n$=0.036\,mas, corresponding to a traveled distance 
     Z=2.0\,mas=13.3\,pc.}
    \end{figure*}
     \begin{figure*}
     \centering
    \includegraphics[width=6cm,angle=-90]{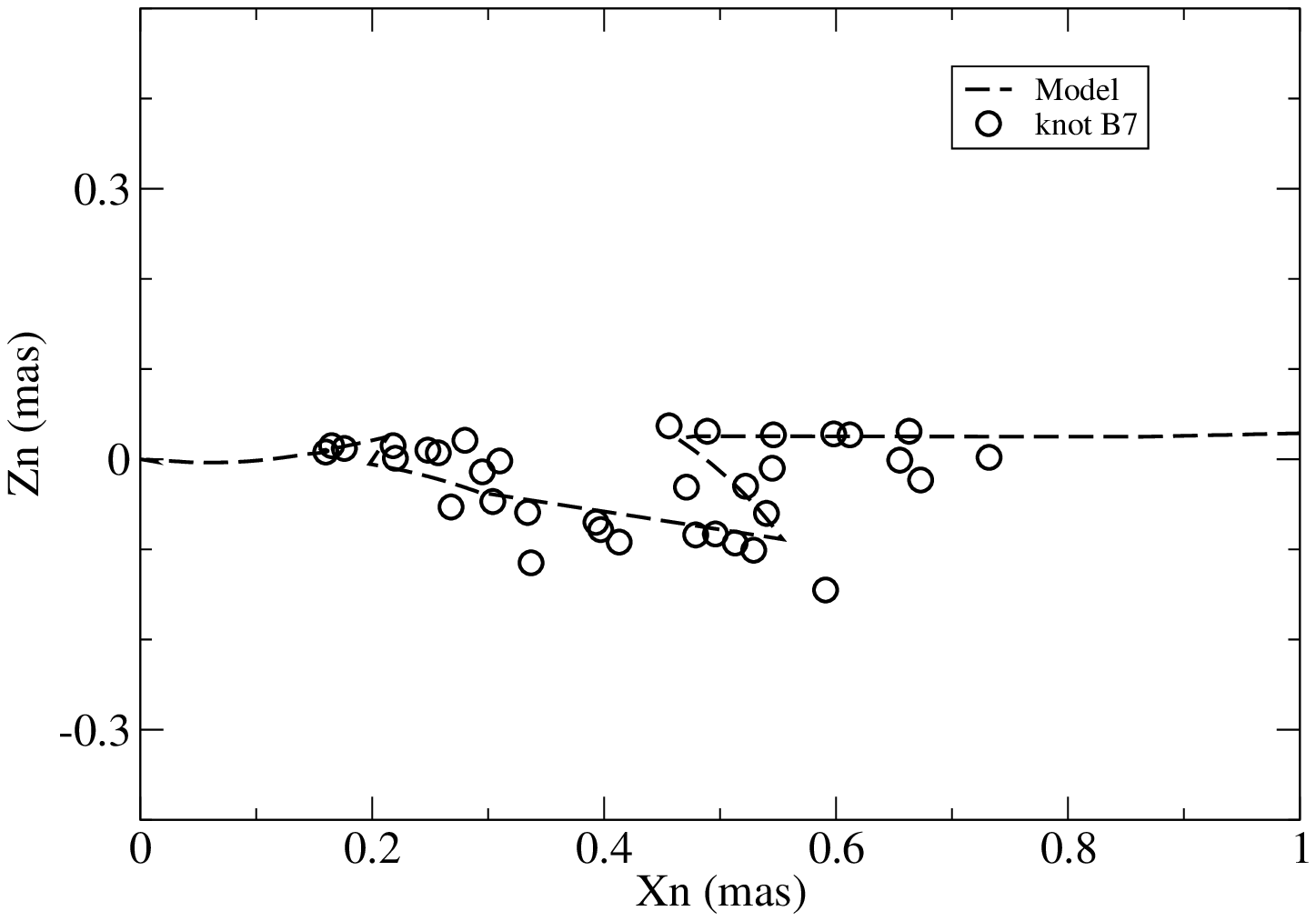}
     \caption{Knot B7: Model fitting of the entire trajectory $Z_n(X_n)$.
     Within $X_n$=0.036\,mas knot B7 moved along the precessing common 
     trajectory, while beyond $X_n$=0.036\,mas it started to move along its own
     individual trajectory. Thus only the individual track was observed and its
      precessing common track was not observed. The entire trajectory was 
     very well fitted, especially the
     trajectory section of its backward motion near $X_n{\sim}$0.5\,mas. }
     \end{figure*}
     \begin{figure*}
      \centering
      \includegraphics[width=5cm,angle=-90]{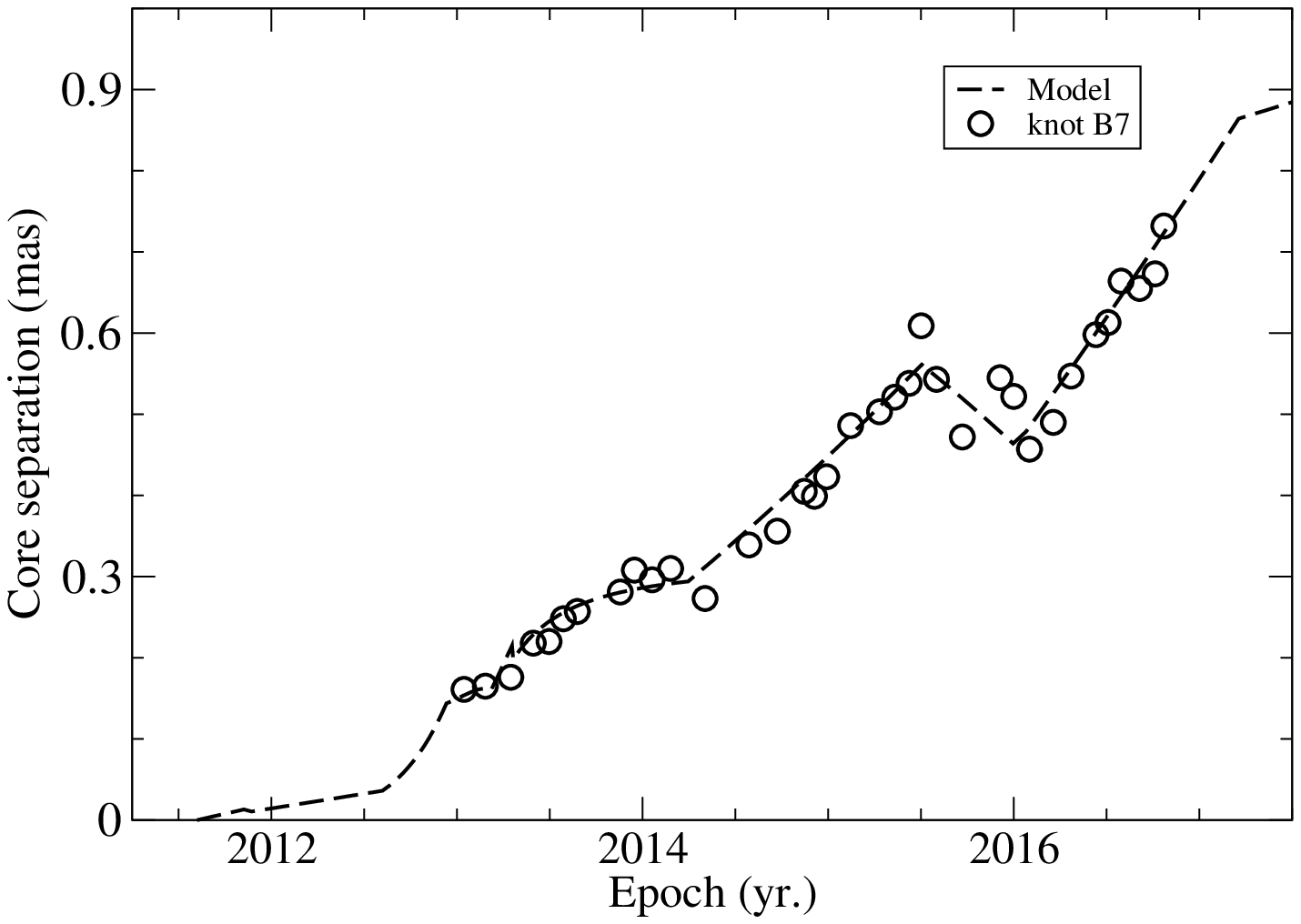}
      \includegraphics[width=5cm,angle=-90]{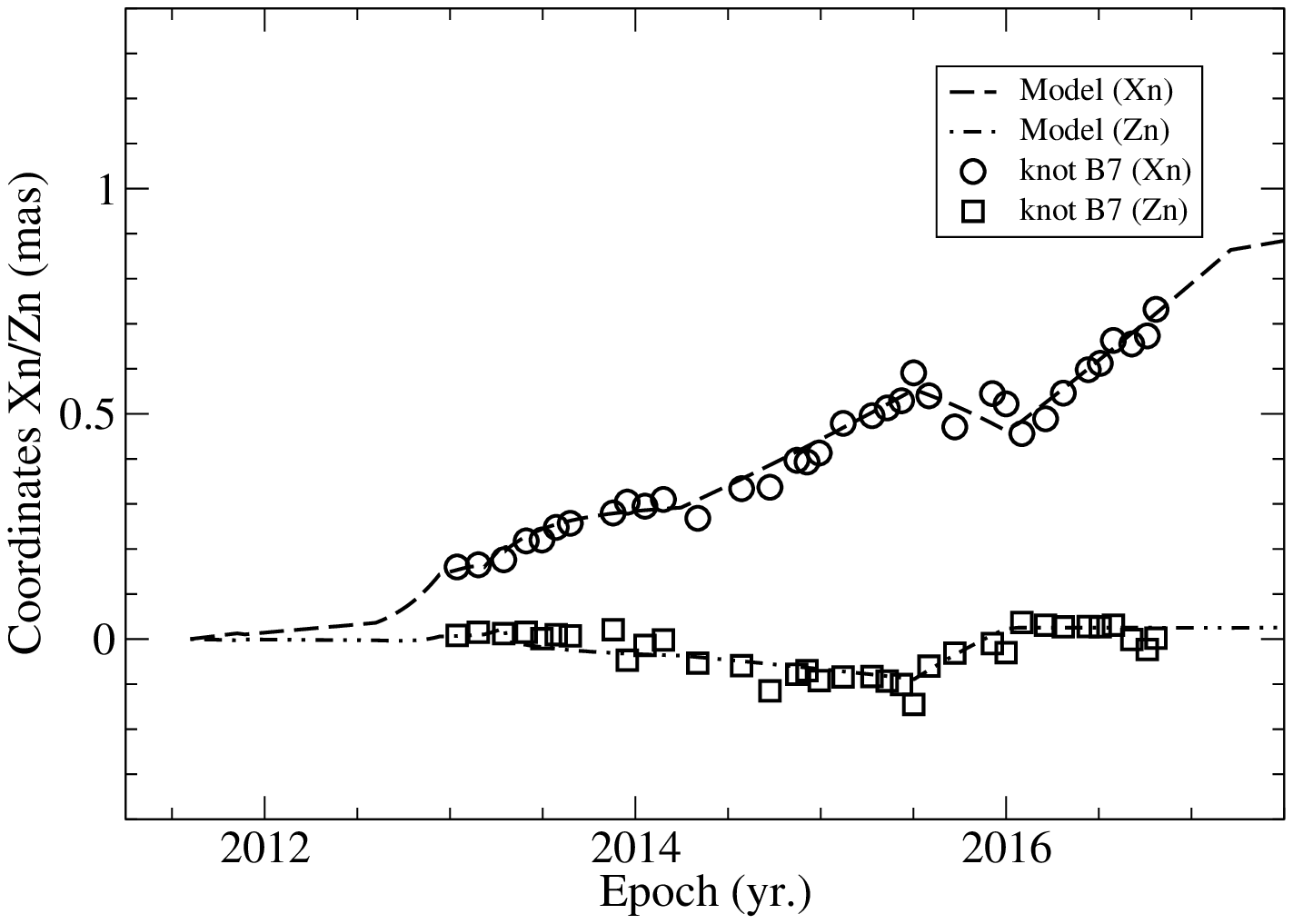}
      \caption{Knot B7: Model fitting of the core separation $r_n(t)$, 
      coordinates $X_n(t)$ and $Z_n(t)$. They are all well fitted,
      especially for its backward motion near epoch 2016.}
      \end{figure*}
      \begin{figure*}
      \centering
       \includegraphics[width=5cm,angle=-90]{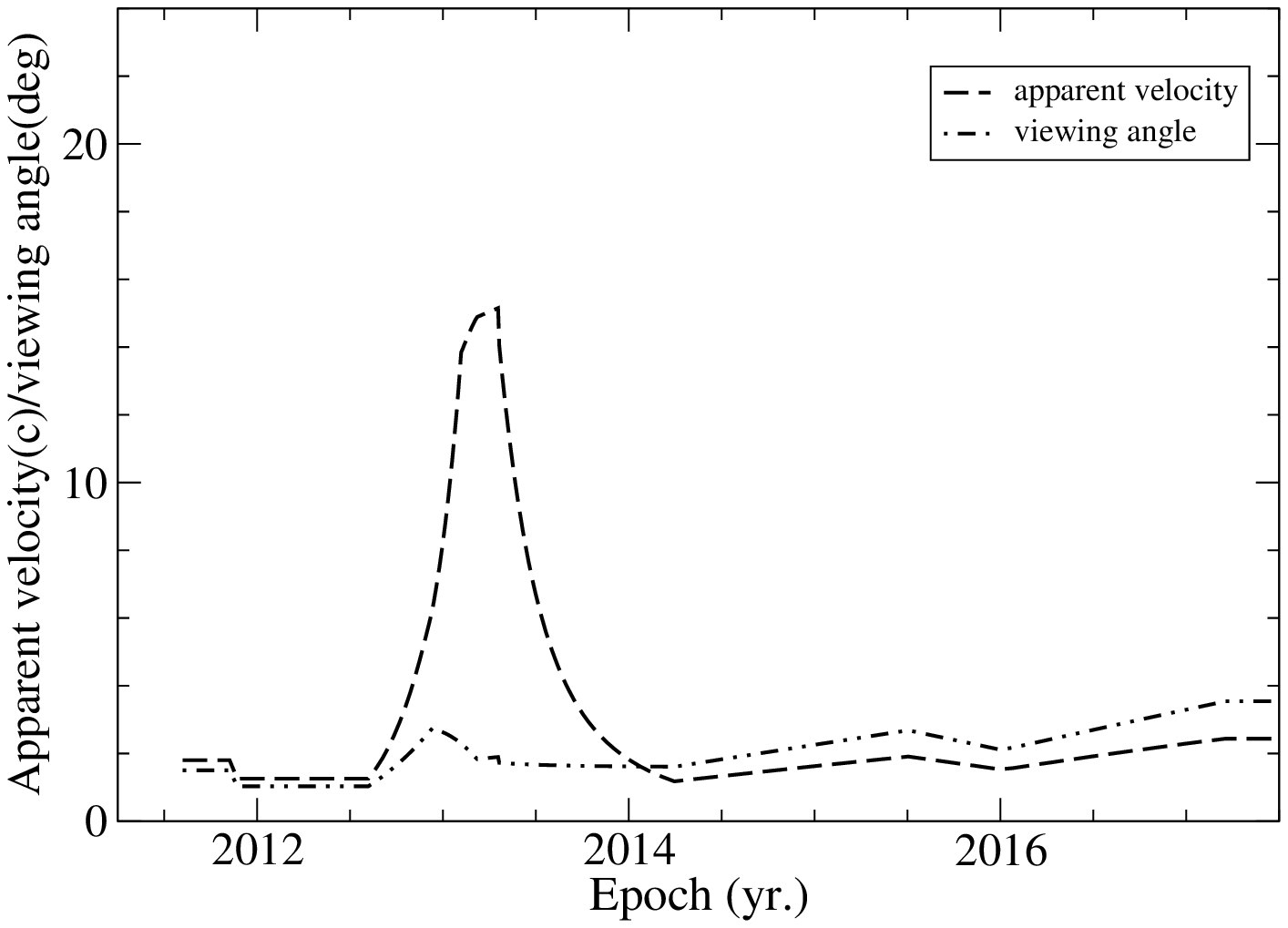}
       \includegraphics[width=5cm,angle=-90]{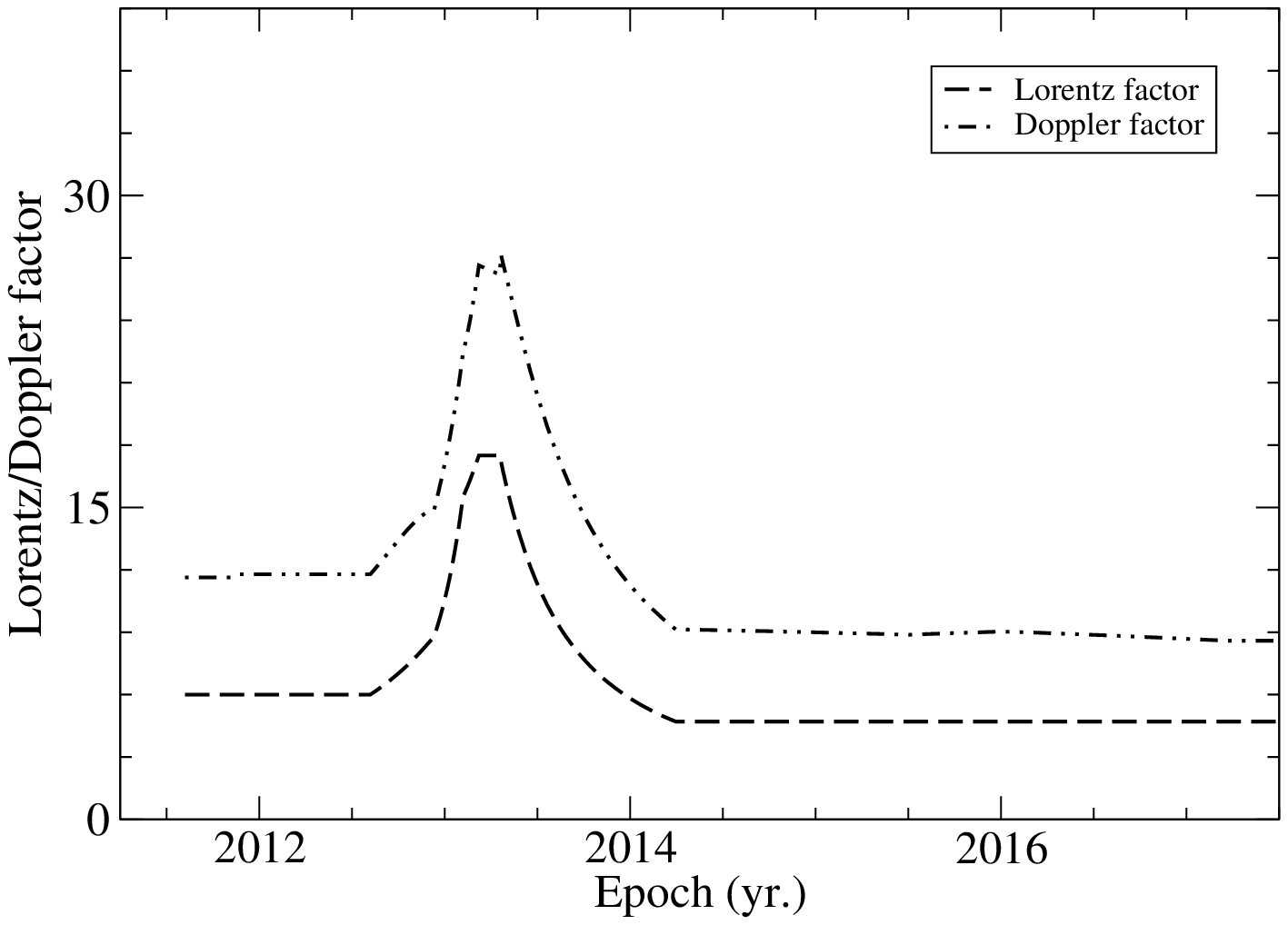}
       \caption{Knot B7. Left panel: the model-derived apparent speed 
      $\beta_{app}(t)$ and viewing angle $\theta(t)$ as  functions 
      of time. The apparent speed has a maximum $\beta_{app,max}$=15.15 at
      2013.30 and the corresponding viewing angle $\theta$=$1.90^{\circ}$
      and Lorentz factor $\Gamma$=17.50. Right panel: the model-derived  bulk
      Lorentz factor $\Gamma(t)$ and Doppler factor $\delta(t)$ as  continuous
      functions of time. The Doppler factor has a maximum  $\delta_{max}$=27.16
      at 2013.30 and the corresponding Lorentz factor $\Gamma$=17.23 and 
      viewing angle $\theta$=$1.72^{\circ}$.}
       \end{figure*}
     \section{Knot B7: Interpretation of kinematics and flux evolution}
     According to the precessing nozzle scenario for jet-B the ejection time
     of knot B7 is $t_0$=2011.60 and its corresponding precession phase
     $\phi_0$=1.30+4$\pi$. Its traveled distance Z(t) and the modeled
     parameters $\epsilon(t)$ and $\psi(t)$ as functions of time are shown in
     Figure 29. Before 2012.60 ($X_n{\leq}$0.036\,mas)
      $\epsilon$=$1.50^{\circ}$ and  $\psi$=$12.0^{\circ}$ knot B7 moved along
      the precessing common  trajectory, extending to the traveled distance 
      Z=2.0\,mas=13.3\,pc. After 2012.60 both parameters started to change and 
      knot B7 started to move along its own individual track, deviating from 
     the precessing common track. That is, the transition from the common track
     to its individual
     track occurred at ${r_n}{\sim}{X_n}$=0.036\,mas. Thus its flaring 
     associated with the Doppler boosting effect completely occurred in the 
     outer jet region. 
     \subsection{Knot B7: Model simulation of kinematics}
     The model fitting results of the entire trajectory $Z_n(X_n)$ and the
     core separation $r_n(t)$, coordinates $X_n(t)$ and $Z_n(t)$ as
     functions of time are shown in Figures 30 and 31. They
     are all well fitted during the four years (2013--2016), especially
     for the prominent decrease  in the core distance $r_n$ near epoch 2016.\\
     The model-derived apparent speed $\beta_{app}(t)$ and viewing angle
     $\theta(t)$ as continuous functions of time are shown in Figure 32 (left
     panel). The maximal apparent speed $\beta_{app,max}$=15.15 at 2013.30.\\
     The model-derived bulk Lorentz factor $\Gamma(t)$ and Doppler factor
     $\delta(t)$ as continuos functions are shown in Figure 32 (right panel).
     The maximal Doppler factor $\delta_{max}$=27.16 occurred at 2013.30,
      coincident with the maximal apparent speed. The 
     corresponding Lorentz factor $\Gamma$=17.5 (a maximum) and 
     $\theta$=$1.89^{\circ}$.
      \begin{figure*}
      \centering
      \includegraphics[width=5cm,angle=-90]{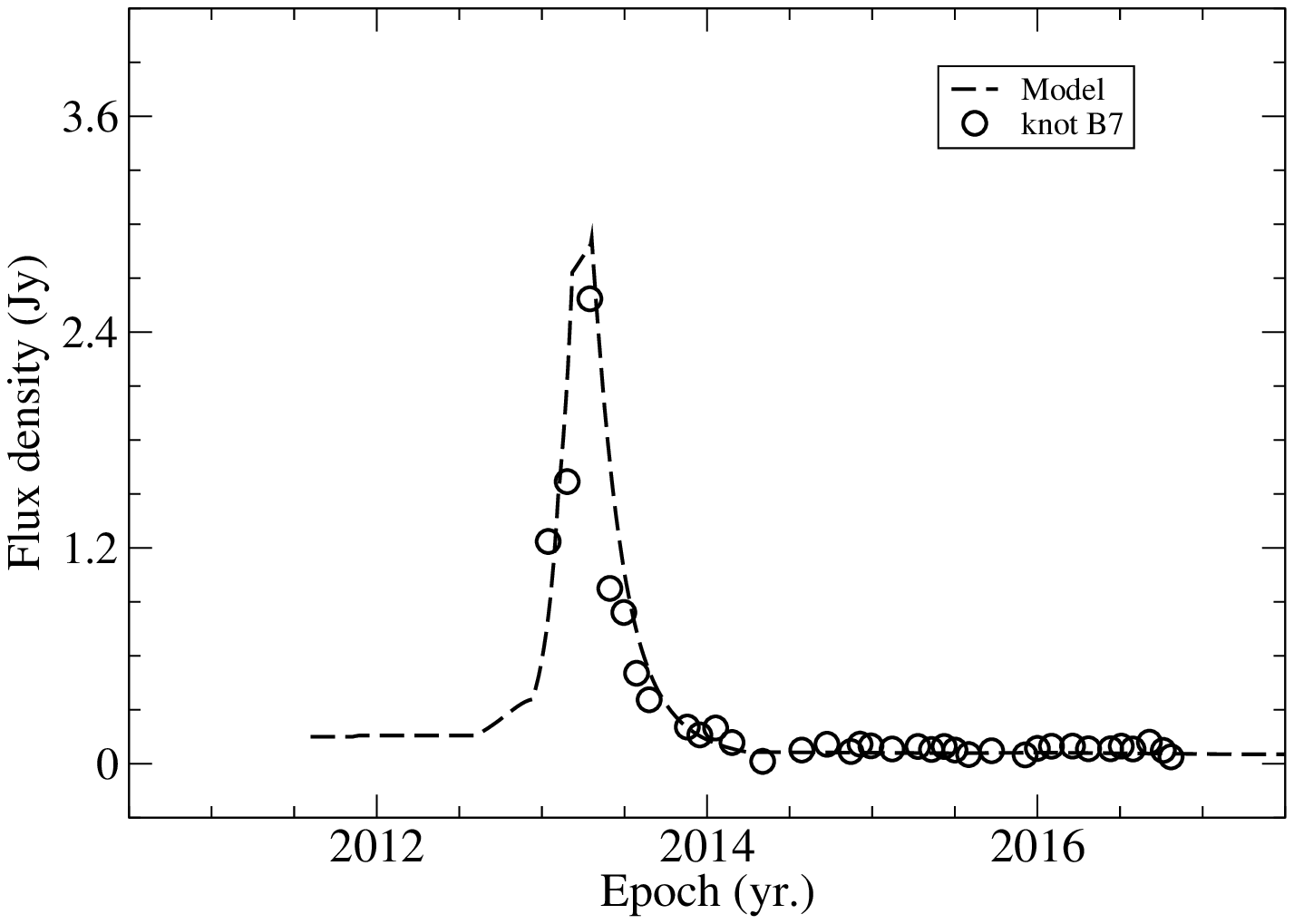}
      \includegraphics[width=5cm,angle=-90]{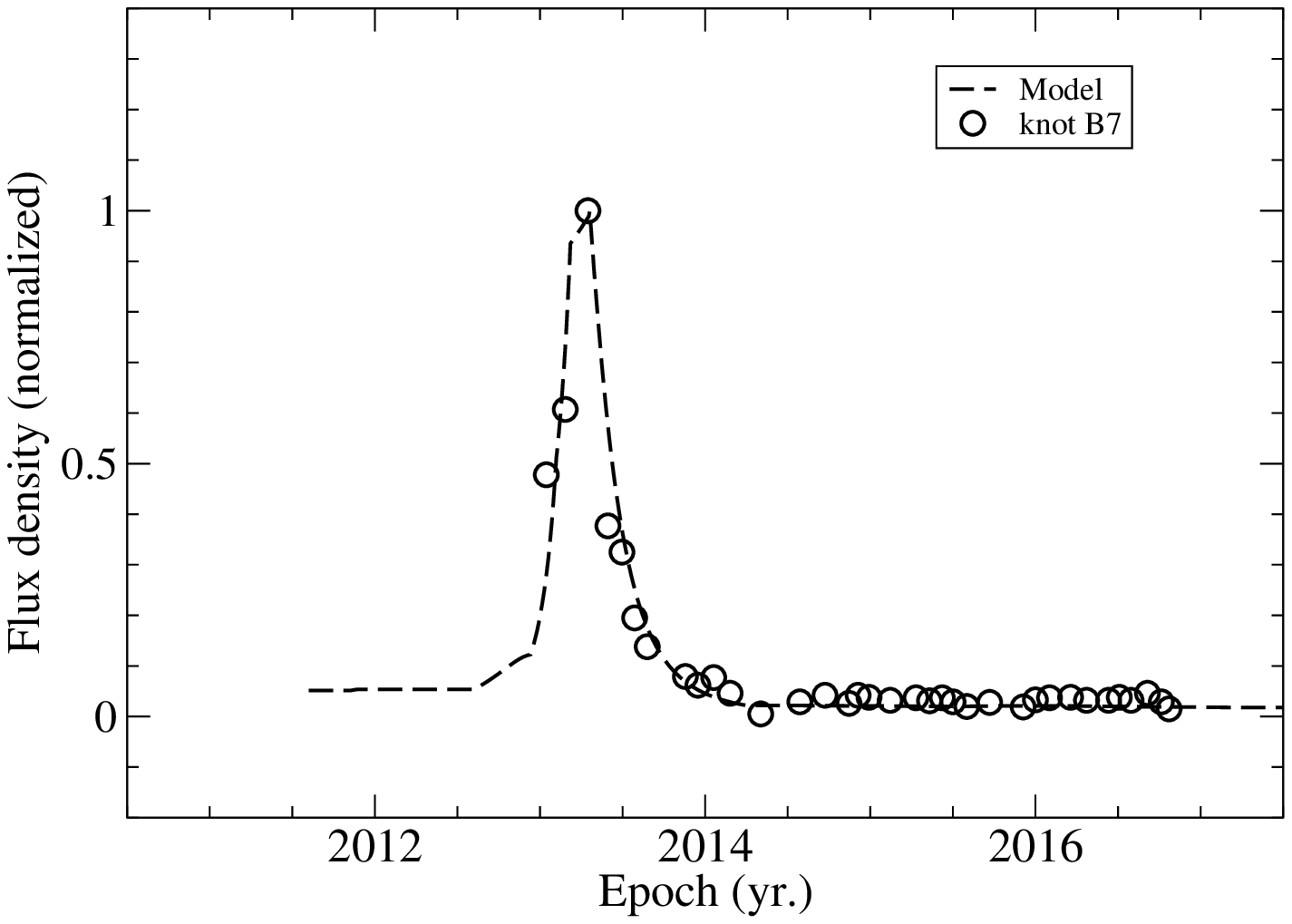}
      \caption{Knot B7: Model fitting of the 43\,GHz light curve (left panel)
     and its corresponding normalized light curve (right panel).
     They are well fitted by the model-derived Doppler boosting profile
    $S_{int}[\delta(t)]^{3+\alpha}$ and $[\delta(t)/\delta_{max}]^{3+\alpha}$,
     respectively. An intrinsic flux density $S_{int}$=28.0$\mu$Jy and a
     spectral index $\alpha$=0.50 are adopted.}
      \end{figure*}
     \subsection{Knot B7: Doppler boosting effect and flux evolution}
     The model fitting results  of the measured 43\,GHz light curve and its
     corresponding normalized light curve are shown in Figure 33. Both are
     well fitted by the Doppler-boosting profile 
     $S_{int}[\delta(t)]^{3+\alpha}$ and $[\delta(t)/\delta_{max}]^{3+\alpha}$, 
     respectively. An intirinsic flux density $S_{int}$=28.0$\mu$Jy and
     a spectral index $\alpha$=0.50 are adopted.
    \section{Conclusion}
    Based on the precessing nozzle scenario  proposed for jet-B in 3C345
    (Qian \cite{Qi22a}) we have model-fitted the flux evolution of five
     superluminal
     components (C19, C20, C21, B5 and B7) associated with their kinematics
    (accelerated/decelerated motion). Combining with the results for the five
     superluminal knots (C4, C5, C9, C10 and C22) of jet-A (Qian \cite{Qi22b}),
     we have interpreted the flux evolution as caused by the Doppler boosting
     effect for ten superluminal knots in 3C345.\\
     In fact, similar studies can be made for more superluminal knots in 3C345:
     e.g., knots C7, C11, C12 and C23 of jet-A  and C15--C18, B8, B11
    and B12 of jet-B. \\
     In addition, we have also applied our precessing nozzle scenario to 
     investigate the flux evolution of superluminal components associated with
     their Doppler boosting effect in QSO B1308+326 (Qian \cite{Qi23a}) and
     blazars 3C454.3 (Qian \cite{Qi23b}) and 3C279 (Qian \cite{Qi22b}). In 
    all these cases the flux evolution of the superluminal components can be 
   interpreted in terms of their Doppler boosting effect combined with their 
    intrinsic flux variations.\\
    We would like to note that there are signicant 
     differences in the application of the
     precessing nozzle scenario to QSOs and blazars. For QSOs [e.g., for
     B1308+326 (Qian et al. \cite{Qi17}, Qian \cite{Qi23a}), PG1302-102
     (Qian et al. \cite{Qi18a}) and NRAO 150 (Qian \cite{Qi16})]
       the precessing nozzle scenario has been 
    applied to study the model-fitting of the kinematics and flux evolution of 
    superluminal components ejected from a single precessing jet, while for
     blazars [e.g., 3C345 (Qian \cite{Qi22a}, \cite{Qi22b} and this paper);
     3C454.3 (Qian et al. \cite{Qi07}, \cite{Qi14}, \cite{Qi21}, 
    Qian \cite{Qi23b}); 3C279 (Qian \cite{Qi11}, \cite{Qi12}, \cite{Qi13}, 
    \cite{Qi22a});  OJ287 (Qian \cite{Qi15}, \cite{Qi18b}, \cite{Qi19},
     \cite{Qi20})] a double jet structure (jet-A plus jet-B)
    has been hypothetically introduced to study the kinematics and flux 
    evolution of their superluminal components, because their knots could be 
    separated into two groups (group-A and group-B) which need different 
    patterns of precessing common trajectory to  fit their kinematics, 
    respectively.\\
     It is worth noting the fact: Analysing the observational data in earlier 
     years (i.e. the observational data on the knots of group-A only), Klare
     (\cite{Kl05}), Lobanov \& Roland (\cite{Lo05}) and Qian et al. 
    (\cite{Qi09}) suggested 3C345 having a precessing jet, while analysing the 
    whole observational data on the knots of both group-A and group-B, 
    Schinzel et al. (\cite{Sc11a}) suggested no precession of the jet in 3C345.
    Thus the separation of the superluminal knots  
    into two groups may be a plausible assumption for investigating the jet 
    precession in 3C345, otherwise we would have to relinquish the assumption
    of jet precession for 3C345.\\
      The full interpretation of the kinematics and flux evolution of
    the superluminal components in 3C345 demonstrates that our precessing 
    nozzle scenario proposed for blazar 3C345 can interpret 
    the kinematic, dynamic and emission properties of its superlumiunal
    components, although the assumption of a double jet structure is only
     a working hypothesis.\\
    We have been trying to propose a scenario with a single precessing
    jet, as an significant alternative, to interpret the kinematics of 
    the superluminal knots in 3C345 as a whole, because there seems a 
    possibility that the knots of group-A and group-B can be combined into
    one group. For example,  we may assume that the trajectory-pattern 
     transition of the knots in the jet-B could occur at core distances 
    $r_n{<}$0.1\,mas and their innermost tracks could connect with
    the precessing common trajectory of jet-A at core distances 
   $r_n{<}$0.10--0.05\,mas. In such a scenario
    the observed tracks of the knots of group-B are only the outer 
    trajectories beyond their trajectory-transitions, while their innermost
    tracks within the transitions ($r_n{\leq}$0.05--0.1\,mas) have not been
    observed. The trajectory-transition observed in the core-distance range of
    $\sim$0.1--0.4\,mas in knot C8 (Qian \cite{Qi22a}) is an extremely
     instructive example. In this case the precession of the jet may be 
     related to the precession of the disk around the black hole in the nucleus
    of 3C345. This kind of scenario with a single precessing jet nozzle
     should be tested by the VLBI observations with higher resolutions in 
   the future. 
   % \acknowledgements
    

\begin{thebibliography}{}
     \bibitem[1995]{Ba95}
     Babadzhanyants M.K., Belokon E.T. \& Gamm N.G., 1995,  Astronomy 
     Report 39, 393
     \bibitem[1986]{Bi86}
     Biretta J.A., Moore R.L. \& Cohen M.H., 1986, ApJ 308, 93
     \bibitem[1987]{Ha87}
     Hardee P.E., 1987, ApJ 318, 78 
     \bibitem[1999]{Ho99}
      Hogg D.W., 1999, astro-ph/9905116
    % \bibitem[2017]{Jo17}
    % Jorstad S.G., Marscher A.P., Morozova D.A., et al., 2017, ApJ 846,
    % article id.98
     \bibitem[2005]{Jo05}
     Jorstad S.G., Marscher A.P., Lister M.L., et al., 2005, AJ 130, 1418
    % \bibitem[1991a]{Qi91a}
    % Qian S.J., Witzel A., Krichbaum T.P., et al. 1991a, Acta Astron. Sin.
    %   32,309
     \bibitem[2003]{Kl03}
     Klare J., 2003, Quasi-Periodicity in the Parsec-Scale Jet of the Quasar
     3C345, PhD Thesis, Rheinische-Friedrich-Wilhelms-Universit\"at Bonn, 
     Bonn, Germany 
     \bibitem[2005]{Kl05}
     Klare J., Zensus J.A., Lobanov A.P., et al., 2005, in "Future Directions
     in High resolution Astronomy: The 10th Anniversary of the VLBA", ASP
     Conference Series , Vol.340 (eds., J.D.~Romney and M.J.~Reid), p.40 
     \bibitem[2005]{Lo05}
     Lobanov A.P. \&  Roland J., 2005, A\&A 431, 831
    \bibitem[1999]{Lo99}
    Lobanov A.P. \& Zensus J.A., 1999, ApJ 521, 509
    \bibitem[1994]{Lo94}
    Lobanov A.P. \& Zenzus J.A., 1994, in Proc. Second EVN/JIVE Symposium, ed.
    A.J.~Kus, R.T.~Schilizzi, K.M.~Borkovski \& L.I.~Gurvits, 93
     \bibitem[1994]{Ma94}
     Malina R.F., et al., 1994, AJ 107, 751
     \bibitem[1984]{Mo84}
     Moore R.L. \& Stockman H.S., 1984, ApJ 279, 465
   %  \bibitem[1987]{Pa87}
   %  Pauliny-Toth I.I.K., Porcas R.W., Zensus J.A., et al., 1987, Nature 328,
   %  778  
     \bibitem[1991a]{Qi91a}
     Qian S.J., Witzel A., Krichbaum T.P., et al., 1991a, Acta Astron. Sin. 32,
     369 (Chin. Astro. Astrophys. 16, 137 (1992))
     \bibitem[1991b]{Qi91b}
     Qian S.J., Krichbaum T.P., Witzel A., et al., 1991b, in: High Energy
     Astrophysics: Compact Stars and Active Galaxies (Proceedings of the 3rd
     Chinese Academy of Sciences and Max-Planck Society Workshop, held 19-23,
     October 1990 in Huangshan, China, edited by Qibin Li, Singapore:
     World  Scientific), p.80
     \bibitem[1996]{Qi96}
     Qian S.J., Krichbaum T.P., Zensus J.A., et al., 1996, A\&A 308, 395
     \bibitem[2007]{Qi07}
     Qian S.J., Kudryavtzva N.A., Britzen S., et al., 2007, Chin. J. Astrophys.
     7, 364
     \bibitem[2009]{Qi09}
     Qian S.J., Witzel A., Zensus J.A., et al., 2009, Research in Astron.
     Astrophys. 9, 137
     \bibitem[2011]{Qi11}
     Qian S.J., 2011, Research in Astron. Astrophys. 11, 43
     \bibitem[2012]{Qi12}
     Qian S.J., 2012, Research in Astron. Astrophys. 12, 46
     \bibitem[2013]{Qi13}
     Qian S.J., 2013,  Research in Astron. Astrophys. 13, 783
    \bibitem[2014]{Qi14}
     Qian S.J., Britzen S., Witzel A., et al., 2014, Research in 
     Astron. Astrophys. 14, 249
    \bibitem[2015]{Qi15}
    Qian S.J., 2015, Research in Astron. Astrophys. 15, 687 
    \bibitem[2016]{Qi16}
    Qian S.J., 2016, Research in Astron. Astrophys., 16, 20
    \bibitem[2017]{Qi17}
    Qian S.J., Britzen S., Witzel A., et al., 2017, A\&A 604, A90
    \bibitem[2018a]{Qi18a}
    Qian S.J., Britzen S., Witzel A., et al., 2018a, A\&A 615, A123
    \bibitem[2018b]{Qi18b}
    Qian S.J., 2018b,  arXiv e-prints, arXiv:1811.11514
    \bibitem[2019]{Qi19}
    Qian S.J., Britzen S., Krichbaum T.P., Witzel A., 2019,  A\&A 621, A11
    \bibitem[2020]{Qi20}
    Qian S.J., 2020, arXiv e-prints, arXiv:2005.05517
    \bibitem[2021]{Qi21}
    Qian S.J., Britzen S., Krichbaum T.P., Witzel A., 2021, A\&A 653, A7
     \bibitem[2022a]{Qi22a}
     Qian S.J., 2022a, arXiv e-prints, arXiv:2202.01915
     \bibitem[2022b]{Qi22b}
     Qian S.J., 2022b, arXiv e-prints, arXiv:2206.14995
     \bibitem[2023a]{Qi23a}
     Qian S.J., 2023a, arXiv e-prints, arXiv:2306.05619
     \bibitem[2023b]{Qi23b}
     Qian S.J., 2023b, arXiv e-prints, arXiv:2306.06863
     \bibitem[2000]{Ro00}
     Ros E., Zensus J.A. \& Lobanov A.P., 2000, A\&A 354, 55
    % \bibitem[2012]{Sc12}
    % Schinzel F.K., Lobanov A.P., Tayor G.B., et al., 2012, A\&A 537, A70
     \bibitem[2011a]{Sc11a}
     Schinzel F.K., 2011a, PhD thesis, University of Cologne
     \bibitem[2011b]{Sc11b}
     Schinzel F.K., Sokolovsky K.V., D'Ammando F., et al., 2011b, A\&A 532,
     A150
     \bibitem[2010a]{Sc10a}
     Schinzel F.K., Lobanov A.P. \& Zensus J.A., 2010a, in: Accretion and
     Ejection in AGNs: A Global View,  ASP Conference Series, Vol.427 (2010),
     eds. L.~Maraschi, G.~Ghisellini, R.~Della Ceca \& F.~Tavecchio
     \bibitem[2010b]{Sc10b}
     Schinzel F.K., Lobanov A.P., Jorstad S.G., et al., 2010b: in
      "Fermi meets Jansky--AGN in Radio and Gamma Rays",
      eds. T.~Savolainen, E.~Ros,  R.W.~Porcas \& J.A.~Zensus
     \bibitem[1993]{Sch93}
     Schramm R.-J., Borgeest U., Camenzind M., et al., 1993, A\&A
     \bibitem[2003]{Sp03}
     Spergel D.N., Verde L., Peilis H.V., et al., 2003, ApJS 148, 145
     \bibitem[1995]{St95}
     Steffen W., Zensus J.A., Kirchbaum T.P., Witzel A., Qian S.J., 1995,
     A\&A 302, 335 
     \bibitem[2004]{Wa04}
     Wang J.M., Luo B., Ho L.C., 2004, ApJ 615, L9
     \bibitem[1997]{Ze97}
     Zensus J.A., 1997, ARA\&A 35, 607
     \bibitem[1995]{Ze95}
     Zensus J.A., Cohen M.H., Unwin S.C., 1995, ApJ 443, 35
     \end{thebibliography}
   \end{document}